\pdfoutput=1
\pdfoutput=1
\documentclass[aps,amsmath,amssymb,preprintnumbers,preprint,nofootinbib,a4paper,11pt]{article}
\usepackage{epsfig}
\usepackage{jheppub,undertilde}
\usepackage{amsthm}
\usepackage{graphicx}
\usepackage{color}
\usepackage{graphicx,textcomp,float,slashed}
\usepackage{arydshln}
\usepackage{empheq}
\newcommand{\be}{\begin{equation}}
\newcommand{\ee}{\end{equation}}
\newcommand\sss{\scriptscriptstyle}
\newcommand{\mW}{m_{\sss W}}
\newcommand{\sW}{s_{\sss W}}
\newcommand{\cW}{c_{\sss W}}
\newcommand{\hc}{+\,\mathrm{h.c.}}
\def\bsp#1\esp{\begin{split}#1\end{split}}
\def\bpm{\begin{pmatrix}}
\def\epm{\end{pmatrix}}

\newcommand{\bea}{\begin{eqnarray}}  
\newcommand{\eea}{\end{eqnarray}}  
 \def\bsp#1\esp{\begin{split}#1\end{split}}
\preprint{LTH 1039}

\title{ \huge Benchmarks for Higgs Effective Theory: Extended Higgs Sectors}

\author[1]{Martin Gorbahn}
\author[2]{Jose Miguel No}
\author[2]{and Ver\'onica Sanz}

\affiliation[1]{Department of Mathematical Sciences, University of Liverpool, Liverpool, L69 7ZL, United Kingdom}
\affiliation[2]{Department of Physics and Astronomy, University of Sussex, Brighton BN1 9QH, United Kingdom}
\emailAdd{Martin.Gorbahn@liverpool.ac.uk}
\emailAdd{J.M.No@sussex.ac.uk}
\emailAdd{v.sanz@sussex.ac.uk}

\abstract{Precise measurements of SM particles properties at the LHC allows to look for heavy New Physics in the context of an 
Effective Field Theory (EFT). These searches, however, often rely on kinematic regions where the validity of the EFT may be compromised. In this paper 
we propose to address this issue by comparing with benchmark models. The connection between models and their manifestations as EFTs at low energies allows us to 
quantify the breakdown of the EFT, and describe ways to combine different sources of constraints beyond Higgs physics. To illustrate these techniques, 
in this paper we propose a set of benchmark models based on extensions of the Higgs sector, namely the inclusion of a singlet, a dilaton and generic 2HDMs. 
We obtain the matching between these models and the EFT involving the Higgs, electroweak bosons and fermions. We then describe current and future indirect and direct 
constraints, consider the effect of correlations among the coefficients within models, and discuss the 
validity of the EFT.}

\begin{document}
\maketitle
\flushbottom

%%%%%%%%%%%%%%%%%%%%%%%%%%%%%%
\section{Introduction}
%%%%%%%%%%%%%%%%%%%%%%%%%%%%%%

The discovery of a Higgs boson \cite{Aad:2012tfa,Chatrchyan:2012ufa} has opened the Era of Higgs Physics. An intense effort is now devoted to measure the properties of this new particle $h$ and 
determine whether it is {\it the} Standard Model (SM) Higgs boson \cite{Higgs:1964pj,Higgs:1964ia,Higgs:1966ev,Weinberg:1967tq}. So far, no evidence for new physics beyond the SM has been observed,
which suggests the use of an Effective Field Theory (EFT) approach, that assumes possible new physics beyond the SM modifying the Higgs interactions to be heavy, 
with a typical scale $\Lambda \sim$ TeV. The effects of new physics are then parametrized through higher-dimensional operators constructed from the SM fields,
the leading operators appearing at dimension-six \cite{Burges:1983zg,Leung:1984ni,Buchmuller:1985jz,Hagiwara:1993ck,Giudice:2007fh,Grzadkowski:2010es}. Some of these operators which affect Higgs properties 
can be measured in Higgs physics only, while others are also related to electroweak (EW) observables since the Higgs scalar excitation is always associated with the EW symmetry breaking order parameter $v$. 
The experimental data from LEP and Tevatron constrain the size of the Wilson coefficients of these operators, and more recently the results from the LHC experiments ATLAS and CMS 
allow also to constrain the Wilson coefficients associated to the operators that impact Higgs physics \cite{Corbett:2012dm,SUSYEFT,Masso:2012eq,Corbett:2012ja,Dumont:2013wma,Pomarol:2013zra, ESY2014a,Gupta:2014rxa,ESY2014b}. 

The LHC energy reach allows to go beyond the analysis of Higgs signal strengths as a probe of new physics beyond the SM, exploiting the information
encoded in kinematical distributions to boost the sensitivity to new physics beyond the SM in the Higgs sector. This has the power to strongly constrain the 
presence of new physics which lead to a significant enhancement of the signal in certain kinematical regions, typically at high energy. However, while the bounds on the Wilson coefficients 
of these dimension-six operators which can be extracted from these measurements are indeed very strong even with the limited amount of present data,
the validity of the EFT approach in the kinematical regions which dominate these constraints is questionable \cite{Biekoetter:2014jwa,Englert:2014cva} (see also \cite{ESY2014b}).
This would render these constraints meaningless in the context of EFTs. It is therefore crucial to assess in detail
the validity of this approach, and a way to do so is by quantitatively studying the breakdown of the EFT by comparing its predictions with 
those of possible UV completions.

Among possible UV completions, extended scalar sectors provide an attractive arena for the use of a SM effective theory description. 
Extensions of the SM Higgs sector constitute a simple yet very well-motivated scenario beyond the SM, with important consequences not only for phenomenology but also for EW cosmology, baryogenesis and dark matter. It is plausible that the new scalar states, which we assume to be significantly heavier than the Higgs boson discovered by ATLAS and CMS, cannot be 
detected directly at LHC either due to them being very heavy or having vanishingly small couplings to SM gauge bosons if the light Higgs is SM-like.
However, it may be possible to measure the effect of these new particles via EW precision observables (EWPO) or using the kinematical information in LHC measurements. 
An EFT of extended Higgs sectors provides a very useful tool to study these effects systematically under the assumption that the new states are significantly heavier than $h$. 
Moreover, since extensions of the SM Higgs sector provide a very simple UV completion to such an EFT, this makes it possible to compare the predictions and bounds derived from the EFT with
those of its UV completion, therefore probing the range of validity of the EFT. This allows us to assess the reliability of the LHC constraints drawn from the high-energy
kinematical regions, as well as those from EWPO, and also to analyze the LHC potential for indirectly probing extended Higgs sectors.

The paper is organized as follows: In Section 2 we introduce the $D=6$ effective operators which are relevant for the analysis. In Section 3 we present the matching of these operators with UV models corresponding to extensions of the SM Higgs sector with an extra scalar singlet, with an extra scalar doublet (a Two-Higgs-Doublet-Model) and with a radion/dilaton, and discuss their experimental constraints both from the UV model and from the EFT point of view. In Section 4 we discuss the validity of the EFT in the light of these scenarios. Finally, in Section 5 we summarize our results and discuss their implications. In the Appendices we describe the connection to the various possible $D=6$ Lagrangian terms after EW symmetry breaking, parametrized as anomalous Higgs couplings, and give details on the 2HDM relevant to our analysis.

%%%%%%%%%%%%%%%%%%%%%%%%%%%%%%
\section{Effective Field Theory for the SM Higgs Field}
\label{section2}
%%%%%%%%%%%%%%%%%%%%%%%%%%%%%%

The Lagrangian for the Standard Model can be supplemented by higher-dimensional operators that parametrize the effects of new physics beyond the SM appearing at energies much larger than an effective scale which may identified with the {\it vev} of the Higgs field $v$. Considering only operators of dimension $D \leq 6$ and assuming baryon and lepton number conservation, the most general 
$SU(3)_C \times SU(2)_L \times U(1)_Y$ gauge invariant Lagrangian ${\cal L}_{\rm Eff}$ constructed out of the SM fields has been known for a long time \cite{Burges:1983zg,Leung:1984ni,Buchmuller:1985jz},
and may be mapped into various alternative bases of $D \leq 6$ independent SM effective operators \cite{Hagiwara:1993ck,Giudice:2007fh,Grzadkowski:2010es} (see also the discussion in \cite{Willenbrock:2014bja}). 

Each experimental measurement will in general constrain only a handful of effective operators of a certain basis. In combining the available constraints from analyses of various physical processes (via {\it e.g.} a global fit), some basis choices will be more appropriate than others\footnote[1]{Depending on the purposes, bases different from the one we adopt here might be more adequate.}. As compared to those of \cite{Hagiwara:1993ck,Grzadkowski:2010es}, the basis of operators ${\cal O}_i$ from \cite{Giudice:2007fh} is well-suited for analyses of Higgs properties in combination with precision measurements of EW observables. In this basis, the SM effective Lagrangian ${\cal L}_{\rm Eff}$ may be expressed as
\be
\label{eq:effL}
  {\cal L}_{\rm Eff} = {\cal L}_{\rm SM}  + \sum_i \bar c_i {\cal O}_i = {\cal L}_{\rm SM} + {\cal L}_{\rm SILH} 
   + {\cal L}_{G} + {\cal L}_{CP} + {\cal L}_{F_1} + {\cal L}_{F_2} + {\cal L}_{F_3}\ .
\ee
We adopt here the decomposition in \cite{Contino:2013kra,benj} and normalize the Wilson coefficients $\bar c_i$ accordingly. 
The term ${\cal L}_{\rm SILH}$ of (\ref{eq:effL}) corresponds to a certain set of $CP$-conserving operators involving the Higgs doublet $\Phi$, inspired by scenarios 
where the Higgs is part of a strongly interacting sector \cite{Giudice:2007fh}
\be
\label{eq:silh}
\bsp
  {\cal L}_{\rm SILH} &=\
    \frac{\bar c_{\sss H}}{2 v^2} \partial^\mu\big[\Phi^\dag \Phi\big] \partial_\mu \big[ \Phi^\dagger \Phi \big]
  + \frac{\bar c_{\sss T}}{2 v^2} \big[ \Phi^\dag {\overleftrightarrow{D}}^\mu \Phi \big] \big[ \Phi^\dag {\overleftrightarrow{D}}_\mu \Phi \big] 
  - \frac{\bar c_{\sss 6} \lambda}{v^2} \big[\Phi^\dag \Phi \big]^3 \\
  &\
  + \frac{i g\ \bar c_{\sss W}}{\mW^2} \big[ \Phi^\dag T_{2k} \overleftrightarrow{D}^\mu \Phi \big]  D^\nu  W_{\mu \nu}^k
  + \frac{i g'\ \bar c_{\sss B}}{2 \mW^2} \big[\Phi^\dag \overleftrightarrow{D}^\mu \Phi \big] \partial^\nu  B_{\mu \nu} \\
  &\
  + \frac{2 i g\ \bar c_{\sss HW}}{\mW^2} \big[D^\mu \Phi^\dag T_{2k} D^\nu \Phi\big] W_{\mu \nu}^k
  + \frac{i g'\ \bar c_{\sss HB}}{\mW^2}  \big[D^\mu \Phi^\dag D^\nu \Phi\big] B_{\mu \nu} \\
  &\
  +\frac{g'^2\ \bar c_{\gamma}}{\mW^2} \Phi^\dag \Phi B_{\mu\nu} B^{\mu\nu}
  +\frac{g_s^2\ \bar  c_{g}}{\mW^2} \Phi^\dag \Phi G_{\mu\nu}^a G_a^{\mu\nu} \\ 
  &\
  -\left[ \frac{\bar c_{u} \, y_u}{v^2} \Phi^\dag \Phi \, {\bar Q}_L \Phi^{\dagger} u_R
  +\frac{\bar c_{d} \, y_d}{v^2} \Phi^\dag \Phi \, {\bar Q}_L \Phi\, d_R 
  +\frac{\bar c_{l} \, y_l}{v^2} \Phi^\dag \Phi \, {\bar L}_L \Phi\, l_R \right]\ .
\esp
\ee
Here $\lambda$ stands for the Higgs quartic coupling, $g'$, $g$ and $g_s$ are respectively the $U(1)_Y$, $SU(2)_L$ and $SU(3)_C$ coupling constants
and $T_{2k} \equiv \sigma_k/2$ are the generators of SU(2) in the fundamental representation ($\sigma_k$ being the Pauli matrices). The Hermitian derivative operator ${\overleftrightarrow D}_\mu$ is defined as 
\be
\Phi^\dag {\overleftrightarrow D}_\mu \Phi = \Phi^\dag D_\mu \Phi - (D_\mu\Phi^\dag) \Phi\, ,
\ee
and our conventions for the gauge-covariant derivatives and field-strength tensors (following \cite{benj}) are
\bea
B_{\mu\nu} & = &\partial_{\mu}B_{\nu} - \partial_{\nu}B_{\mu} \nonumber \\
W^k_{\mu\nu} & = &\partial_{\mu}W^k_{\nu} - \partial_{\nu}W^k_{\mu} + g\,\epsilon_{ij}^{\,\,\,\,k} \,W^i_{\mu} \,W^j_{\nu}  \nonumber \\
G^a_{\mu\nu} & = &\partial_{\mu}G^a_{\nu} - \partial_{\nu}G^a_{\mu} + g_s\,f_{bc}^{\,\,\,\,a} \,G^b_{\mu} \,G^c_{\nu}  \nonumber \\
D_{\mu}\Phi & = &\partial_{\mu}\Phi - i\, g\,T_{2k} W^k_{\mu} \Phi - \frac{i}{2} \, g' \, B_{\mu} \Phi  \\
D_{\rho}W^k_{\mu\nu} & = &\partial_{\rho}\partial_{\mu}W^k_{\nu} - \partial_{\rho}\partial_{\nu}W^k_{\mu}  + g\,\epsilon_{ij}^{\,\,\,\,k} \,\partial_{\rho}\left( W^i_{\mu} \,W^j_{\nu}\right) \nonumber \\
& & + g\,\epsilon_{ij}^{\,\,\,\,k} \,W^i_{\rho}\left( \partial_{\mu}W^j_{\nu} - \partial_{\nu}W^j_{\mu}\right) - g^2\,W_{\rho\, i} \left( W^i_{\mu} \,W^k_{\nu} - W^i_{\nu} \,W^k_{\mu}\right) \nonumber \\
D_{\rho}G^a_{\mu\nu} & = &\partial_{\rho}\partial_{\mu}G^a_{\nu} - \partial_{\rho}\partial_{\nu}G^a_{\mu}  + g_s\,f_{bc}^{\,\,\,\,a} \,\partial_{\rho}\left( G^b_{\mu} \,G^c_{\nu}\right) \nonumber \\
& & + g_s\,f_{bc}^{\,\,\,\,a} \,G^b_{\rho}\left( \partial_{\mu}G^c_{\nu} - \partial_{\nu}G^c_{\mu}\right) - g_s^2\,G_{\rho\, b} \left( G^b_{\mu} \,G^a_{\nu} - G^b_{\nu} \,G^a_{\mu}\right)\, ,  \nonumber
\eea
with $\epsilon_{ij}^{\,\,\,\,k}$ and $f_{bc}^{\,\,\,\,a}$ being respectively the structure constants for $SU(2)$ and $SU(3)$. The term ${\cal L}_{\rm G}$ consists of operators not directly connected to Higgs physics, but that affect the gauge sector through modifications of the gauge boson self-energies and self-interactions,
\be
\bsp
  {\cal L}_{G} = &\
    \frac{g^3\ \bar c_{\sss 3W}}{\mW^2} \epsilon_{ijk} W_{\mu\nu}^i W^\nu{}^j_\rho W^{\rho\mu k}
  + \frac{g_s^3\ \bar c_{\sss 3G}}{\mW^2} f_{abc} G_{\mu\nu}^a G^\nu{}^b_\rho G^{\rho\mu c}
  + \frac{g^2\ \bar c_{\sss 2W}}{\mW^2} D^\mu W_{\mu\nu}^k D_\rho W^{\rho\nu}_k \\
 &\
  + \frac{g'^2 \bar c_{\sss 2B}}{\mW^2} \partial^\mu B_{\mu\nu} \partial_\rho B^{\rho\nu}
  + \frac{g_s^2\ \bar c_{\sss 2G}}{\mW^2} D^\mu G_{\mu\nu}^a D_\rho G^{\rho\nu}_a\ .
\esp\label{eq:lagG}
\ee
The term ${\cal L}_{\rm CP}$ in (\ref{eq:effL}) supplements ${\cal L}_{\rm SILH}$ and ${\cal L}_{G}$ with a set of CP-violating operators
\be
\label{eq:silhCPodd}
\bsp
  {\cal L}_{CP} = &\
    \frac{i g\ \tilde c_{\sss HW}}{\mW^2}  D^\mu \Phi^\dag T_{2k} D^\nu \Phi {\widetilde W}_{\mu \nu}^k
  + \frac{i g'\ \tilde c_{\sss HB}}{\mW^2} D^\mu \Phi^\dag D^\nu \Phi {\widetilde B}_{\mu \nu}
  + \frac{g'^2\  \tilde c_{\gamma}}{\mW^2} \Phi^\dag \Phi B_{\mu\nu} {\widetilde B}^{\mu\nu}\\
 &\
  +\!  \frac{g_s^2\ \tilde c_{g}}{\mW^2}      \Phi^\dag \Phi G_{\mu\nu}^a {\widetilde G}^{\mu\nu}_a
  \!+\!  \frac{g^3\ \tilde c_{\sss 3W}}{\mW^2} \epsilon_{ijk} W_{\mu\nu}^i W^\nu{}^j_\rho {\widetilde W}^{\rho\mu k}
  \!+\!  \frac{g_s^3\ \tilde c_{\sss 3G}}{\mW^2} f_{abc} G_{\mu\nu}^a G^\nu{}^b_\rho {\widetilde G}^{\rho\mu c} \ ,
\esp
\ee
with the dual field strength tensors defined by
\be
  \widetilde B_{\mu\nu} = \frac12 \epsilon_{\mu\nu\rho\sigma} B^{\rho\sigma} \ , \quad
  \widetilde W_{\mu\nu}^k = \frac12 \epsilon_{\mu\nu\rho\sigma} W^{\rho\sigma k} \ , \quad
  \widetilde G_{\mu\nu}^a = \frac12 \epsilon_{\mu\nu\rho\sigma} G^{\rho\sigma a} \ .
\ee
Finally, there are further operators contained in ${\cal L}_{F_1}$ -operators involving two Higgs fields and a pair of quarks/leptons-, 
${\cal L}_{F_2}$ -operators involving one Higgs field, a gauge boson and a pair of quarks/leptons- and ${\cal L}_{F_3}$ -four-fermion operators- (see {\it e.g.} \cite{Contino:2013kra,benj}).
These effective operators are nevertheless not present at leading order in the extensions of the SM we consider in the present work\footnote[2]{An exception are certain operators in ${\cal L}_{F_3}$ 
which do get generated in Two-Higgs-Doublet-Model extensions of the SM. These are however proportional to $Y_a^2$ (with $Y_a$ a fermion Yukawa coupling), being then negligible for $1^{\mathrm{st}}$ and 
$2^{\mathrm{nd}}$ fermion generations, and so are essentially unconstrained.}, and so we do not discuss them in the following. 
Furthermore, we consider CP conserving scenarios for the time being, leaving an analysis of ${\cal L}_{CP}$ for the future \cite{GNS_CP}. 

%We are primarily interested in effects involving the Higgs field $\Phi$ and the SM gauge bosons, but no fermions, and thus we may disregard 
%the last three operators in (\ref{eq:silh}), as well as ${\cal L}_{F_1}$ (operators involving two Higgs fields and a pair of quarks/leptons) 
%${\cal L}_{F_2}$ (operators involving one Higgs field, a gauge boson and a pair of quarks/leptons) and ${\cal L}_{F_3}$ 
%(four-fermion operators)\footnote[2]{We stress that these operators may be disregarded for the physical processes we analyze due to the specific basis chosen. As explained {\it e.g.}, in \cite{Willenbrock:2014bja}, certain operators involving fermions may be very relevant for the analysis of EW observables in other bases, such as \cite{Grzadkowski:2010es}.}. We will
%nevertheless comment on their impact in the context of extended Higgs sectors in section \ref{Sec_MSSM}.

\vspace{3mm}

After EW symmetry breaking, we can write the SM effective Lagrangian ${\cal L}_{\rm Eff}$ in the unitarity gauge and in the mass basis, with 
\be
W^{\pm}_{\mu} = \frac{1}{\sqrt{2}}(W^{1}_{\mu} \mp i\,W^{2}_{\mu}) \quad \quad 
\left( 
\begin{array}{c}
Z_{\mu} \\
A_{\mu}
\end{array}
\right) = 
\left( 
\begin{array}{cr}
c_{\sss W} & - s_{\sss W}\\
s_{\sss W} & c_{\sss W}
\end{array}
\right)
\left( 
\begin{array}{c}
W^3_{\mu} \\
B_{\mu}
\end{array}
\right)
\quad \quad \Phi = \frac{1}{\sqrt{2}} \left( 
\begin{array}{c}
0 \\
v + h
\end{array}
\right)
\ee
with $s_{\sss W}$ ($c_{\sss W}$) being the sine (cosine) of the weak mixing angle at tree level. 
In Appendix~\ref{Appendixc}, we detail the relation of the $D=6$ terms presented here with the possible anomalous Higgs couplings after EW symmetry breaking.  

\vspace{3mm}

We are now ready to move onto relating the EFT to specific UV completions. In this work we have chosen to focus on the matching of the Wilson coefficients onto models with extended Higgs sectors, where the effects of New Physics are more apparent in deviations of the couplings of the Higgs and electroweak bosons. We describe the results of the matching, as well as the relation between
these models and the EFT in the next section.

%%%%%%%%%%%%%%%%%%%%%%%%%%%%%%   
 \section{Higgs EFT from Extended Higgs Sectors}
%%%%%%%%%%%%%%%%%%%%%%%%%%%%%%

As discussed in the Introduction, an important aspect of the $D \leq 6$ SM effective theory from the previous section is its energy range of validity. 
This is a key issue if one is to reliably confront the predictions of the effective theory with experimental data, in particular those sensitive to energy scales 
$E \gg v$. In this sense, extended (non-minimal) scalar sectors provide a very simple renormalizable completion to the $D \leq 6$ SM effective theory, and allow 
for a quantitative assessment of the EFT's energy range of validity. 

Moreover, extensions of the SM scalar sector provide an attractive arena for the use of the SM effective theory: they are a simple scenario beyond the SM, well-motivated 
from the point of view of EW cosmology and baryogenesis, and may also be regarded as part of a complete theory beyond the Standard Model at the TeV scale, such as Composite 
Higgs scenarios or low-energy Supersymmetry. 
Assuming that the new scalar states are significantly heavier than $m_h$, the effective theory of extended Higgs sectors provides a way to study these effects systematically, as 
has already been shown e.g. in \cite{deBlas:2014mba}. 
In particular, it is plausible that the new scalar states from the extended Higgs sector are very hard to probe at LHC (they might be very heavy, or have no
decay branching fractions to SM gauge bosons), but their effect might be possible to detect either in precision EW measurements or using differential information in LHC measurements. 

Below we construct the $D \leq 6$ SM effective theory for various extensions of the SM scalar sector: a singlet extension of the 
SM (the so-called ``Higgs portal"), a Two Higgs Doublet Model and an extension of the SM by a dilaton/radion. In each case, we obtain the Wilson 
coefficients for the $D = 6$ effective operators by matching to the UV theory, and perform an analysis of their current experimental bounds and future prospects. 
The results of this construction for the different scenarios is briefly summarized in Table \ref{TableHiggs}. 

In the respective matching procedure we demand that the effective action for the UV theory  and the $D \leq 6$ EFT agree after an expansion in the light degrees of 
freedoms over the new physics mass scale. This results in the matching of the one-light-particle irreducible (1LPI) Green's functions in the full and effective 
theory  -- see \cite{Buras:1998raa} for a pedagogical introduction for these type of calculations. In the following we will perform an off-shell matching, where 
we expand in external momenta and the EW mass-scale over the new physics mass scale and keep equation of motion vanishing operators in the calculation until the final 
projection. Accordingly all propagators of SM fields will be massless after the expansion and we can perform the calculation in the $SU(2)_L \times U(1)_Y$ symmetric phase, 
quite analogous to the matching calculation performed in \cite{Gorbahn:2009pp}. The light degrees of freedoms of the $SU(2)_L \times U(1)_Y$ symmetric phase 
comprise the gauge singlet $B$, triplet 
$W^a$ and octet $G^a$ fields as well as the scalar doublet $\Phi$ and the fermionic doublets and singlets. In practice we compute 1PI Green's 
functions with up to 6 Higgs and 3 gauge boson fields where the total number of fields does not exceed 8. The resulting expressions are obviously 
related via $SU(2)_L \times U(1)_Y$ gauge invariance of the operators in \ref{eq:silh}, \ref{eq:lagG}, \ref{eq:silhCPodd} and \ref{eq:dim-8}, which provides 
a useful consistency check of our calculation. Moreover, we do cross-check with an explicit calculation in the broken phase, detailed in Appendix \ref{Appendixc}.

\begin{table}[h] 
\begin{center}
\renewcommand{\arraystretch}{1.5} 
\begin{tabular}{c|c|c|c | c | c | c | c |c |c | c |} 
\, & $ \bar c_{H}$ & $ \bar c_{6}$ & $ \bar c_{T}$ &  $ \bar c_{\sss W}$ & $ \bar c_{\sss B}$  & $ \bar c_{\sss HW}$ & $ \bar c_{\sss HB}$ & $ \bar c_{\sss 3W}$ & $ \bar c_{\gamma}$ & $ \bar c_{g}$\\
\hline
Higgs Portal ($G$)&  L &  L  & X  &  X & X  & X  & X  & X & X & X \\
Higgs Portal (\begin{small}Spontaneous\end{small} $G$\hspace{-2.5mm}/)& T  & L  & RG  & RG  & RG  & X  & X  & X & X & X \\
Higgs Portal (\begin{small}Explicit\end{small} $G$\hspace{-2.5mm}/)&  T &  T  & RG  & RG  & RG  & X  & X  & X & X & X \\
 \hdashline
2HDM Benchmark A ($c_{\beta -\alpha} = 0$)& L  &  L  & L  & L  & L  & L  & L  & L & L & X \\
2HDM Benchmark B ($c_{\beta -\alpha}  \neq 0$)& T  & T  & L  & L  & L  & L  & L  & L  & L & X \\
\hdashline
Radion/Dilaton & T  & T  & RG  & T  & T  & T  & T  & L & T & T \\
\hline
\end{tabular}
\end{center}
%\etb
%\captionsetup{singlelinecheck=off}
\caption[ . ]{Leading order at which the various Wilson coefficients for the $D = 6$ SM effective field theory are generated in each of the scenarios under consideration. In each case, the operator can be generated at Tree-Level (T) or 1-Loop (L). If some operators are generated at Tree-Level, this may lead to the generation of others via operator mixing under 1-loop Renormalization Group evolution (see {\it e.g.} \cite{Elias-Miro:2013mua,Jenkins:2013zja}), which we denote by RG. Operators which are generated at higher order in RG and EFT expansion are denoted with an X.}
\label{TableHiggs}
\end{table} 
 
\subsection{The Singlet Higgs Portal: Doublet-Singlet Mixing}
\label{singlet_section}

The addition of a singlet (real or complex) scalar field is arguably the simplest possible extension of the SM. 
Despite its minimality, this extension of the SM can have important consequences for the stability of the EW vacuum at high energies \cite{Lebedev:2012zw,EliasMiro:2012ay}, 
and it could at the same time constitute a ``Higgs portal" into a dark/hidden sector \cite{Schabinger:2005ei,Patt:2006fw}.
It may also have important consequences for Cosmology, potentially accounting for the dark matter relic density \cite{Silveira:1985rk,McDonald:1993ex,O'Connell:2006wi} 
or yielding a first order EW phase transition in the early 
Universe \cite{Anderson:1991zb,Espinosa:1993bs,Profumo:2007wc,Espinosa:2008kw,Espinosa:2011ax,Profumo:2014opa,Farzinnia:2014yqa,Curtin:2014jma} that could 
explain the matter-antimatter asymmetry in the Universe through baryogenesis. In addition, it may give rise to interesting collider phenomenology 
(see {\it e.g.} \cite{Barger:2007im,Dolan:2012ac,No:2013wsa,Efrati:2014uta}).
Altogether, the singlet scalar extension of the SM constitutes a well-motivated scenario, and the interplay of different present and future 
experimental data to probe it has been widely studied (see \cite{Pruna:2013bma,Martin-Lozano:2015dja,Falkowski:2015iwa} for up-to-date analyses).  
Let us then consider the SM scalar potential extended by a singlet scalar field $s$ 
\be
\label{Vsinglet1}
V(\Phi,s) = - \mu^2_{\sss \mathrm{H}} \left|\Phi\right|^2 + \lambda \left|\Phi\right|^4 - \frac{\mu^2_{\sss \mathrm{S}}}{2}\, s^2 + \frac{\lambda_{\sss \mathrm{S}}}{4}\, s^4 +  
\frac{\lambda_{\mathrm{m}}}{2} \left|\Phi\right|^2 s^2\, .
\ee
We assume initially that linear and cubic terms in $s$ are absent
from $V(\Phi,s)$, which may be achieved by means of a discrete/continuous symmetry $G$ in the hidden sector, and focus on the scenario in which the field $s$ develops a 
{\it vev}\footnote[3]{The spontaneous breaking of a discrete or global continuous symmetry $G$ would respectively lead to domain wall formation in the early Universe or the existence of massless Goldstone bosons, both features being undesirable in a realistic model. Possible solutions are to consider 
$G$ to be a spontaneously broken gauge symmetry, or to allow for a small explicit breaking of the symmetry. We will disregard these issues in the following discussion.}, 
$s \to  s + v_{s}$. This generates linear and cubic terms in $s$, which specific relations among these and the rest of parameters in the potential. After EW symmetry breaking, the scalar potential reads
\bea
\label{Vsinglet2}
V(h,s) = \frac{m_h^2}{2} h^2  +  \frac{m_s^2}{2} s^2 + m^2_{hs} h s + v\lambda \, h^3  + v_{s}\lambda_{\sss \mathrm{S}} \,s^3 + \frac{\lambda_{\mathrm{m}}\,v}{2} \, h s^2 + \frac{\lambda_{\mathrm{m}} v_s}{2}\, h^2 s \nonumber \\
+ \frac{\lambda_{\mathrm{m}}}{4} h^2 s^2 + \frac{\lambda}{4} h^4 + \frac{\lambda_{\sss \mathrm{S}}}{4} s^4 
\eea
with $m_h^2 = 2 \,\lambda \, v^2$, $ m^2_s = 2\, \lambda_{\sss \mathrm{S}} \, v_s^2$ and $m^2_{hs} = \lambda_{\mathrm{m}}\, v \,v_s$, and where we have 
used the minimization conditions $\mu^2_{\sss \mathrm{H}} = \lambda\, v^2 + \frac{\lambda_{\mathrm{m}}}{2} v^2_{s} $ and $\mu^2_{\sss \mathrm{S}} = \lambda_{\sss \mathrm{S}} v_{s}^2 + \frac{\lambda_{\mathrm{m}}}{2} v^2 $ in $V(h,s)$ to trade the mass parameters $\mu^2_{\sss \mathrm{H}},\mu^2_{\sss \mathrm{S}}$ for the {\it vevs}.
The term $m^2_{hs} h s$ in (\ref{Vsinglet2}) induces doublet-singlet mixing, leading to two mass eigenstates $h_{1,2}$, the lighter of which ($h_1$) we identify with the discovered $125$ GeV Higgs particle. 
The mixing angle $\theta$ and masses are given by
\be
s^2_{\theta} = \frac{4\, m^4_{hs}}{4\, m^4_{hs} + \left(m_s^2 - m_h^2 + \sqrt{(m_s^2-m_h^2)^2 + 4 \,m^4_{hs}}\right)^2}
= \frac{4\, y^2}{4\, y^2 + \left(1 - x^2 + \sqrt{(1-x^2)^2 + 4 \,y^2}\right)^2}
\label{mix}
\ee
\be
m^2_{1,2} = \frac{1}{2} \left(m_h^2 + m_s^2 \mp \sqrt{(m_s^2-m_h^2)^2 + 4 \,m^4_{hs}}\right) = \frac{m_s^2}{2} \left(1 + x^2 \mp \sqrt{(1-x^2)^2 + 4 \,y^2}\right)
\label{mass}
\ee
with $s_{\theta} \equiv \mathrm{sin}(\theta)$, $x \equiv m_h/m_s \sim v/v_s$ and $y \equiv m_{hs}^2/m_s^2 \sim v/v_s$. The limit $m_{s} \gg v$ corresponds then to $v_s \gg v$, with $x,y \ll 1$ and so
\be
s^2_{\theta} \simeq y^2 \quad , \quad m^2_{1} \simeq  m^2_s\,(x^2-y^2) = m_h^2 - s^2_{\theta}\,m_s^2 \quad , \quad m^2_{2} \simeq  m^2_s\,(1+y^2) = m^2_s\,(1+s^2_{\theta})   
\label{massmixing}
\ee
where we have neglected terms of $\mathcal{O}(x^4,x^2y^2,y^4)$. From (\ref{Vsinglet2}), the relevant scalar self-interactions $h^2_1\, h_2$ and $h^3_1$ read
\bea
\label{SingletScalarInt}
V(h_1, h_2) \supset \left[  \frac{m^2_{1}}{2\,v} + \mathcal{O}(x^2,y^2)\right] h^3_1 + \left[ \frac{m^2_s\, y}{2\,v} + \mathcal{O}(x,y)\right] h^2_1 \, h_2 
\eea
Neglecting $\mathcal{O}(y^4)$ corrections, the couplings of $h_{1,2}$ to the $W^{\pm}$ and $Z$ bosons read 
\bea
\label{SingletGaugeInt}
\left[ g\, m_W \left(1 - y^2/2\right) h_1 + \frac{g^2}{4} \left(1 - y^2\right) h_1^2  -  (g m_W y)\, h_2 + ...\right] \,W^{+}_{\mu} W^{\mu-} \nonumber \\
+ \left[ \frac{g m_Z}{2 \,c_\mathrm{W}} \left(1 - y^2/2\right) h_1 + \frac{g^2}{8 \,c^2_\mathrm{W}} \left(1 - y^2\right) h_1^2  -  \left(\frac{g m_Z}{2 \,c_\mathrm{W}} y\right)\, h_2 + ...\right] \,Z_{\mu} Z^{\mu}
\eea
Noting that $y \equiv m_{hs}^2/m_s^2 = m_{hs}^2/m_2^2 + \mathcal{O}(y^3)$, we may integrate out the heavy, singlet-like state $h_2$. At leading order, this generates an $\mathcal{O}(y^2)$ contribution to 
$h_1^2 \,V_{\mu} V^{\mu}$, which precisely cancels the $\mathcal{O}(y^2)$ correction in (\ref{SingletGaugeInt}). The Higgs-gauge interactions then read
\be
\label{SingletGaugeInt2}
\left[ g\, m_W \left(1 - y^2/2\right) h_1 + \frac{g^2}{4} \, h_1^2  \right] \,W^{+}_{\mu} W^{\mu-} + \left[ \frac{g m_Z}{2 \,c_\mathrm{W}} \left(1 - y^2/2\right) h_1 + \frac{g^2}{8 \,c^2_\mathrm{W}}  h_1^2  \right] \,Z_{\mu} Z^{\mu}\, ,
\ee
leading to a $g_V^2 y^2/2$ mismatch (with $g_V = \frac{g}{\sqrt{2}},\,\frac{g}{\sqrt{2}\,\cW}$ for $V = W,\, Z$ respectively) between $g^{(3)}_{\sss hhVV}$ and $\frac{g\,g^{(3)}_{\sss hVV}}{2\,\mW}$.
We turn now to the fermionic couplings, focusing on the Higgs-top quark interactions 
\be
\frac{Y_t}{\sqrt{2}}\left(1 - \frac{y^2}{2}\right) h_1\,\bar{t}_L t_R - \frac{Y_t\,y}{\sqrt{2}}\, h_2\,\bar{t}_L t_R\,,
\label{topq}
\ee
with $Y_t$ the top-quark Yukawa coupling. This affects the gluon fusion Higgs effective coupling for $h_1$, which gets modified w.r.t the SM one
\be
- \frac{g_{\sss hgg}}{4}  \left(1 - y^2/2\right) G^{a}_{\mu\nu} G_{a}^{\mu\nu} h_1 %- \frac{g_{\sss hhgg}}{8}  \left(1 - y^2\right) - \frac{g_{s \sss gg}(\tau)\,y^2}{4\,v} 
%\, G^{a}_{\mu\nu} G_{a}^{\mu\nu} h_1^2 
\, .
\label{gf1}
\ee
The coupling $G^{a}_{\mu\nu} G_{a}^{\mu\nu} h_1^2$ (relevant for di-Higgs production) also gets modified, both through an $\mathcal{O}(y^2)$ correction due to the singlet-doublet mixing 
and directly via the presence of $h_2$, which can mediate the process $g\,g \to h_1 h_1$ (for $\hat{\mathrm{s}}/m^2_s \ll 1$, being $\hat{\mathrm{s}}$ the partonic center of mass energy for the process, this contribution is however small).  

\vspace{4mm}

The above discussion may be directly mapped into an $SU(2)_L \times U(1)_Y$ invariant effective field theory for the SM. We consider (\ref{Vsinglet1}) after $s$ develops a {\it vev}  
\be
\label{Vsinglet3}
V(\Phi,s) = - \tilde \mu^2_{\sss \mathrm{H}}\left|\Phi\right|^2 + \lambda \left|\Phi\right|^4 + \frac{m_s^2}{2} s^2 + v_{s}\lambda_{\sss \mathrm{S}} \,s^3 + \frac{\lambda_{\sss \mathrm{S}}}{4}\, s^4 +\lambda_{\mathrm{m}} v_{s}\, \left|\Phi\right|^2 s +  \frac{\lambda_{\mathrm{m}}}{2} \left|\Phi\right|^2 s^2\, ,
\ee

\begin{figure}[ht!]
\begin{center}
\includegraphics[width=0.45\textwidth]{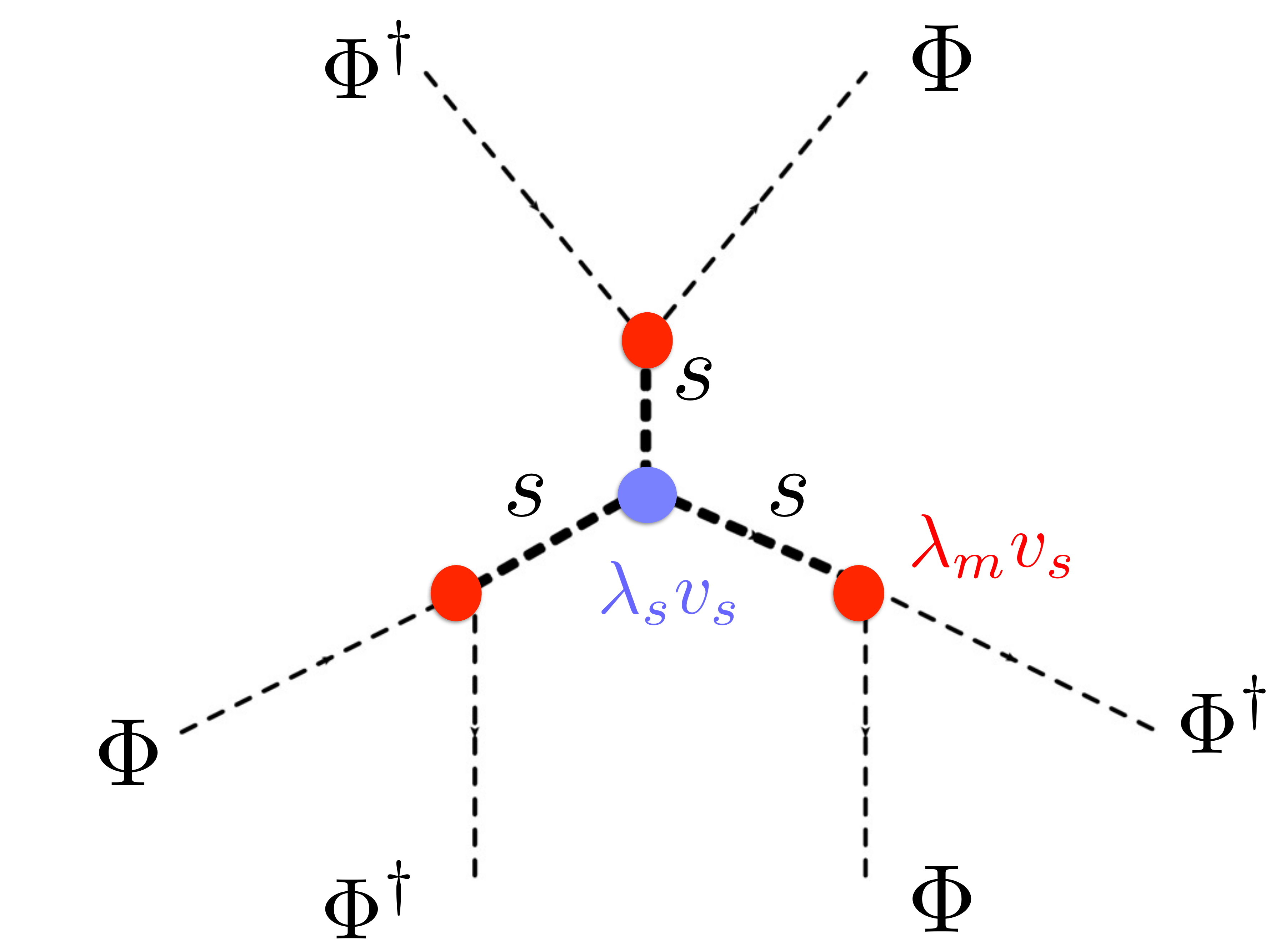} \hspace{5mm}
\includegraphics[width=0.45\textwidth]{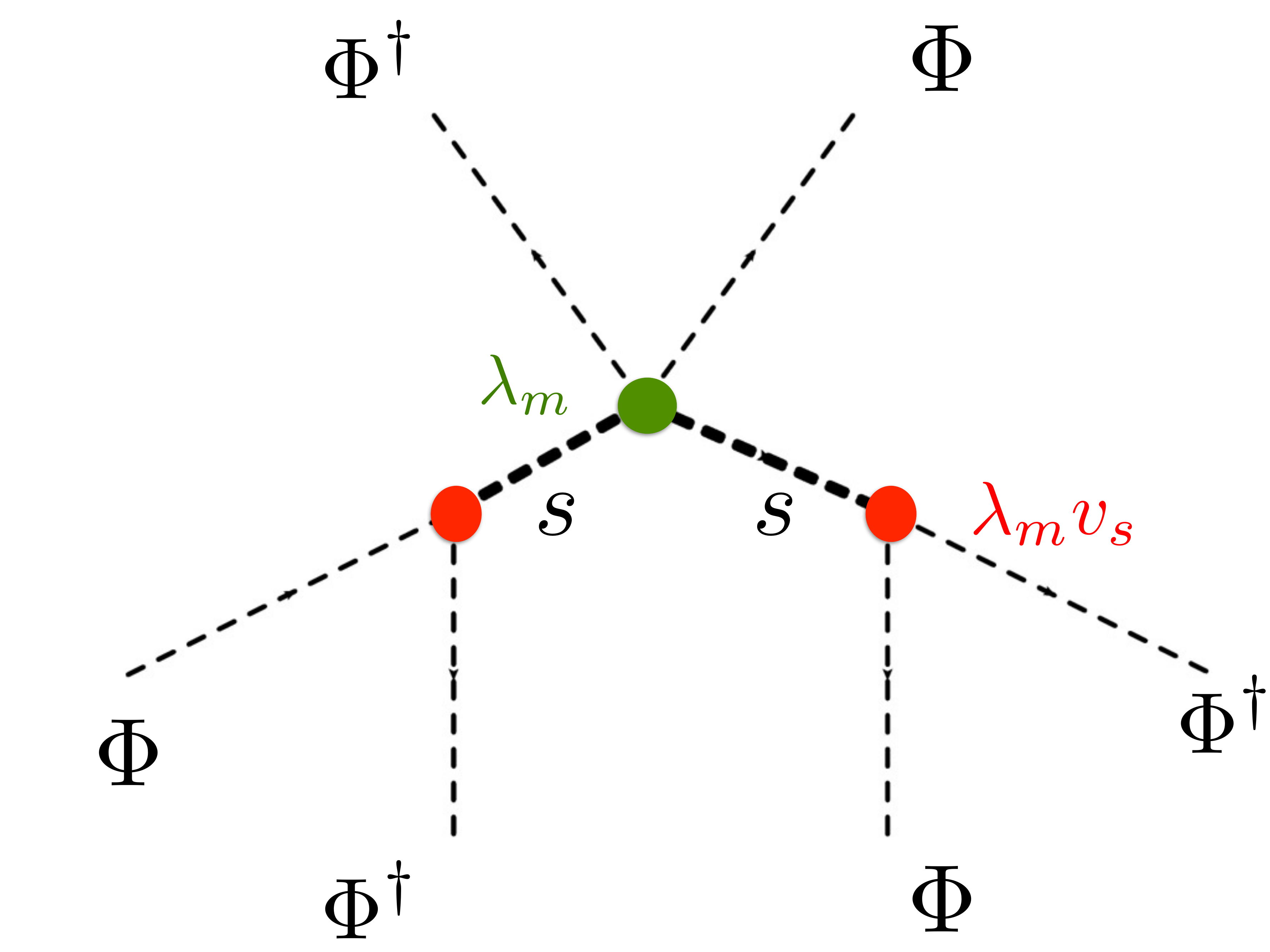} 
\caption{The two Feynman diagrams leading to the effective operator $\bar c_6$ in the singlet extension of the SM. Their respective combinatorial factors are 6 (Left) and 3 (Right).}
\label{fig:c6singlet}
\end{center}
\end{figure}
\noindent with $\tilde \mu^2_{\sss \mathrm{H}} =  \mu^2_{\sss \mathrm{H}} - (\lambda_{\mathrm{m}} v_{s}^2) /2$. We then integrate-out the field $s$, which 
yields the effective Lagrangian
\be
\label{Effsinglet}
\mathcal{L}_{\mathrm{Eff}} \supset \frac{\lambda^2_{\mathrm{m}} v^2_{s}}{2 m_s^4} \partial^\mu \left|\Phi\right|^2 \partial_\mu \left|\Phi\right|^2 
  - \left(6\,\frac{\lambda^3_{\mathrm{m}} \lambda_{\sss \mathrm{S}} v^4_{s}}{m_s^6} - 3\,\frac{\lambda^3_{\mathrm{m}}\,v^2_{s}}{m_s^4} \right) \left|\Phi\right|^6 = 
  \frac{y^2}{2\,v^2} \partial^\mu \left|\Phi\right|^2 \partial_\mu \left|\Phi\right|^2 \, .
%  - \frac{\lambda_{\mathrm{m}}\, y^2}{2 v^2} \left|\Phi\right|^6\,.
\ee
From (\ref{eq:silh}), this corresponds to $\bar c_H = y^2$. The two contributions to $\bar c_6$ are shown in Figure \ref{fig:c6singlet}, and cancel by means of $m_s^2 = 2 \,\lambda_{\sss \mathrm{S}}\, v_{s}^2$ (yielding $\bar c_6 = 0$). Upon EW symmetry breaking, $\bar c_H$ modifies the wave function of the Higgs $h$ 
\be
\mathcal{L}_{\mathrm{Eff}} \supset \left( 1 + \bar c_H \right) \frac{1}{2} (\partial_{\mu} h)^2
\ee
which universally reduces all couplings of $h$: {\it e.g.} the coupling between $h$ and the weak bosons simply read
\bea
g^{(3)}_{\sss hVV} = \frac{g^2_V\,v}{\sqrt{ 1 + y^2}} \simeq g^2_V\,v \, \left(1 - \frac{y^2}{2}\right) \, ,
%\quad , \quad 
%g^{(3)}_{\sss hhVV} = \frac{g^2_V}{(1 + y^2)} \simeq  g^2_V \left(1 - y^2\right)
\eea
matching as expected the result from (\ref{SingletGaugeInt2}). 
By comparing the EFT results with (\ref{SingletGaugeInt2}) we observe that the EFT 
appproach reproduces the effect of scalar mixing on interactions involving one Higgs scalar $h$, but 
fails to do so for the case of two scalars $h h$. This 
suggests that di-Higgs production (see e.g. \cite{Dolan:2012ac,No:2013wsa,Efrati:2014uta,Gouzevitch:2013qca,Chen:2014ask}) could be used as a tool to discriminate 
between a pure linear realization of the SM Higgs doublet and the presence of mixing effects, although the prospects seem challenging and we will not discuss them here.

\vspace{2mm}

The previous results for $\bar c_H$ are left unchanged by the inclusion of explicit linear and cubic terms for $s$ in (\ref{Vsinglet1}). These can however generate a non-zero value for 
$\bar c_6$, since now the cancellation among the diagrams in Figure \ref{fig:c6singlet} is not exact. Let us illustrate this by adding a term $\mu_{m}\, \left|\Phi\right|^2 s$ 
to (\ref{Vsinglet3}).
This term does not alter the minimization conditions, but contributes to the singlet-doublet mixing upon EW symmetry breaking, so that now the mixing is given by 
$y \equiv m_{hs}^2/m_s^2 =  (\lambda_{\mathrm{m}}\,v_s + \mu_{m})\, v/m_s^2$. Upon integrating-out $s$, this now yields 
\bea
\label{Effsinglet2}
\mathcal{L}_{\mathrm{Eff}} &\supset& \frac{(\lambda_{\mathrm{m}} v_{s} + \mu_{m})^2}{2\, m_s^4} \partial^\mu \left|\Phi\right|^2 \partial_\mu \left|\Phi\right|^2 
  - \left(6\,\frac{(\lambda_{\mathrm{m}}v_{s} + \mu_{m})^3 \lambda_{\sss \mathrm{S}} v_{s}}{m_s^6} - 3\frac{\lambda_{\mathrm{m}}(\lambda_{\mathrm{m}}v_{s} + \mu_{m})^2}{m_s^4} \right) 
  \left|\Phi\right|^6 \nonumber \\
 & = &  \frac{y^2}{2\,v^2} \partial^\mu \left|\Phi\right|^2 \partial_\mu \left|\Phi\right|^2 - \frac{3\,\mu_{m}}{v_s}\,\frac{y^2}{v^2}  \left|\Phi\right|^6\,,
\eea
leading again to $\bar c_H = y^2$, and now to $\lambda\,\bar c_6 = 3\,y^2\,\mu_{m}/v_s$. The parameter $\mu_{m}/v_s \equiv \delta$ measures the relative importance of explicit {\it vs} spontaneous 
symmetry breaking in (\ref{Vsinglet1}). 

\vspace{2mm}

Let us also comment on the case where the field $s$ does not develop a {\it vev} and still the terms linear and cubic in $s$ are absent from (\ref{Vsinglet1}) (the symmetry $G$ from (\ref{Vsinglet1}) remains unbroken). In this case there is no Higgs-singlet mixing, but nevertheless the operator $\partial^\mu \left|\Phi\right|^2 \partial_\mu \left|\Phi\right|^2$
is generated at 1-loop \cite{Craig:2013xia} with a Wilson coefficient $\bar c_H = \frac{n_s\,\lambda^2_{\mathrm{m}}\,v^2}{96\,\pi^2\,m^2_s}$ (being $n_s$ the number of singlet scalar degrees of freedom). 
The universal suppresion of Higgs couplings can then be defined in a similar fashion to the previous case, $y^2 = \bar c_H$.

\begin{table}[ht]
\hspace{2.5cm}
\begin{tabular}{l| c |}
& $\mu_{\mathrm{obs}}/\mu_{\mathrm{SM}} $ \\
\hline 
ATLAS $\gamma \gamma$ 7+8 TeV: $\mu_{\mathrm{ggF}}$ & $1.32\pm 0.38$ \\ 
ATLAS $\gamma \gamma$ 7+8 TeV: $\mu_{\mathrm{VBF}}$& $0.8\pm 0.7$ \\ 
ATLAS $WW^*$ 7+8 TeV: $\mu_{\mathrm{ggF}}$  & $0.82\pm 0.36$\\ 
ATLAS $WW^*$ 7+8 TeV: $\mu_{\mathrm{VBF}}$  & $1.66\pm 0.79$\\ 
ATLAS $ZZ^*$ 7+8 TeV (inclusive)  & $1.44^{+0.40}_{-0.33}$\\ 
ATLAS $b\bar{b}$ 7+8 TeV: $\mu_{\mathrm{VH}}$  & $0.2^{+0.7}_{-0.6}$\\ 
ATLAS $\tau \tau$ 7+8 TeV: $\mu_{\mathrm{ggF}}$ & $1.2^{+0.8}_{-0.6}$\\ 
ATLAS $\tau \tau$ 7+8 TeV: $\mu_{\mathrm{VBF}}$ & $1.6^{+0.6}_{-0.5}$\\ 
CMS $\gamma \gamma$ (Mass Fit) 7+8 TeV: $\mu_{\mathrm{ggF+ttH}}$ & $1.13^{+0.37}_{-0.31}$ \\ 
CMS $\gamma \gamma$ (Mass Fit) 7+8 TeV: $\mu_{\mathrm{VBF+VH}}$& $1.16^{+0.63}_{-0.58}$ \\ 
CMS $WW^*$ 7+8 TeV (0/1-jet) & $0.74^{+0.22}_{-0.20}$\\ 
CMS $WW^*$ 7+8 TeV (2-jets, VBF tag)  & $0.60^{+0.57}_{-0.46}$ \\ 
CMS $ZZ^*$ 7+8 TeV (inclusive) & $0.93^{+0.29}_{-0.24}$ \\ 
CMS $b\bar{b}$ 7+8 TeV: $\mu_{\mathrm{VH}}$  & $1.0\pm 0.5$\\ 
CMS $\tau \tau$ 7+8 TeV (0-jet) & $0.34\pm 1.09$\\ 
CMS $\tau \tau$ 7+8 TeV (1-jet) & $1.07\pm 0.46$\\ 
CMS $\tau \tau$ 7+8 TeV (2-jets, VBF tag) & $0.94\pm 0.41$\\ 
\hline
\end{tabular}
\caption{\small ATLAS and CMS measured Higgs Signal Strengths $\mu_i \equiv \mu^i_{\mathrm{obs}}/\mu^i_{\mathrm{SM}}$ in $h \to \gamma \gamma$ \cite{Aad:2014eha,Khachatryan:2014ira}, $h \to Z Z^*$ \cite{Aad:2014eva,Chatrchyan:2013mxa}, $h \to W W^*$ \cite{ATLAS:2013wla,Chatrchyan:2013iaa}, $h \to \bar{b}\,b$ \cite{TheATLAScollaboration:2013lia,Chatrchyan:2013zna} and $h \to \tau\,\tau$ \cite{ATLAS_tau,Chatrchyan:2014nva} final states.}
\label{table:ATLAS_CMS}
\end{table}

\vspace{2mm}

\begin{figure}[ht!]
\begin{center}
\includegraphics[width=0.75\textwidth]{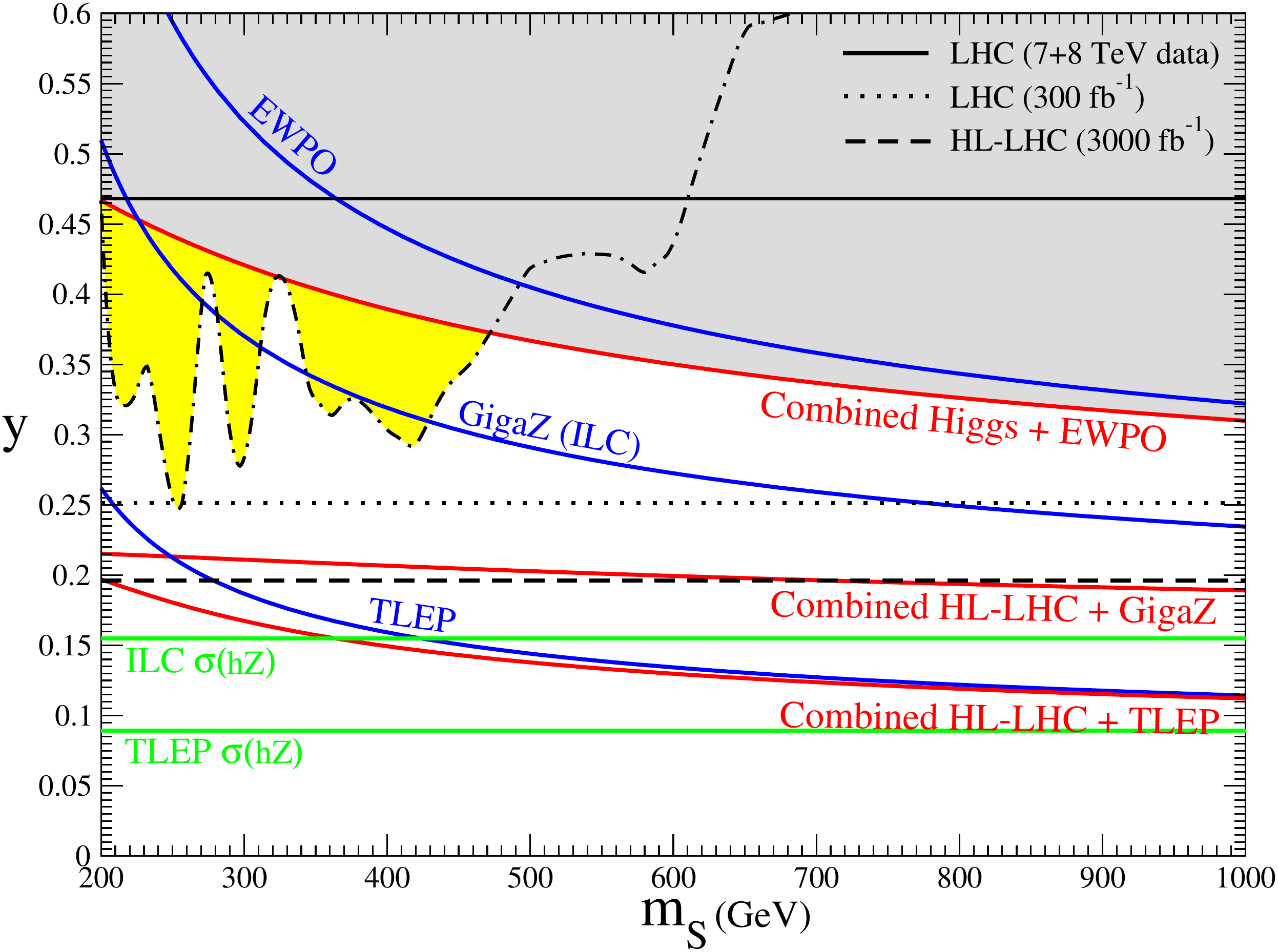} 
\caption{Present $95\%\, \mathrm{C.L.}$ exclusion limits in the $(m_s,\,y)$ plane arising from ATLAS and CMS measurements of Higgs signal strengths as shown in Table \ref{table:ATLAS_CMS} (horizontal solid-black) and from EWPO (blue). The shaded grey region is excluded at $95\%\, \mathrm{C.L.}$ by the combination (EWPO and Higgs signal strengths). The shaded yellow region may also be excluded by constraints from heavy scalar searches at LHC (dashed-dotted black), but these limits can be evaded in the presence of non-SM decays (see text). Also shown is the projected exclusion reach
from Higgs signal strengths at the 14 TeV run of LHC with $\mathcal{L}=300 \, \,\mathrm{fb}^{-1}$ (horizontal dotted-black) and at HL-LHC with $\mathcal{L}=3000 \, \,\mathrm{fb}^{-1}$ (horizontal dashed-black), from measurements of the $S$ and $T$ oblique parameters with ILC-{\it GigaZ} and TLEP (blue) and their combination with the HL-LHC exclusion reach (red), and from a precision measurement of the Higgs production cross section in association with a $Z$ boson $\sigma(hZ)$ (horizontal solid-green lines) at ILC ($\sqrt{s} = 250$ GeV, $\mathcal{L}=1150\, \, \mathrm{fb}^{-1}$) and TLEP ($\sqrt{s} = 240$ GeV, $\mathcal{L}=500\, \, \mathrm{fb}^{-1}$).}
\label{fig:Singlet}
\end{center}
\end{figure}

We now discuss the current experimental constraints on the mass of the singlet-like state and the parameter $y$, coming from the latest measurements of Higgs boson
signal strengths from ATLAS and CMS, from oblique corrections to EWPO and from direct searches of heavy Higgs scalars at LHC.
For the analysis of Higgs signal strengths, we consider the up-to-date measurements by ATLAS and CMS in $h \to \gamma \gamma$ \cite{Aad:2014eha,Khachatryan:2014ira}, $h \to Z Z^*$ \cite{Aad:2014eva,Chatrchyan:2013mxa}, $h \to W W^*$ \cite{ATLAS:2013wla,Chatrchyan:2013iaa}, $h \to \bar{b}\,b$ \cite{TheATLAScollaboration:2013lia,Chatrchyan:2013zna} and $h \to \tau\,\tau$ \cite{ATLAS_tau,Chatrchyan:2014nva} final states, shown in Table \ref{table:ATLAS_CMS}.
We then perform a combined $\chi^2$ fit
\be
\label{chi_Higgs}
\chi^2_h = \sum_i \left(\frac{\mu_i - (1-y^2)}{\Delta \mu_i}\right)^2\, ,
\ee
with the various $\mu_i \equiv \mu^i_{\mathrm{obs}}/\mu^i_{\mathrm{SM}}$ and $\Delta \mu_i$ taken from Table \ref{table:ATLAS_CMS}, and with potential correlations among the different signal strength measurements not included in the fit. This yields an ATLAS and CMS combined limit $y < 0.468$ at $95\%\, \mathrm{C.L.}$ via $\Delta\chi^2_h(y) = \chi^2_h(y) - \chi^2_{\mathrm{min}} = 4$, as shown in 
Figure \ref{fig:Singlet} (horizontal solid-black). In addition, we obtain the $95\%\, \mathrm{C.L.}$ exclusion prospects for LHC at 14 TeV with $\mathcal{L}=300 \, \,\mathrm{fb}^{-1}$ (horizontal dotted-black) and $\mathcal{L}=3000\, \, \mathrm{fb}^{-1}$ (HL-LHC, horizontal dashed-black). In doing so, we assume that future measurements of Higgs signal strengths will yield $\mu_i = 1$, and use the projected CMS 
sensitivities\footnote[4]{These assume that theoretical uncertainties improve by a factor $1/2$ compared to their present values, while all other systematic uncertainties improve by a factor $1/\sqrt{\mathcal{L}}$.} \cite{CMS:2013xfa} $\Delta \mu^{\gamma\gamma}_{300} = 0.06$, $\Delta \mu^{WW}_{300} = 0.06$, $\Delta \mu^{ZZ}_{300} = 0.07$, $\Delta \mu^{\tau\tau}_{300} = 0.08$, $\Delta \mu^{bb}_{300} = 0.11$, $\Delta \mu^{\mu\mu}_{300} = 0.40$, $\Delta \mu^{\gamma\gamma}_{3000} = 0.04$, $\Delta \mu^{WW}_{3000} = 0.04$, $\Delta \mu^{ZZ}_{3000} = 0.04$, $\Delta \mu^{\tau\tau}_{3000} = 0.05$, $\Delta \mu^{bb}_{3000} = 0.05$, $\Delta \mu^{\mu\mu}_{3000} = 0.20$. 

\vspace{2mm}

Turning to EWPO, we perform a fit to the oblique parameters $S, T, U$ using the best-fit values and standard deviations from the global analysis of the 
GFitter Group \cite{Baak:2014ora}, with a SM reference point with $m_t = 173$ GeV and a $126$ GeV Higgs mass. Under the assumption $U = 0$, this yields
\be
\label{chi_EWPO1}
\begin{array}{c}
\Delta S \equiv S - S_{\mathrm{SM}} = 0.06 \pm 0.09 \\
\Delta T \equiv T - T_{\mathrm{SM}} = 0.10 \pm 0.07
\end{array} \quad \quad \quad \quad 
\rho_{ij} = \left(\begin{array}{cc}
             1 & 0.91\\
             0.91 &1
            \end{array}\right)
\ee
being $\rho_{ij}$ the covariance matrix in the $S-T$ plane. The BSM corrections to $S$ and $T$ in the case of singlet-doublet mixing are given by
\bea
\label{ST}
\Delta S = \frac{1}{\pi}\,y^2 \left[ - H_S\left(\frac{m^2_h}{m^2_Z}\right) + H_S\left(\frac{m^2_s}{m^2_Z}\right)\right] \nonumber \\
\Delta T = \frac{g^2}{16 \,\pi^2 \,\cW^2 \,\alpha_{\mathrm{EM}}}\,y^2 \left[ - H_T\left(\frac{m^2_h}{m^2_Z}\right) + H_T\left(\frac{m^2_s}{m^2_Z}\right)\right]
\eea
with the functions $H_S(x)$ and $H_T(x)$ defined in Appendix C of \cite{Hagiwara:1994pw}. 
%In the absence of mixing, $\Delta S$ and $\Delta T$ will only receive the first correction in (\ref{ST}). 
%
We then define
\be
\label{chi_EWPO2}
\Delta\chi^2_{EW}(m_s,y) = \sum_{i,j} \left(\Delta\mathcal{O}_{i}(m_s,y) - \Delta\mathcal{O}^0_{i}\right) (\sigma^2)_{ij}^{-1}
\left(\Delta\mathcal{O}_{j}(m_s,y) - \Delta\mathcal{O}^0_{j}\right)\, ,
\ee
where $\Delta\mathcal{O}^0_{i}$ denote the central values in (\ref{chi_EWPO1}) and $(\sigma^2)_{ij} \equiv \sigma_i \rho_{ij} \sigma_j$, being $\sigma_i$ the $S$ and $T$ standard deviation from (\ref{chi_EWPO1}). We show in Figure \ref{fig:Singlet} the $95\%\, \mathrm{C.L.}$ exclusion limit $y(m_s)$ from $\Delta\chi^2_{EW}(m_s,y)$ (blue), together with the $95\%\, \mathrm{C.L.}$ exclusion limit $y(m_s)$ from the combination of $\Delta\chi^2_h(y)$ and $\Delta\chi^2_{EW}(m_s,y)$ (red). 
We also study the future exclusion reach that can be derived from prospects of measurements of EW precision observables by planned $e^+e^-$ colliders (see {\it e.g.} \cite{Fan:2014vta}): Assuming a SM best-fit value, the ILC {\it GigaZ} program's expected precision is $\sigma_S = 0.017$ and $\sigma_T = 0.022$ \cite{Baak:2014ora}, while measurement of EWPO at TLEP could yield 
$\sigma_S = 0.007$ and $\sigma_T = 0.004$ \cite{Gomez-Ceballos:2013zzn,Mishima}. Figure \ref{fig:Singlet} includes the $95\%\, \mathrm{C.L.}$ exclusion reach for $y(m_s)$ both for ILC and TLEP (blue), as well as the respective $95\%\, \mathrm{C.L.}$ exclusion reach when combined with HL-LHC ($\mathcal{L}=3000\, \, \mathrm{fb}^{-1}$). Figure \ref{fig:Singlet} also shows the potential $95\%\, \mathrm{C.L.}$ exclusion reach from a precise measurement of the Higgs production cross section in association with a $Z$ boson $\sigma(hZ)$ at ILC with $\sqrt{s} = 250$ GeV and $\mathcal{L}=1150\, \, \mathrm{fb}^{-1}$ (with $\Delta \sigma(hZ)/\sigma(hZ) \sim 0.012$) \cite{Asner:2013psa} and at TLEP with $\sqrt{s} = 240$ GeV and $\mathcal{L}=500\, \, \mathrm{fb}^{-1}$ (with 
$\Delta \sigma(hZ)/\sigma(hZ) \sim 0.004$) \cite{Gomez-Ceballos:2013zzn}. From Figure \ref{fig:Singlet} it is evident that this precision measurement is potentially the most powerful probe of a 
non-SM Higgs admixture, significantly surpassing the reach of EWPO measurements.

\vspace{1mm}

It is worth stressing that the contributions to the $S$ and $T$ oblique parameters in the Higgs portal scenario could be re-derived in the EFT approach via operator mixing under 
renormalization group evolution (RGE), as shown {\it e.g.} in \cite{Henning:2014gca}. Specifically, the running of $ \bar c_H$ from the matching scale $m_s$ down to $m_Z$ generates a contribution to $S$ and $T$, given at leading order by \cite{Elias-Miro:2013eta,Henning:2014gca} 
\be
\label{Ren_Group_EWPO}
\Delta S = \frac{1}{6\,\pi} \, \bar c_H (m_s) \, \mathrm{log} \left(\frac{m_s}{m_Z}\right) \quad \quad \quad
\Delta T = -\frac{3}{8\,\pi\,c^2_W} \, \bar c_H (m_s) \, \mathrm{log} \left(\frac{m_s}{m_Z}\right)
\ee
which correspond to the leading order contributions in (\ref{ST}) for $m_s \gg m_Z$.

\vspace{2mm}

Finally, Figure \ref{fig:Singlet} includes the latest $95\%\, \mathrm{C.L.}$ limits on $y(m_s)$ (dotted-dashed black line and yellow region) obtained from CMS searches of a heavy neutral scalar decaying to $ZZ\to4\ell$, $ZZ\to 2\ell2j$ and $ZZ\to 2\ell2\nu$ final states \cite{CMS:2013pea} (searches in decays to $WW \to 2\ell2\nu$ final states \cite{CMS:2012bea} are found not to be as sensitive). These constitute at present the most stringent constraint on $y$ for $m_s \lesssim 500$ GeV, but we stress that for $m_s > 2 m_h = 250$ GeV these limits may be weakened/avoided for a significant branching fraction 
$\mathrm{Br}(h_2 \to h_1 h_1)$. 

\begin{figure}[ht!]
\begin{center}
\includegraphics[width=0.477\textwidth]{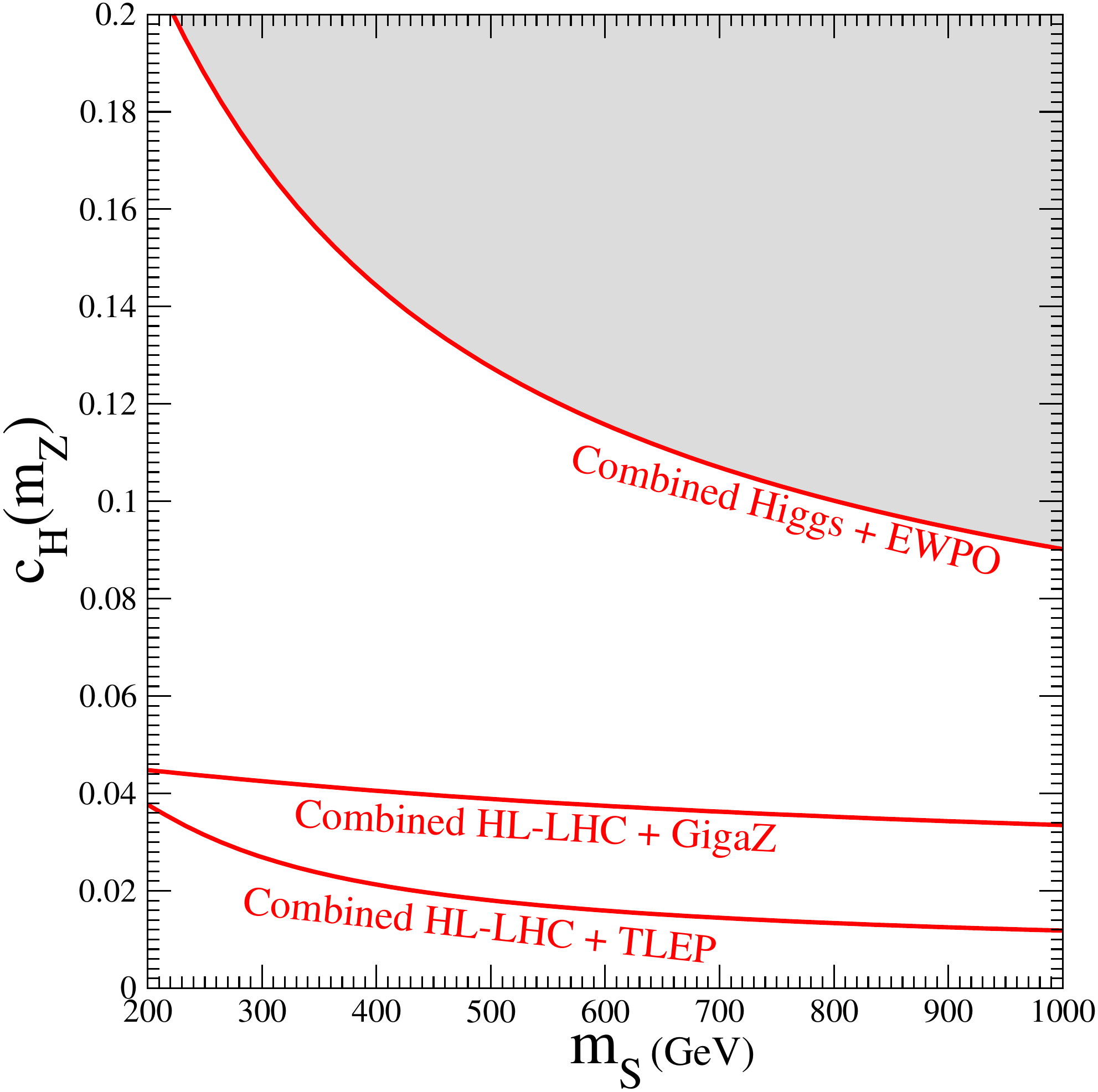} \hspace{3mm}
\includegraphics[width=0.482\textwidth]{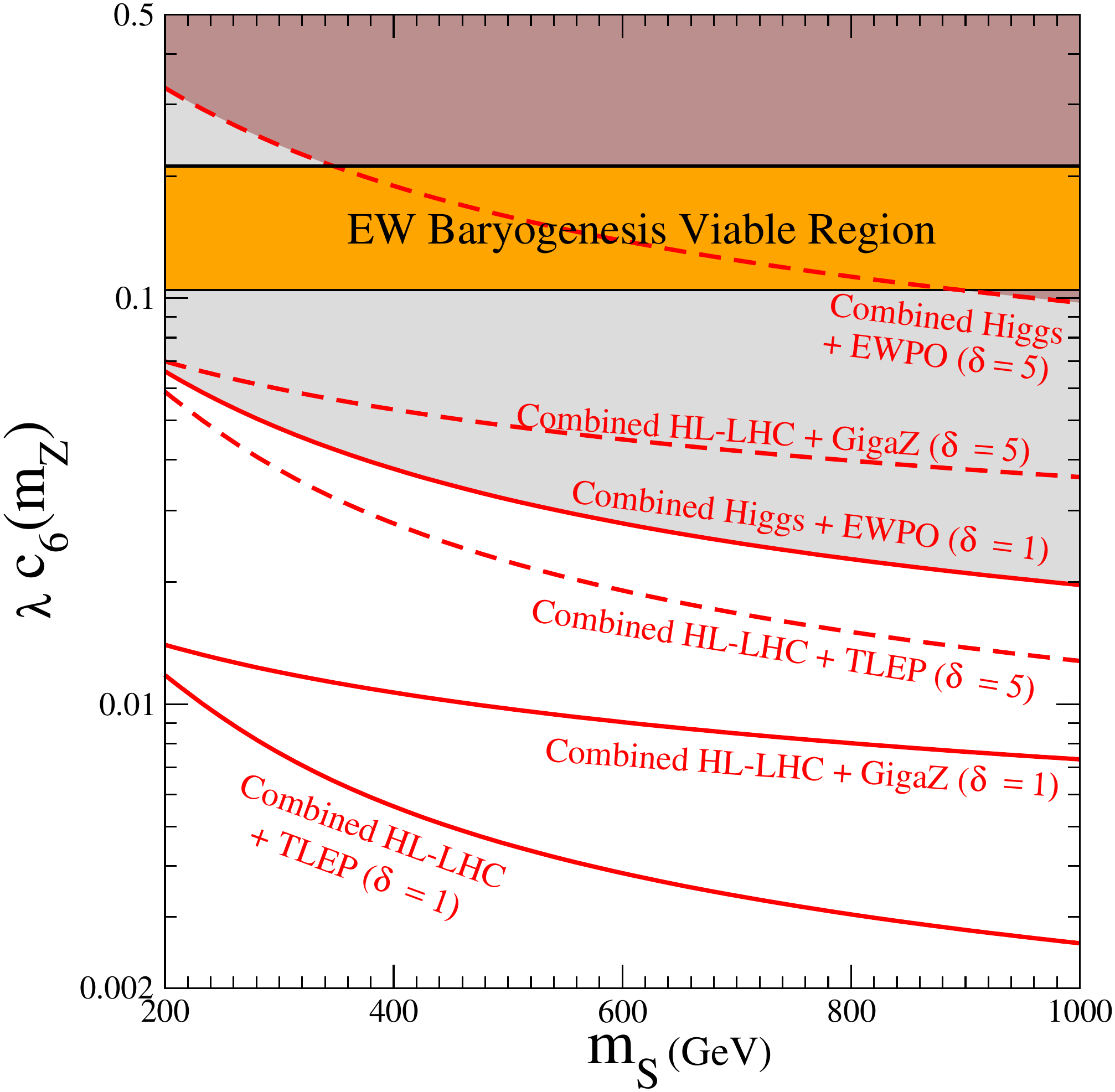} 
\caption{Present and future (as defined in Figure \ref{fig:Singlet}) $95\%\, \mathrm{C.L.}$ exclusion limits on the size of $ \bar c_H (m_Z)$ (Left) and $\lambda\, \bar c_6 (m_Z)$ (Right) as a function of $m_s$, from the combination of Higgs signal strengths and EW oblique parameters. For the case of $\lambda\, \bar c_6 (m_Z)$, the region leading to a strongly first order EW phase transition as required by EW baryogenesis is shown as a horizontal orange band, while the grey (brown) region is currently excluded at $95\%\, \mathrm{C.L.}$ for $\delta \equiv \mu_{m}/v_s = 1$ ($\delta = 5$).}
\label{fig:Singlet_CH}
\end{center}
\end{figure}

\vspace{2mm}

From the present limits and future prospects discussed above, we can derive current/projected bounds on the size of the Wilson coefficients $ \bar c_H (m_Z)$ and $ \bar c_6 (m_Z)$ in this scenario. 
From  $\bar c_H (m_s) = y^2$ and $ \lambda \, \bar c_6 (m_s) = 3\,y^2 \mu_{m}/v_s$, we perform a 1-loop RGE \cite{Elias-Miro:2013mua} (see also
\cite{Jenkins:2013zja})
\be
\label{Ren_Group}
\bar c_H (m_s) - \bar c_H (m_Z) \simeq \frac{\left[-\frac{9}{2}g^2 - 3 g'^2 + 24 \lambda + 12 y_t^2\right] }{16\,\pi^2}\, \bar c_H (m_s) \, \mathrm{log} \left(\frac{m_s}{m_Z}\right) 
\ee
\bea
\label{Ren_Group2}
\lambda \, \bar  c_6 (m_s) - \lambda\, \bar c_6 (m_Z) \simeq \frac{\left[-\frac{27}{2}g^2 - \frac{9}{2} g'^2 + 108 \lambda + 18 y_t^2\right]}{16\,\pi^2}\, \lambda \, \bar c_6 (m_s) \, \mathrm{log} \left(\frac{m_s}{m_Z}\right)\\
+ \frac{\left[-3g^2 +40 \lambda \right]}{8\,\pi^2} \, \lambda \, \bar c_H (m_s) \, \mathrm{log} \left(\frac{m_s}{m_Z}\right)
\eea
Current bounds on $\bar c_H (m_Z)$ and $\bar c_6 (m_Z)$ as a function of the new physics scale $m_s$, as well as the projected $95\%\, \mathrm{C.L.}$ exclusion sensitivity from the combination of HL-LHC and {\it GigaZ}, and HL-LHC and TLEP EWPO ({\it Tera}Z) measurements, are shown in Figure \ref{fig:Singlet_CH}. 

\vspace{2mm}

As recently noted in \cite{Henning:2014gca}, a sizeable $\bar c_6 (m_Z)$ could in this scenario lead to a strongly first order EW phase transition, potentially allowing for EW baryogenesis in the Early Universe \cite{Grojean:2004xa,Bodeker:2004ws,Delaunay:2007wb}. Requiring a sufficiently strong first order phase transition imposes a lower bound $\lambda\,\bar c_6 (m_Z) \gtrsim 0.105$\footnote[5]{The EW phase transition occurs at $T = T_c \simeq 100$ GeV, and we can approximate $T_c \simeq m_Z$ for $\bar c_6$.}, while the successful completion of the phase transition sets the upper bound $\lambda\,\bar c_6 (m_Z) \lesssim 0.211$ \cite{Bodeker:2004ws}. As shown in Figure \ref{fig:Singlet_CH} (Right), the present combination of measured Higgs signal strengths and EW precision data
rules out a strong EW phase transition, except for very large values $\delta \equiv \mu_{m}/v_s \gg 1$ combined with relatively low singlet masses $m_s$. 
These results are somewhat stronger than those of \cite{Henning:2014gca}, the reason being that the analysis of the EW phase transition in the presence of $\bar c_6$ from \cite{Grojean:2004xa} (with reference \cite{Henning:2014gca} follows) considers thermal masses as the only thermal effects in the effective potential, and tends to overestimate the strength of the EW phase transition as compared to a full 1-loop analysis \cite{Bodeker:2004ws}. Finally, while we stress that our results certainly do not rule out a strong EW phase transition in the singlet scalar extension of the SM (see {\it e.g.} \cite{Profumo:2014opa}), they imply that having it originate from the presence of $\bar c_6$ is currently challenging, and the entire region of parameter space will be covered by the combination of HL-LHC and {\it GigaZ}. 

\vspace{3mm}

Let us finish this section on the extension of the SM Higgs sector via a singlet scalar by summarizing the results. We have seen that the EFT leads to just two non-zero operators at tree-level,

\begin{empheq}[box=\fbox]{align}
 \bar c_H  &= y^2  \textrm{  (mixing) } \nonumber \\
  \bar c_H  &=  \frac{n_s}{96 \pi^2} \, \left( \frac{\lambda_m v}{m_s} \right)^2 \textrm{  (no mixing) } \nonumber \\ 
  \lambda \,\bar c_6 & = 3\, \,\delta \,  \bar c_H \textrm{  (only w. explicit symmetry breaking) }
 \end{empheq}

%%%%%%%%%%%%%%%%%%%%%%%%%
\subsection{Two Higgs Doublet Models}
%%%%%%%%%%%%%%%%%%%%%%%%%%
\label{2HDMsec}

Theories with two scalar doublets appear in a wide variety of scenarios, ranging from the MSSM and its extensions \cite{Djouadi:2005gj,Ellwanger:2009dp} to 
models of dark matter \cite{Barbieri:2006dq,LopezHonorez:2006gr,Dolle:2009fn} and neutrino masses (see {\it e.g.} \cite{Ma:2006km}). 
The phenomenology of Two Higgs Doublet Model (2HDM) scenarios has been widely studied in the literature (for a general review of 2HDMs, see \cite{Branco:2011iw}), including
the impact of measured properties of the observed Higgs boson on the 2HDM 
parameter space \cite{Celis:2013rcs,Krawczyk:2013gia,Grinstein:2013npa,Chen:2013rba,Eberhardt:2013uba,Belanger:2013xza, Dumont:2014wha,Kanemura:2014bqa}, the possible connection 
to the EW phase transition and baryogenesis \cite{Cline:1996mga,Kanemura:2004ch,Fromme:2006cm,Dorsch:2013wja} and its potential 
LHC signatures \cite{Kanemura:2009mk,Craig:2013hca,Baglio:2014nea,Coleppa:2014hxa,Dorsch:2014qja,Hespel:2014sla,Altmannshofer:2012ar}. 

We consider here 2HDM scenarios with a $\mathbb{Z}_2$-symmetry which is at most softly broken, avoiding tree-level Flavor Changing Neutral Currents (FCNCs) in the Yukawa sector \cite{Glashow:1976nt}. 
This $\mathbb{Z}_2$-symmetry leads to four types of 2HDM (see {\it e.g.} \cite{Branco:2011iw} for details), according to the way in which $\Phi_{1,2}$ are coupled to the different SM fermion
species, as shown in Table \ref{Table_2HDMTypes}. 
\begin{table}[h] 
\begin{center}
\renewcommand{\arraystretch}{1.5} 
\begin{tabular}{c|c|c|c |} 
\, & Up-type Quarks & Down-type Quarks & Charged Leptons \\
\hline
$\mathrm{Type}-\mathrm{I}$& $\Phi_{2}$  & $\Phi_{2}$  & $\Phi_{2}$   \\
 \hdashline
$\mathrm{Type}-\mathrm{II}$& $\Phi_{2}$ &  $\Phi_{1}$  &  $\Phi_{1}$  \\
 \hdashline
$\ell-\mathrm{Specific}$& $\Phi_{2}$  &  $\Phi_{2}$  & $\Phi_{1}$  \\
\hdashline
$\mathrm{Flipped}$& $\Phi_{2}$  & $\Phi_{1}$  & $\Phi_{2}$  \\
\hline
\end{tabular}
\end{center}
%\etb
%\captionsetup{singlelinecheck=off}
\caption[ . ]{Classification of 2HDM with a $\mathbb{Z}_2$-symmetry in the Yukawa sector. For each 2HDM-type, 
we indicate which scalar doublet couples to up-type quarks, down-type quarks and charged 
leptons.}
\label{Table_2HDMTypes}
\end{table} 

The scalar potential for the two scalar doublets $\Phi_{1,2}$ reads
\begin{eqnarray}	
\label{2HDM_potential}
V_{\rm tree}(\Phi_1,\Phi_2)= \mu^2_1 \left|\Phi_1\right|^2 + \mu^2_2\left|\Phi_2\right|^2 - \left[\mu^2\,\Phi_1^{\dagger}\Phi_2+\mathrm{h.c.}\right] +\frac{\lambda_1}{2}\left|\Phi_1\right|^4\nonumber \\
 +\frac{\lambda_2}{2}\left|\Phi_2\right|^4 + \lambda_3 \left|\Phi_1\right|^2\left|\Phi_2\right|^2
+\lambda_4 \left|\Phi_1^{\dagger}\Phi_2\right|^2+
 \frac{1}{2}\left[\lambda_5\left(\Phi_1^{\dagger}\Phi_2\right)^2+\mathrm{h.c.}\right].
\end{eqnarray}
In the following, we consider CP-conserving scenarios and set $\lambda_5$ and $\mu^2$ to be real%(we briefly comment on the CP-violating scenario in section \ref{conclusion})
. The doublets and their {\it vevs} at the EW minimum can be written as
\begin{equation}
\label{2HDMminima}
\Phi_{k} = \left( \begin{array}{c} \varphi_k^+ \\ \frac{h_k+i\eta_k}{\sqrt{2}}  \end{array} \right),\hspace{0.5cm}
\langle\Phi_1\rangle = \left( \begin{array}{c} 0 \\ \frac{v_1}{\sqrt{2}} \end{array} \right),\hspace{0.5cm}
\langle\Phi_2\rangle = \left( \begin{array}{c} 0 \\ \frac{v_2}{\sqrt{2}} \end{array} \right),
\end{equation}
with $v_1^2+v_2^2 = v = 246$~GeV and $v_2/v_1 = \tan\beta$. The angle $\beta$ also parametrizes the rotation to the mass eigenbasis for the charged states $G^{\pm},H^{\pm}$ and neutral CP-odd $G^0,A^0$ states, with $G^{\pm}, G^0$ being the Goldstone bosons and $H^{\pm}, A^0$ the physical states. We likewise define $\alpha$ to be the mixing angle parametrizing the rotation to the mass eigenbasis for the CP-even neutral states $h, H^0$ (see Appendix \ref{Appendix1}). The $\mu$ parameter is responsible for the soft-breaking of the $\mathbb{Z}_2$-symmetry in (\ref{2HDM_potential}). After EW symmetry breaking, the parameters $\mu_i$, $\lambda_i$ in (\ref{2HDM_potential}) may be written in terms of the masses of the physical states $m_{h}$, $m_{H^0}$, $m_{A^0}$, $m_{H^\pm}$, the mixing angles $\alpha$, $\beta$ and the $\mu$ parameter, as shown in Appendix \ref{Appendix1}. 

\vspace{3mm}

In order to obtain an EFT for the SM Higgs, we may perform an $SU(2)$ rotation from the field basis 
$\Phi_1, \Phi_2$ to a basis $H_1, H_2$ in which $\langle H_1\rangle = \frac{v}{\sqrt{2}}$ and $\langle H_2\rangle = 0$ (the so-called {\it Higgs basis}). This rotation is precisely parametrized by the 
angle $\beta$. After the field rotation, the scalar potential for $H_{1,2}$ reads
\begin{eqnarray}	
\label{2HDM_potentialbis}
V_{\rm tree}(H_1,H_2)= \tilde{\mu}^2_1 \left|H_1\right|^2 + \tilde{\mu}^2_2\left|H_2\right|^2 - \tilde{\mu}^2 \left[H_1^{\dagger}H_2+\mathrm{H.c.}\right] +\frac{\tilde{\lambda}_1}{2}\left|H_1\right|^4 \nonumber \\
 +\frac{\tilde{\lambda}_2}{2}\left|H_2\right|^4 + \tilde{\lambda}_3 \left|H_1\right|^2\left|H_2\right|^2
+\tilde{\lambda}_4 \left|H_1^{\dagger}H_2\right|^2+
 \frac{\tilde{\lambda}_5}{2}\left[\left(H_1^{\dagger}H_2\right)^2+\mathrm{H.c.}\right] \nonumber \\
 + \tilde{\lambda}_6 \left[  \left|H_1\right|^2 H_1^{\dagger}H_2 +\mathrm{H.c.}\right] + \tilde{\lambda}_7 \left[  \left|H_2\right|^2 H_1^{\dagger}H_2 +\mathrm{H.c.}\right]
\end{eqnarray}
with $\tilde{\mu}_i$, $\tilde{\mu}$ and $\tilde{\lambda}_i$ being functions of the original parameters in (\ref{2HDM_potential}). We note that the field rotation may generate $\tilde{\lambda}_{6,7}$ even if initially absent from (\ref{2HDM_potential}) due to the $\mathbb{Z}_2$-symmetry.

\vspace{3mm}

We may then construct an effective Lagrangian for $H_1$ by matching to the theory with the second doublet $H_2$ at the scale $\tilde{\mu}^2$, assuming $\tilde{\mu}^2_2 \gg v^2$. We stress however that 
the doublet $H_1$ can only be fully identified with the SM Higgs doublet in the {\it alignment limit} $\mathrm{cos}(\beta-\alpha) \equiv c_{\beta-\alpha} \to 0$, where mixing in the CP-even sector
is absent (see the discussion in Appendix \ref{Appendix1}). Away from alignment ($\alpha \neq \beta - \pi/2$), these mixing effects lead to tree-level modifications of SM Higgs couplings (see {\it e.g.} the discussion in \cite{Englert:2014uua}). For a general 2HDM, $c_{\beta-\alpha} = 0$ is possible at tree-level\footnote[6]{In contrast, for the MSSM $c_{\beta-\alpha} \to 0$ is only obtained at tree-level in the {\it decoupling limit} $\tilde{\mu}^2_2 \to \infty$. Moreover, $c_{\beta-\alpha} = 0$ at loop level is only possible in a very small portion of parameter space \cite{Gunion:2002zf,Carena:2013ooa}.} \cite{Gunion:2002zf,Carena:2013ooa}, and indeed the latest analyses of Higgs data by ATLAS and CMS strongly prefer $c_{\beta-\alpha}^2 \ll 1$, as shown in section \ref{Sec_2HDM_Constraints}. 

\vspace{3mm}

In the following, we introduce and discuss two benchmark scenarios for the use of Higgs Effective Theory in 2HDM: An exact alignment scenario $c_{\beta-\alpha} = 0$ and a scenario with 
$\left|c_{\beta-\alpha}\right| \ll 1$ (MSSM-like scenario). We then analyze the constraints on the 2HDM parameter space obtained from measurement of Higgs signal strengths and EW Oblique Parameters,
and the corresponding constraint on the values of the Wilson coefficients of the Higgs EFT. 

\subsubsection{Benchmark A: Exact Alignment $c_{\beta-\alpha} = 0$}
\label{BenchmarkA}

As discussed above, in this scenario mixing effects are absent and $H_1$ is precisely the SM Higgs doublet $\Phi$. The scalar potential (\ref{2HDM_potentialbis}) 
simplifies in this limit, since $\tilde{\lambda}_6$ and $\tilde{\mu}$ are both $\propto c_{\beta-\alpha}$ and vanish as shown
in (\ref{modifiedmassesalignment})-(\ref{modifiedcouplingsalignment}). We then match the 2HDM to an $SU(2)_L \times U(1)_Y$ invariant EFT with the field content of the SM,
being $H_2$ the only massive field which will be decoupled in the matching calculation. The $D=6$ effective operators from (\ref{eq:effL}) are generated first at 1-loop order, with the corresponding Wilson coefficients given by

\vspace{2mm}

\begin{empheq}[box=\fbox]{align}
\label{Cgamma_UB}
\bar{c}_{H} &=-\left[-4 \tilde{\lambda}_3 \tilde{\lambda}_4 + \tilde{\lambda}_4^2 + \tilde{\lambda}_5^2 - 
    4 \tilde{\lambda}_3^2 \right] \frac{v^2}{192\, \pi^2\, \tilde \mu_2^2 } \nonumber \\
\bar{c}_{6} &=-\left(\tilde{\lambda}_4^2 + \tilde{\lambda}_5^2 \right) \frac{v^2}{192\, \pi^2\, \tilde \mu_2^2 } \nonumber \\
\bar{c}_{T} &=(\tilde{\lambda}_4^2  - \tilde{\lambda}_5^2) \, \frac{v^2}{192 \, \pi^2 \,\tilde  \mu_2^2 } \nonumber \\
\bar{c}_{\gamma} &= \frac{m^2_W\,\tilde{\lambda}_{3}}{256\, \pi^2\, \tilde{\mu}_2^2} \nonumber \\
\bar{c}_{W} &= - \bar{c}_{HW} = \frac{m^2_W\,(2\,\tilde{\lambda}_{3}+ \tilde{\lambda}_{4})}{192\, \pi^2\, \tilde{\mu}_2^2} = \frac{8}{3} \, \bar{c}_{\gamma} + \frac{m^2_W\, \tilde{\lambda}_{4}}{192\, \pi^2\, \tilde{\mu}_2^2}\nonumber \\
\bar{c}_{B} &= - \bar{c}_{HB} = \frac{m^2_W\,(- 2\,\tilde{\lambda}_{3}+ \tilde{\lambda}_{4})}{192\, \pi^2\, \tilde{\mu}_2^2} = - \frac{8}{3} \, \bar{c}_{\gamma} + \frac{m^2_W\,\tilde{\lambda}_{4}}{192\, \pi^2\, \tilde{\mu}_2^2} \nonumber \\
\bar{c}_{3W} &= \frac{\bar{c}_{2W}}{3} = \frac{m^2_W\,}{1440\, \pi^2\, \tilde{\mu}_2^2}
 \end{empheq}

\vspace{2mm}

\noindent from which we immediately obtain
\be
\label{CW+CB}
\bar{c}_{W} + \bar{c}_{B} = \frac{m^2_W\, \tilde{\lambda}_{4}}{96\, \pi^2\, \tilde{\mu}_2^2} \quad \quad , \quad \quad \bar{c}_{W} - \bar{c}_{B} = \frac{16}{3} \, \bar{c}_{\gamma} =  \frac{m^2_W\, \tilde{\lambda}_{3}}{48\, \pi^2\, \tilde{\mu}_2^2}
\ee
Before we continue, let us note that $\bar{c}_{\gamma}$, $\bar{c}_{W}$ and $\bar{c}_{B}$ may take positive and negative values, as the bounded-from-below conditions (\ref{Stability}) 
do not restrict either possibility. The various relations among the Wilson coefficients $\bar{c}_{HW}$, $\bar{c}_{W}$, $\bar{c}_{HB}$, $\bar{c}_{B}$ and $\bar{c}_{\gamma}$ 
in (\ref{Cgamma_UB}) imply that the $D =6$ effective operators $\mathcal{O}_{HW}$, $\mathcal{O}_{W}$, $\mathcal{O}_{HB}$, $\mathcal{O}_{B}$ and $\mathcal{O}_{\gamma}$ can in 
fact be re-casted in terms of three operators
\be
\label{eq:silh2HDM}
 \frac{g^2\,\bar c_{\sss WW}}{\mW^2} | H_1|^2 \, W_{\mu \nu}^k W^{\mu \nu}_k  +  \frac{2 g\,g'\,\bar c_{\sss WB}}{\mW^2} \big[ H_1^\dag  T_{2k}\, H_1 \big] W_{\mu \nu}^k B^{\mu \nu} 
 +\bar c_{\gamma} \, \mathcal{O}_{\gamma}
\ee
with
\bea
\label{eq:silh2HDM_operator1}
\frac{g^2}{\mW^2} | H_1|^2 \, W_{\mu \nu}^k W^{\mu \nu}_k  &\equiv& \mathcal{O}_{WW} = 4\,(\mathcal{O}_{W} - \mathcal{O}_{B} + \mathcal{O}_{HB} - \mathcal{O}_{HW}) + \mathcal{O}_{\gamma} \\
\label{eq:silh2HDM_operator2}
\frac{2 g\,g'}{\mW^2} \big[ H_1^\dag  T_{2k}\, H_1 \big] W_{\mu \nu}^k B^{\mu \nu}  &\equiv& \mathcal{O}_{WB}  =  4\,(\mathcal{O}_{B} - \mathcal{O}_{HB}) - \mathcal{O}_{\gamma}
\eea
%
%
%%% V. I would remove the next
%The value of $\bar c_{\sss WW}$ and $\bar c_{\sss WB}$ is given by virtue of (\ref{Cgamma_UB}) and (\ref{eq:silh2HDM_operator1}-\ref{eq:silh2HDM_operator2}) by
%
%\be
%\label{eq:silh2HDM_Wilson}
 %\bar c_{\sss WW} = - \frac{125 m^2_W\,\tilde{\lambda}_{3}}{768\, \pi^2\, \tilde{\mu}_2^2}\quad , \quad \quad \bar c_{\sss WB} = 
%t\ee
%
%
The three operators in (\ref{eq:silh2HDM}) share the common feature that they do not involve derivatives of the Higgs field $H_1$. This very property is in fact responsible for 
the relations among Wilson coefficients in (\ref{Cgamma_UB}), since the Feynman diagrams involved in the EFT matching (see Figure \ref{Figure:Feynman_fermion}-Left)
do not involve Higgs field derivatives.

\subsubsection{Benchmark B: Departure from Alignment $c_{\beta-\alpha} \ll 1$ (MSSM-Like)}
\label{Sec_MSSM}

Upon departure from the alignment limit, mixing in the CP-even sector leads to several effects that are absent for $c_{\beta-\alpha} = 0$: First, there 
is a modification of the couplings of the Higgs boson $h$ to gauge bosons and fermions at tree-level, parametrized in terms of
$\kappa$-factors for vector bosons $\kappa_V \equiv g_{hV_{\mu} V_{\nu}}/g^{\mathrm{SM}}_{hV_{\mu} V_{\nu}}$ and fermions 
$\kappa_f \equiv g_{h\bar f f}/g^{\mathrm{SM}}_{h\bar f f}$ (for up-type quarks $\kappa_u$, down-type quarks $\kappa_d$ and leptons $\kappa_{\ell}$). For the various 
Types of 2HDM from Table \ref{Table_2HDMTypes}, these are given in terms of $c_{\beta-\alpha}$ and $t_{\beta} \equiv \mathrm{tan}\beta$ as 
\begin{eqnarray}
\label{t:kappaI}
\mathrm{Type}-\mathrm{I}: & \kappa_V = s_{\beta-\alpha} \,\,; & \kappa_u = \kappa_d = \kappa_{\ell} = \frac{c_{\beta-\alpha}}{t_{\beta}} + s_{\beta-\alpha}\\
\label{t:kappaII}
\mathrm{Type-II:} & \kappa_V = s_{\beta-\alpha}\,\,; & \kappa_u = \frac{c_{\beta-\alpha}}{t_{\beta}} + s_{\beta-\alpha} \,\,; \kappa_d = \kappa_{\ell} = s_{\beta-\alpha} - t_{\beta}\,c_{\beta-\alpha}  \\
\label{t:kappaLS}
\ell-\mathrm{Specific:} & \kappa_V = s_{\beta-\alpha} \,\,;& \kappa_u = \kappa_d = \frac{c_{\beta-\alpha}}{t_{\beta}} + s_{\beta-\alpha} \,\,;\kappa_{\ell} = s_{\beta-\alpha} - t_{\beta}\,c_{\beta-\alpha}\\
\label{t:kappaF}
\mathrm{Flipped:} & \kappa_V = s_{\beta-\alpha} \,\,;& \kappa_u = \kappa_{\ell} = \frac{c_{\beta-\alpha}}{t_{\beta}} + s_{\beta-\alpha} \,\,;\kappa_{d} = s_{\beta-\alpha} - t_{\beta}\,c_{\beta-\alpha}
\end{eqnarray}
In addition, away from alignment $g_{H^0hh}$ (coupling between $H^0$ and two Higgs bosons $h$), $g_{H^0V_{\mu} V_{\nu}}$ (coupling 
between $H^0$ and two gauge bosons $V_{\mu}$) and $g_{\phi V_{\mu} h}$ (couplings among $h$, a gauge boson $V_{\mu}$ and $\phi = A^0,H^{\pm}$) are non-zero, giving rise to 
1-loop diagrams contributing to the vertices $V_{\mu} V_{\nu} h$ and $V_{\mu} V_{\nu} V_{\rho}$ with both heavy ($H^0,A^0,H^{\pm}$) and light ($h,V_{\mu}$) states running in the loop.

%These will appear in three ways: first, there will be mixing effects at tree-level that will
%affect the couplings of $h$ to gauge bosons and fermions. Second, since now the $g_{\phi V_{\mu} V_{\nu}}$ and $g_{\phi hh}$ couplings (with $\phi = H^0,A^0,H^{\pm}$)
%are non-zero, integrating-out the new heavy states $\phi$ will yield further tree-level modifications of the SM interactions $V_{\mu} V_{\nu} V_{\rho} V_{\sigma}$ 
%and $V_{\mu} V_{\nu} h\,h$. Third, since also the couplings $g_{\phi V_{\mu} h}$ are non-zero, this will give rise to extra 1-loop diagrams contributing to the cubic 
%vertices $V_{\mu} V_{\nu} h$ and $V_{\mu} V_{\nu} V_{\rho}$, with both heavy states $\phi$ and $h/V^{\mu}$ states entering the loop.

\vspace{3mm}

Let us discuss all these effects in an $SU(2)_L \times U(1)_Y$ invariant Higgs EFT approach. 
We first note that $\langle H_1\rangle = \frac{v}{\sqrt{2}}$ and $\langle H_2\rangle = 0$ in (\ref{2HDM_potentialbis}) imply the relations 
\be
\label{mini}
\tilde{\mu}^2_1 = - \frac{\tilde{\lambda}_1\, v^2}{2} \quad \quad , \quad \quad \tilde{\mu}^2 = \tilde{\lambda}_6\, v^2
\ee
which may be simply regarded as minimization conditions. The mass matrix for the neutral CP-even states is non-diagonal, brought into a diagonal form via the rotation matrix $U$
\be
\label{rotation_Nonalignment}
U = \left( \begin{array}{cc} 1 & \epsilon \\ 
- \epsilon & 1 \end{array} \right) \hspace{2cm} \epsilon = -\frac{\tilde{\mu}^2}{2\,\tilde{\mu}_2^2} + \frac{3}{2}\frac{\tilde{\lambda}_6\,v^2}{\tilde{\mu}_2^2} = \frac{\tilde{\lambda}_6\,v^2}{\tilde{\mu}_2^2}
\ee
The deviation from alignment is then parametrized by $(\tilde{\lambda}_6 v^2)/\tilde{\mu}_2^2$, which matches the corresponding expression for $c_{\beta-\alpha}$ after EW symmetry breaking (see
Appendix \ref{Appendix1})  
\be
c^2_{\beta-\alpha} \sim \frac{(\tilde{\lambda}_6 v^2)^2}{\tilde{\mu}_2^4} \sim \frac{v^4}{\tilde{\mu}_2^4}  \ll 1\, ,
\ee
recovering the well-known scaling result for $c^2_{\beta-\alpha}$ in 2HDMs as the decoupling limit is approached \cite{Gunion:2002zf}.

At tree-level, the EFT matching generates effective operators suppressed at least by $1/\tilde{\mu}_2^4$
(except for $\bar c_6$, which receives a further contribution suppressed only by $1/\tilde{\mu}_2^2$), with operators of different $D$ contributing to order $v^4/\tilde{\mu}_2^4$: 
\bea
  \label{eq:dim-8}
    \mathcal{L}_{\mathrm{Eff}}^{\mathrm{tree}} & \supset &
    \frac{\bar c_{2h}\,\tilde{\mu}^4}{\tilde{\mu}_2^4}\big( D_{\mu} H_1^{\dagger} \big) \big( D^{\mu} H_1 \big) +
    \frac{\bar c_{4h}\, \tilde{\mu}^2}{\tilde{\mu}_2^4}    \partial^\mu| H_1|^2 \partial_\mu | H_1|^2  \nonumber \\
    &+ &  \frac{\bar c_{6h}}{\tilde{\mu}_2^4}  D_{\mu} \big( H_1^{\dagger} | H_1|^2  \big)
     D^{\mu} \big( H_1 | H_1|^2 \big) - \left( \frac{\tilde \lambda_6 \, \tilde{\mu}^2}{\tilde{\mu}_2^4} - \frac{\tilde \lambda_6^2}{\lambda\, \tilde{\mu}_2^2} \right)\, |H_1|^6\, .
\eea
The matching procedure yields $\bar c_{6h} = \tilde \lambda_6^2$, $\bar c_{4h} = \bar c_H = 0$ and $\bar c_{2h} = 1$, and use of the relation $\tilde{\mu}^2 = \tilde{\lambda}_6\, v^2$
results in both the first and the third term in (\ref{eq:dim-8}) contributing at order $(\tilde{\lambda}_6 v^2)^2/\tilde{\mu}_2^4$. 
The first term rescales both the SM Higgs kinetic term and its couplings to gauge boson by the same amount, so it does not have a net effect. The third term does however include
a rescaling of the SM Higgs kinetic term $ \propto c^2_{\beta-\alpha}$ that is not compensated by a similar one in the gauge boson interactions, leading to the well-known tree-level 
deviation from the SM Higgs couplings to gauge bosons away from alignment, proportional to $c^2_{\beta-\alpha}$.

\vspace{2mm}

Regarding $\bar c_6$, the presence of a non-zero $\tilde \lambda_6$ also yields an extra contribution at 1-loop as compared to (\ref{Cgamma_UB}). The full result for $\bar c_6$
away from alignment is 
\be
\bar{c}_{6} =- \frac{\left(\tilde{\lambda}_4^2 + \tilde{\lambda}_5^2 + 12\,\tilde{\lambda}_6^2 \right)v^2}{192\, \pi^2\, \tilde \mu_2^2 } - \left(\frac{\tilde \lambda_6^2\, v^2}{\lambda\, \tilde{\mu}_2^2} 
-\frac{\tilde \lambda_6 \, \tilde{\mu}^2\, v^2}{\tilde{\mu}_2^4}\right)
\ee

\vspace{2mm}

Let us also stress that similarly to the tree-level effects just discussed, the extra 1-loop diagrams appearing away from alignment (involving both light and heavy degrees of freedom) 
are proportional to $c^2_{\beta-\alpha}$. Thus the contribution of these diagrams to the various Wilson coefficients from (\ref{eq:silh}) and (\ref{eq:lagG}) is at least of 
order $(v^4/\tilde{\mu}_2^4) \times${\sl1-loop} and can be safely neglected. As a result, the values of the Wilson coefficients $\bar{c}_T$, $\bar{c}_{HW}$, $\bar{c}_W$, $\bar{c}_{HB}$,
$\bar{c}_B$, $\bar{c}_{\gamma}$ for $c^2_{\beta-\alpha} \ll 1$ remains unchanged {\it w.r.t.} the {\it alignment} scenario.

\begin{figure}[ht!]
\begin{center}
\includegraphics[width=0.35\textwidth]{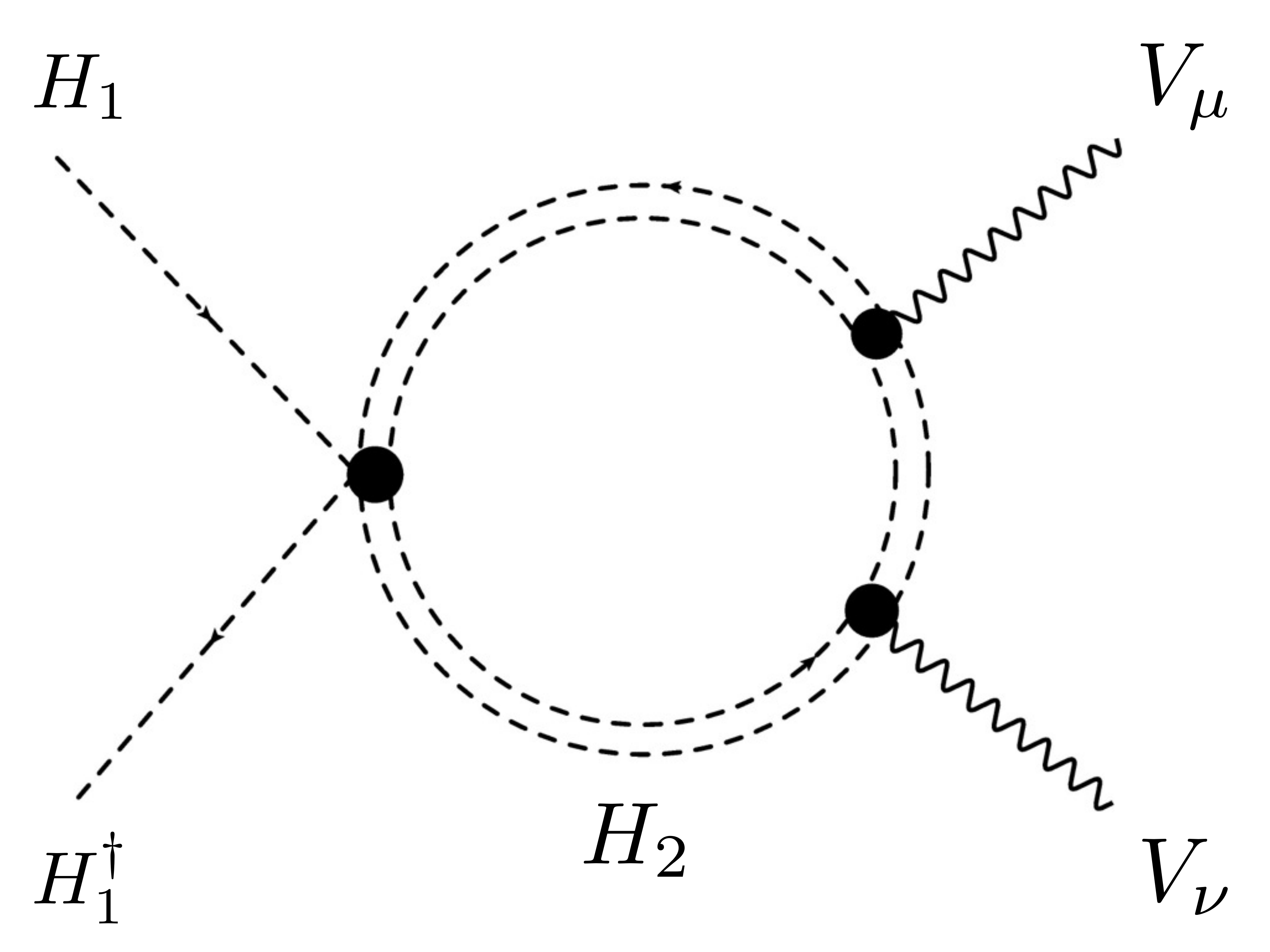} \hspace{9mm}
\includegraphics[width=0.35\textwidth]{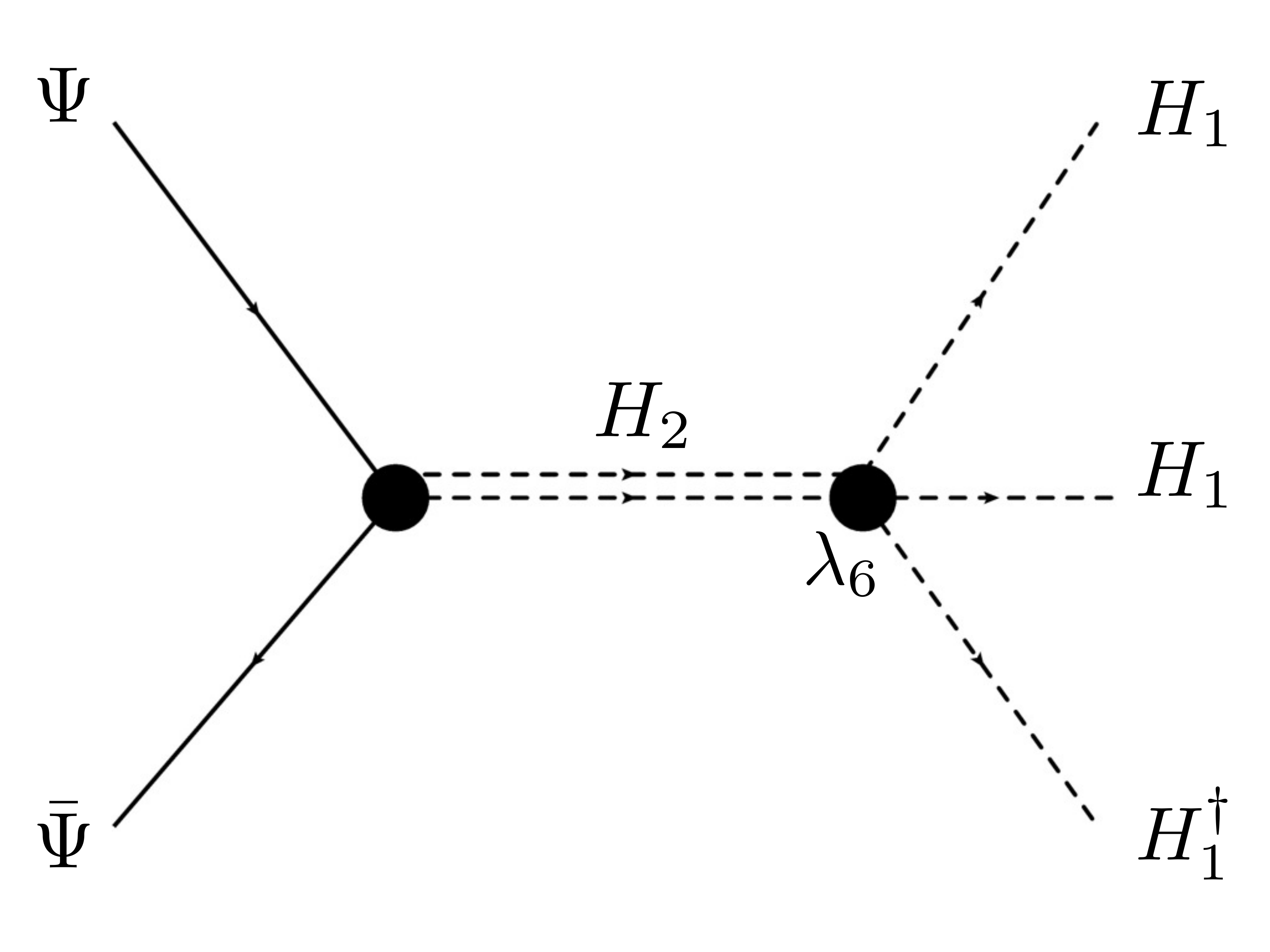}
\caption{Left: Feynman diagram responsible for anomalous Higgs couplings to gauge boson at 1-loop. Right: Feynman diagram responsible for anomalous Higgs couplings to fermions at tree-level.}
\label{Figure:Feynman_fermion}
\end{center}
\end{figure}

The interactions between the SM Higgs and fermions receive tree-level corrections of the form shown in Figure \ref{Figure:Feynman_fermion}-Right away from the alignment limit, encoded in the 
$D = 6$ effective operators 
\be
\mathcal{L}_{\mathrm{SILH}}^{(6)} = \left[ \frac{\bar c_{u} \, y_u}{v^2} H_1^\dag H_1 \, {\bar Q}_L H_1^{\dagger} u_R
  +\frac{\bar c_{d} \, y_d}{v^2} H_1^\dag H_1 \, {\bar Q}_L H_1\, d_R 
  +\frac{\bar c_{\ell} \, y_{\ell}}{v^2} H_1^\dag H_1 \, {\bar L}_L H_1\, \ell_R \right],
\ee
with 
\be
\bar c_{u} \, y_u = \frac{m_u\,f_u}{v} \, \frac{\tilde{\lambda}_6\, v^2}{\tilde{\mu}_2^2} \quad 
\bar c_{d} \, y_d = \frac{m_d\,f_d}{v} \, \frac{\tilde{\lambda}_6\, v^2}{\tilde{\mu}_2^2} \quad 
\bar c_{\ell} \, y_{\ell} = \frac{m_{\ell}\, f_{\ell}}{v} \, \frac{\tilde{\lambda}_6\, v^2}{\tilde{\mu}_2^2}
\ee
and $f_{u,d,\ell}$ depending of the 2HDM-Type under consideration, and given by
\begin{eqnarray}
\label{t:fs}
\mathrm{Type}-\mathrm{I}: & f_u = f_d = f_{\ell} = t^{-1}_{\beta} \\
\mathrm{Type-II:} & f_u = t^{-1}_{\beta} \,\,; f_d = f_{\ell} =  - t_{\beta}  \\
\ell-\mathrm{Specific:} & f_u = f_d = t^{-1}_{\beta} \,\,;f_{\ell} =  - t_{\beta}\\
\mathrm{Flipped:} & f_u = f_{\ell} = t^{-1}_{\beta} \,\,;f_{d} = - t_{\beta}
\end{eqnarray}
These results reproduce the $\kappa$-factors from (\ref{t:kappaI}-\ref{t:kappaF}) to leading order in $c_{\beta-\alpha}$.

%%%%%%%%%%%%%%%%%%%%%%%%%%%%%%
\subsubsection{2HDM Constraints: Higgs Signal Strengths \& EW Oblique Parameters}
\label{Sec_2HDM_Constraints}
%%%%%%%%%%%%%%%%%%%%%%%%%%%%%% 

The ATLAS and CMS measurements of Higgs boson signal strengths from Table \ref{table:ATLAS_CMS} constrain the allowed parameter region of the 2HDM in the plane 
($c_{\beta-\alpha}$, $\mathrm{tan} \, \beta$) (see {\it e.g.} the analyses from \cite{Craig:2013hca,Celis:2013rcs,Grinstein:2013npa,Chen:2013rba,Eberhardt:2013uba,Dumont:2014wha})
through a combined $\chi^2$ fit, which in this case treats the different signal strength measurements as independent (not correlated) 
\be
\label{chi_Higgs2HDM}
\chi^2_h(c_{\beta-\alpha},t_{\beta}) = \sum_i \left(\frac{\mu_i - \bar \mu_i (c_{\beta-\alpha},t_{\beta})}{\Delta \mu_i}\right)^2\, .
\ee
Each $\bar \mu_i$ (the expected signal strength for each channel) may be expressed in terms of rescaling $\kappa$-factors for the coupling of the Higgs boson to vector bosons $\kappa_V$, 
up-type quarks $\kappa_u$, down-type quarks $\kappa_d$ and leptons $\kappa_{\ell}$. The expression of each $\bar \mu_i$ corresponding to each process considered in Table \ref{table:ATLAS_CMS}
in terms of these $\kappa$-factors is given in table \ref{table:kappas}
\begin{table}[ht]
\hspace{2.5cm}
\begin{tabular}{|l| c |}
& $\bar \mu \equiv \mu(\kappa)/\mu_{\mathrm{SM}}$ \\
\hline 
$\gamma \gamma$: $\mu_{\mathrm{ggF}}$ & $\kappa^2_u\times \kappa_{\gamma}(\kappa_u,\kappa_V)^2\times\Gamma(\kappa_d,\kappa_V,\kappa_u,\kappa_{\ell})^{-1} $ \\ 
$\gamma \gamma$: $\mu_{\mathrm{VBF}}$& $\kappa^2_V\times \kappa_{\gamma}(\kappa_u,\kappa_V)^2\times\Gamma(\kappa_d,\kappa_V,\kappa_u,\kappa_{\ell})^{-1}$ \\ 
$WW^*$: $\mu_{\mathrm{ggF}}$-0/1-jet  & $\kappa_u^2 \times \kappa^2_V\times\Gamma(\kappa_d,\kappa_V,\kappa_u,\kappa_{\ell})^{-1}$\\ 
$WW^*$: $\mu_{\mathrm{VBF}}$  & $\kappa^4_V\times\Gamma(\kappa_d,\kappa_V,\kappa_u,\kappa_{\ell})^{-1}$\\ 
$ZZ^*$: Inclusive  & $\kappa_u^2\times \kappa_V^2\times\Gamma(\kappa_d,\kappa_V,\kappa_u,\kappa_{\ell})^{-1}$\\ 
$b\bar{b}$: $\mu_{\mathrm{VH}}$  & $\kappa_V^2\times \kappa^2_d\times\Gamma(\kappa_d,\kappa_V,\kappa_u,\kappa_{\ell})^{-1}$\\ 
$\tau \tau$: $\mu_{\mathrm{ggF}}$-0/1-jet & $\kappa_u^2 \times \kappa_{\ell}^2\times\Gamma(\kappa_d,\kappa_V,\kappa_u,\kappa_{\ell})^{-1}$\\ 
$\tau \tau$: $\mu_{\mathrm{VBF}}$ & $\kappa_V^2\times \kappa_{\ell}^2\times\Gamma(\kappa_d,\kappa_V,\kappa_u,\kappa_{\ell})^{-1}$\\
\hline
\end{tabular}
\caption{\small Modified (by $\kappa$-factors) Higgs Signal Strengths $\bar \mu_i$.}
\label{table:kappas}
\end{table}

\noindent with $\kappa_i$ defined in (\ref{t:kappaI}-\ref{t:kappaF}) and $\kappa_{\gamma}(\kappa_u,\kappa_V) = 1.26\, \kappa_V - 0.26\, \kappa_u$, and $\Gamma(\kappa_d,\kappa_V,\kappa_u,\kappa_{\ell})$ given by
\begin{equation}
\Gamma(\kappa_d,\kappa_V,\kappa_u,\kappa_{\ell}) = 0.2427 \, \kappa^2_V + 0.1124 \, \kappa_u^2 + 0.578 \, \kappa_d^2 + 0.0637 \, \kappa^2_{\ell}
\end{equation}

\begin{figure}[ht!]
\begin{center}
\includegraphics[width=0.485\textwidth]{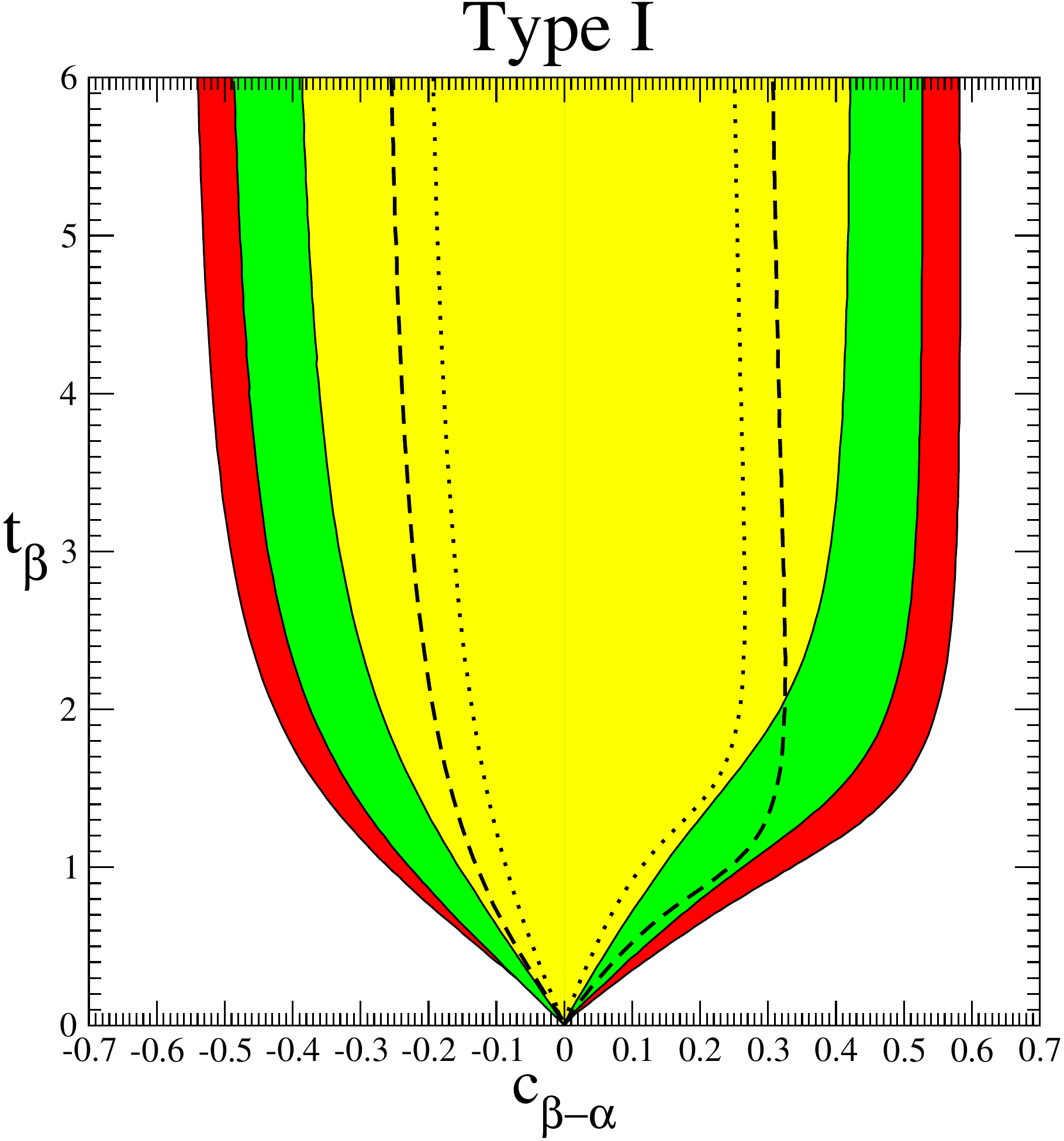} \hspace{1mm}
\includegraphics[width=0.485\textwidth]{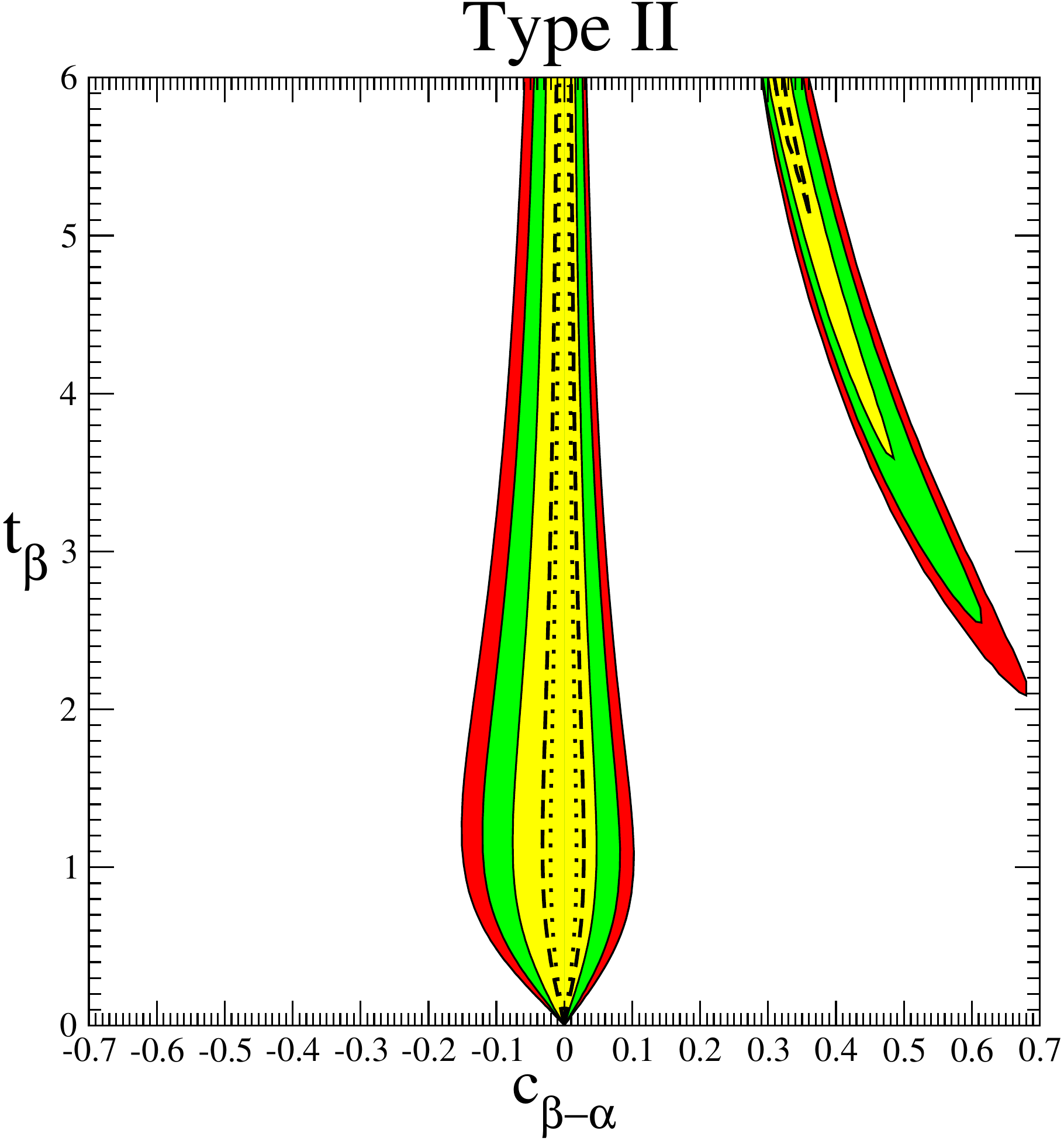}
\includegraphics[width=0.485\textwidth]{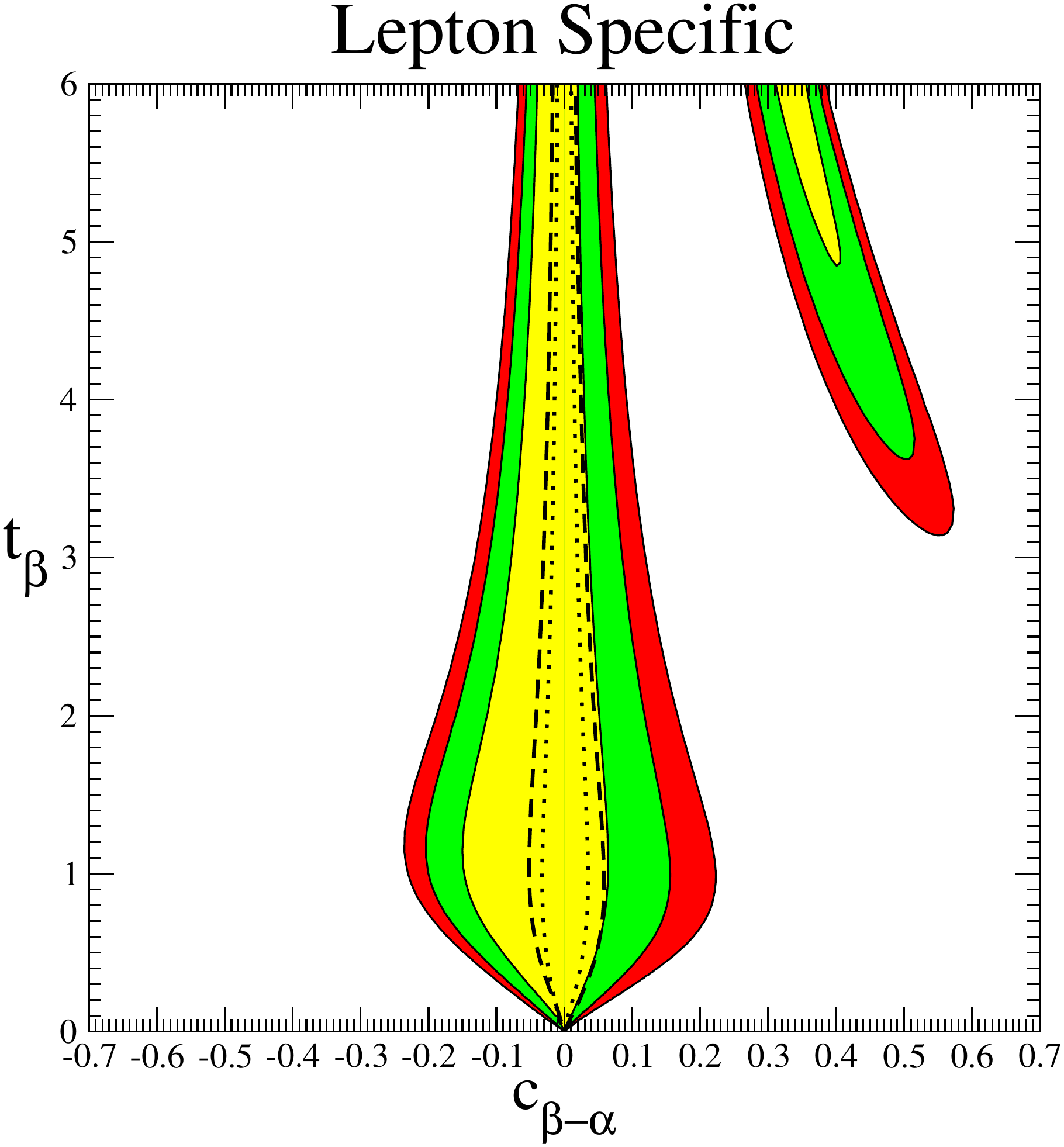} \hspace{1mm}
\includegraphics[width=0.485\textwidth]{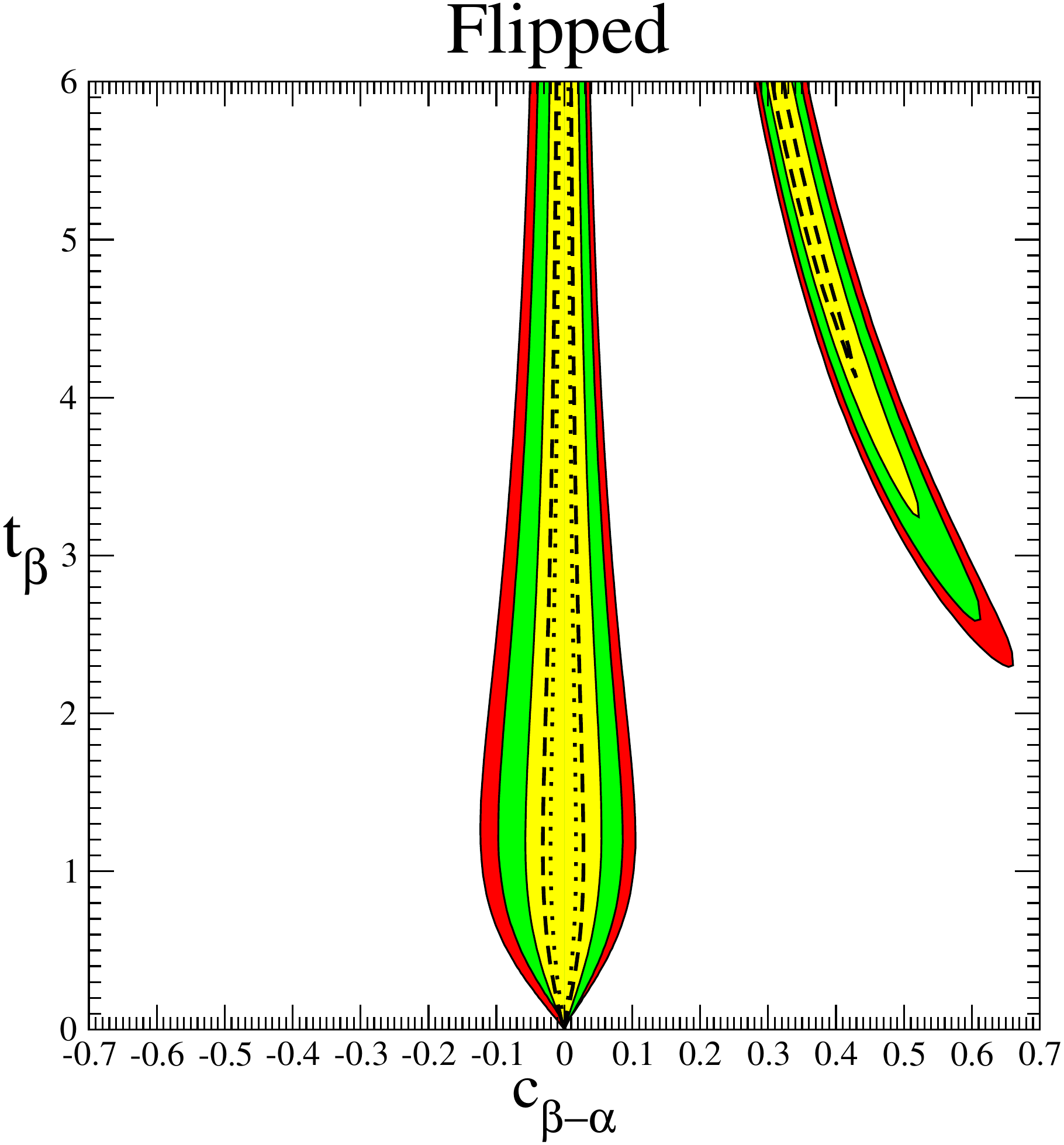}
\caption{68\% C.L. (yellow), 95\% C.L. (green) and 99\% C.L. (red) allowed regions in the $(\mathrm{cos} (\beta-\alpha),\, \mathrm{tan}\,\beta)$ plane from up-to-date measurements of Higgs boson 
couplings (see text for details), shown respectively for 2HDM of Type I (Upper-Left), Type II (Upper-Right), Lepton-Specific (Lower-Left) and Flipped (Lower-Right). In each case, 
the projected $95\%\, \mathrm{C.L.}$ exclusion limit for LHC at 14 TeV with $\mathcal{L}=300 \, \,\mathrm{fb}^{-1}$ (dashed-black) and $\mathcal{L}=3000\, \, \mathrm{fb}^{-1}$ (dotted-black), assuming the measured Higgs signal strengths being $\mu_i = 1$, is also shown.}
\label{fig:a-b}
\end{center}
\end{figure}

We present the $68\%$, $95\%$ and $99\%\, \mathrm{C.L.}$ limits on the $(c_{\beta-\alpha},t_{\beta})$ plane from global fits of the light Higgs boson couplings, given respectively by $\Delta\chi^2_h(c_{\beta-\alpha},t_{\beta}) \equiv \chi^2_h(c_{\beta-\alpha},t_{\beta}) - \chi^2_{\mathrm{min}} = 2.27$, $\Delta\chi^2_h(c_{\beta-\alpha},t_{\beta}) = 5.99$ and $\Delta\chi^2_h(c_{\beta-\alpha},t_{\beta}) = 9.21$, for the different Types of 2HDM in Figure \ref{fig:a-b}. These are particularly stringent for Type II and Flipped as compared to 
Lepton-Specific and specially to Type I, due to the different nature of the Higgs coupling to down-type quarks. Figure \ref{fig:a-b} also includes the projected
$95\%\, \mathrm{C.L.}$ exclusion limits for LHC at 14 TeV with $\mathcal{L}=300 \, \,\mathrm{fb}^{-1}$ (dashed-black) and $\mathcal{L}=3000\, \, \mathrm{fb}^{-1}$ (HL-LHC, dotted-black) assuming the measured Higgs signal strengths being $\mu_i = 1$ and using the projected CMS sensitivities from Section \ref{singlet_section}. 

\begin{figure}[ht!]
\begin{center}
\includegraphics[width=0.48\textwidth]{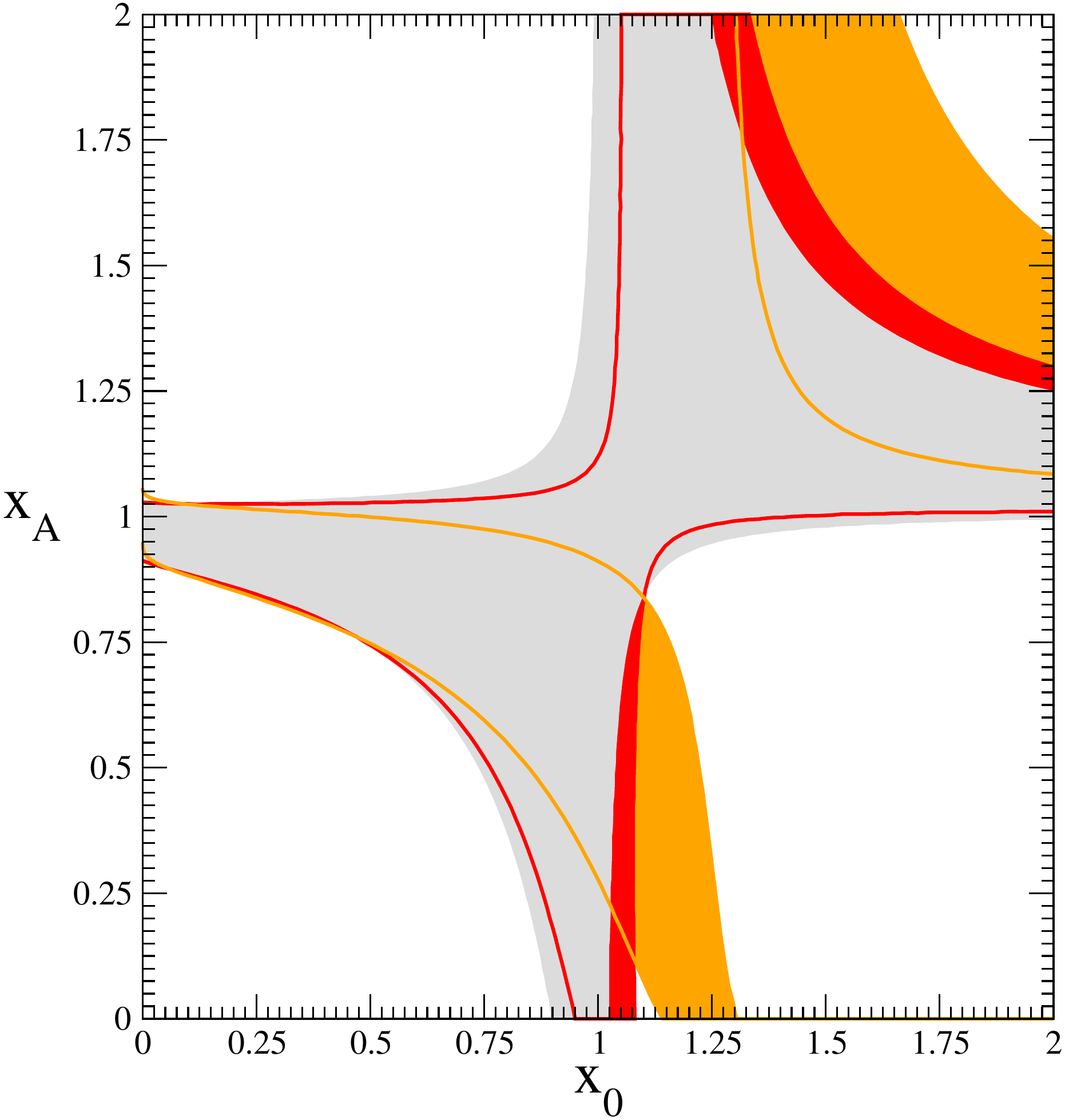} \hspace{2mm}
\includegraphics[width=0.48\textwidth]{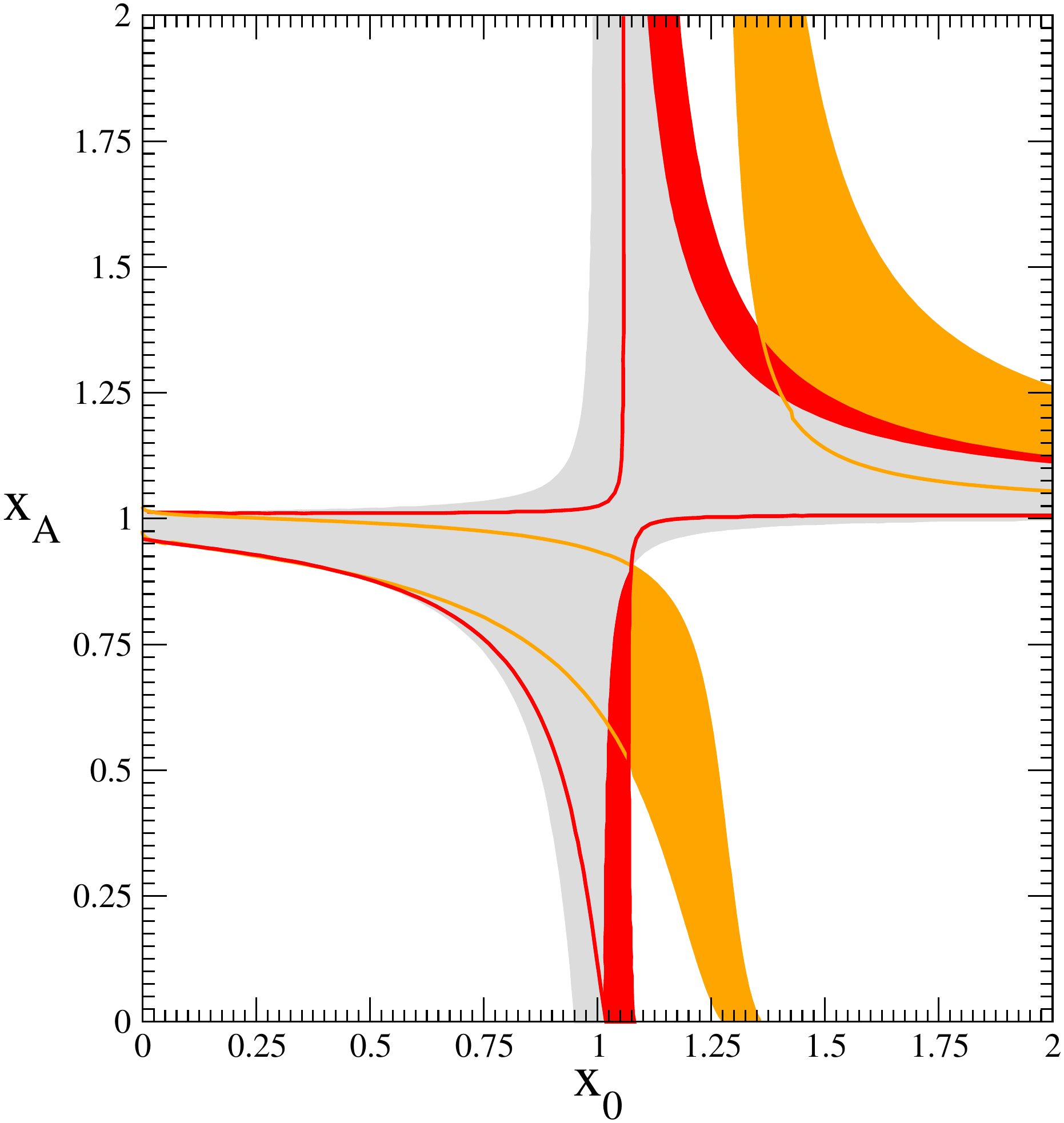}
%\hspace{3mm} \includegraphics[width=0.45\textwidth]{EWPO_Scan_2HDM2.pdf}
\caption{95\% C.L. allowed region from the 2HDM fit to $S,T$ oblique parameters (\ref{chi_EWPO_2HDM}) in the plane ($x_0$, $x_A$) for $m_{H^{\pm}} = 400$ GeV (Left) and $m_{H^{\pm}} = 600$ GeV (Right). The allowed regions correspond respectively to $c_{\beta-\alpha} = 0$ (grey, light), $c_{\beta-\alpha} = 0.2$ (red, dark) and $c_{\beta-\alpha} = 0.4$ (orange, medium).}
\label{fig:ST}
\end{center}
\end{figure}

We now analyze the constraints from EWPO, performing a fit to the oblique parameters $S,T$ under the assumption $U = 0$ using 
the best-fit values and standard deviations from the global analysis of the GFitter Group \cite{Baak:2014ora} ($m_t = 173$ GeV and $m_h = 126$ GeV), as shown in 
(\ref{chi_EWPO1}). The 2HDM contributions to $\Delta S$ and $\Delta T$ are given by \cite{Grimus:2008nb}
\bea
\label{EWPO_2HDM_S}
\Delta S &=& \frac{g^2\,\sW^2}{96\, \pi^2 \,\alpha_{\mathrm{EM}}} \left[ (1-2\sW^2)^2 \,G_{H^{\pm},H^{\pm}, Z} + G_{A^0,H^0, Z} + \mathrm{log}\left(m^2_{A^0}m^2_{H^0}/m^4_{H^{\pm}}\right) \right. \nonumber \\
  &+& \left. c^2_{\beta-\alpha} \left( G_{A^0,h,Z} - G_{A^0,H^0, Z} + \hat G_{H^0,Z} - \hat G_{h, Z}\right) \right] 
\eea
% 
% s^2_{\beta-\alpha}\left( 
\bea
\label{EWPO_2HDM_T}
\Delta T &=& \frac{F_{H^{\pm},A^{0}} +F_{H^{\pm},H^{0}} - F_{A^0,H^0} }{16\, \pi^2 \,v^2 \,\alpha_{\mathrm{EM}}} + c^2_{\beta-\alpha} 
\frac{F_{H^{\pm},h} - F_{H^{\pm},H^{0}} + F_{A^0,H^0} - F_{A^{0},h}}{16\, \pi^2 \,v^2 \,\alpha_{\mathrm{EM}}}
\nonumber \\
  &+& c^2_{\beta-\alpha} \frac{ 3 \left(F_{H^{0},Z} - F_{H^{0},W} -F_{h,Z} + F_{h,W}\right)}{16\, \pi^2 \,v^2 \,\alpha_{\mathrm{EM}}}
\eea
with $G_{A,B,C}$, $\hat G_{A,B}$ and $F_{A,B}$ given in Appendix \ref{Appendix1bis}. We note that both $\Delta S$  and $\Delta T $ are independent of $t_{\beta}$. We define 
\bea
\label{x0A}
x_0 \equiv \frac{m^2_{H^0}}{m^2_{H^{\pm}}} \ , \, x_A \equiv \frac{m^2_{A^0}}{m^2_{H^{\pm}}} \ ,
\eea 
and construct a $\Delta\chi^2_{EW}$ function as 
\be
\label{chi_EWPO_2HDM}
\Delta\chi^2_{EW}(m^2_{H^{\pm}},x_0,x_A,c_{\beta-\alpha}) = \sum_{i,j} \left(\Delta\mathcal{O}_{i} - \Delta\mathcal{O}^0_{i}\right) (\sigma^2)_{ij}^{-1}
\left(\Delta\mathcal{O}_{j} - \Delta\mathcal{O}^0_{j}\right)\, ,
\ee
with $\Delta\mathcal{O}^0_{i}$ and $(\sigma^2)_{ij}$ defined after (\ref{chi_EWPO2}). The results of the fit are shown in Figure~\ref{fig:ST} for various values of 
$c_{\beta-\alpha}$ and $m_{H^\pm} = 400,600$ GeV. 
%The preferred region corresponds to $x_{0,A}\simeq$ 1.
%

The calculation of the oblique $S$ and $T$ parameters in the 2HDM matches with the $D = 6$ effective operators present in the EFT. Considering first the alignment limit and 
expanding (\ref{EWPO_2HDM_S}) and (\ref{EWPO_2HDM_T}) around $x_{0},x_A \sim 1$ (as indicated by the present 95\% C.L. experimental limits), $\Delta S$ and $\Delta T$ in the 2HDM read
\bea
\Delta S &=&  \frac{g^2\,\sW^2}{96\, \pi^2 \,\alpha_{\mathrm{EM}}}  \, \log(x_0 \, x_A) \simeq - \frac{g^2\,\sW^2 (1- x_A + 1- x_0)}{96\, \pi^2 \,\alpha_{\mathrm{EM}}}  \nonumber \\
\Delta T &=& \frac{m_{H^\pm}^2 (1-x_A) (1-x_0) }{48 \, \pi^2 \,v^2 \,\alpha_{\mathrm{EM}}} 
\eea
where we note that the functions $G_{A,B,C}$ scale as $(1-x_{A,0})^2$ and can be neglected with respect to the $\log(x_0/x_A)$ term in $\Delta S$.
By means of (\ref{x0xA1}) and (\ref{x0xA2}) in Appendix \ref{Appendix1}, the above relations can be rewritten as 
\bea
\frac{\alpha_{EM}}{4 \, s_W^2} \Delta S &=& \frac{m_W^2}{96 \pi^2} \, \frac{\tilde \lambda_4}{m_{H^\pm}^2} \nonumber \\  
\alpha_{\mathrm{EM}} \Delta T & =& \frac{v^2}{192 \pi^2 }  \frac{\tilde{\lambda}_4^2-\tilde{\lambda}_5^2}{m_{H^\pm}^2}
\eea
For $m_{H^\pm}^2 \gg v^2$, and upon the replacement $m_{H^\pm}^2 \to \tilde{\mu}^2_2$, these match the EFT results (\ref{Cgamma_UB})
\bea
\bar c_W + \bar c_B &=& \frac{m_W^2}{96 \pi^2} \, \frac{\tilde \lambda_4}{\tilde{\mu}^2_2} = \frac{\alpha_{EM}}{4 \, s_W^2} \, \Delta S =  \epsilon_3\nonumber \\
\bar c_T &=& \frac{v^2}{192 \pi^2 }  \frac{\tilde{\lambda}_4^2-\tilde{\lambda}_5^2}{\tilde{\mu}^2_2} = \alpha_{\mathrm{EM}} \, \Delta T = \epsilon_1
\label{STcT}
\eea
where we  have also translated into the Altarelli-Barbieri parametrization~\cite{barbieri_epsilons} for reference. The bounds on these effective operators are \cite{Baak:2014ora}
\be
  \bar c_{T}(m_Z) \in [-1.5,2.2] \times 10^{-3} \qquad\text{and}\qquad
  \Big(\bar c_{W}(m_Z) + \bar c_{B}(m_Z)\Big) \in [-1.4,1.9] \times 10^{-3} \ .
\ee

Away from alignment, $\Delta S$ and $\Delta T$ receive further contributions $\propto c^2_{\beta-\alpha}$. In an EFT language, these are captured via RG evolution: various operators from 
(\ref{eq:dim-8}) may mix with $\mathcal{O}_W$, $\mathcal{O}_B$ and $\mathcal{O}_T$ via RG running, as is the case of {\it e.g.} $\mathcal{O}_H$ \cite{Elias-Miro:2013mua}, whose mixing at leading 
order would yield 
\bea
\label{1loopRG_2HDM1}
  \bar c_{T}(m_Z) &\simeq& \bar c_{T}(\tilde{\mu}_2) - \frac{3 \,g'^2}{32\,\pi^2} \, \bar c_{H}(\tilde{\mu}_2)\, \mathrm{log} \left(\frac{\tilde{\mu}_2}{m_Z}\right) \\
\label{1loopRG_2HDM2}
  \bar c_{W}(m_Z) + \bar c_{B}(m_Z) &\simeq& c_{W}(\tilde{\mu}_2) + \bar c_{B}(\tilde{\mu}_2)  
  + \frac{1}{24\,\pi^2} \, \bar c_{H}(\tilde{\mu}_2)\, \mathrm{log} \left(\frac{\tilde{\mu}_2}{m_Z}\right) \, .
\eea
However, for the 2HDM $\bar c_H(\tilde{\mu}_2) = 0$ at tree-level, and thus the Wilson coefficient responsible for the contribution to $\Delta S$ and $\Delta T$  
dependant on $c^2_{\beta-\alpha}$ is $\bar c_{6h}$ \cite{GNS_CP}.

\begin{figure}[ht!]
\begin{center}
\includegraphics[width=0.65\textwidth]{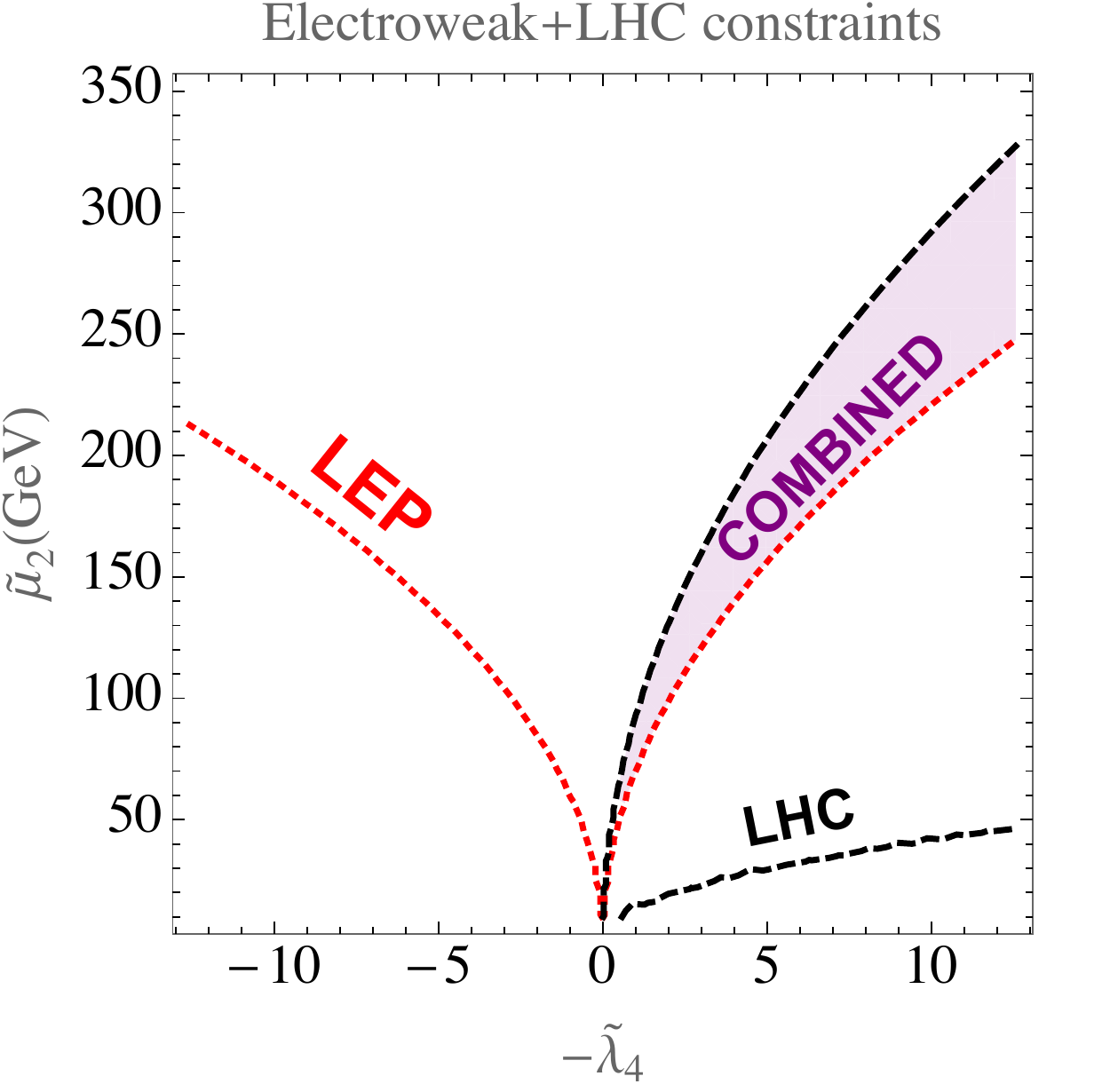}
%\hspace{3mm} \includegraphics[width=0.45\textwidth]{EWPO_Scan_2HDM2.pdf}
\caption{95\% C.L. limits on the 2HDM parameters $\tilde \lambda_4$ and $\tilde \mu_2$. The red-dotted lines enclose the region allowed by constraints on the $S$ parameter, with the black-dashed lines the corresponding LHC limits from \cite{ESY2014b}. The purple-solid region corresponds to the combined limits.}
\label{fig:combinedl4}
\end{center}
\end{figure}

%%%%%%%%%%%%%%%%%%%%%%%%%%%%%%
\subsubsection{Alignment limit: complementarity of EWPTs and LHC limits}
%%%%%%%%%%%%%%%%%%%%%%%%%%%%%%
 
 In the previous section we discussed how LEP electroweak constraints translate into the full theory (\ref{Cgamma_UB}), and in turn in the effective theory (\ref{STcT}) which affected the operators $\bar c_T$ and a combination $\bar c_W+\bar c_B$. It is time to move onto the constraints from LHC, the most stringent ones on operators affecting the decay of the Higgs to photons ($\bar c_\gamma$) and production through gluon fusion ($\bar c_g$). Other operators are better determined by looking at other production and decay channels. In particular, three combinations~\cite{ESY2014a}
 \bea
 \bar c_W -  \bar c_B  \ , \, c_{HW} \textrm{ and  } \bar c_{HB} \ ,
 \eea 
 are constrained with the help of Higgs production in association with a vector boson, and with diboson data~\cite{ESY2014b}. In the 2HDM, the global fit is more constraining than a fit for a general EFT at dimension-six level. In particular, by varying all the free parameters in the 2HDM simultaneously in a fit to Higgs plus diboson data, one obtains the following 95\% C.L. regions~\cite{ESY2014b}
\bea
{\bar c}_{HW} & \in &  (0.0004, 0.02) \nonumber \\
{\bar c}_g & \in & - (0.00004, 0.000003) \nonumber \\
{\bar c}_\gamma & \in &  (-0.0006, 0.00003)
\label{2HDMfit}
\eea

 Whereas the constraint on the $T$ parameter favours a $\tilde \lambda_5$ close to  $\tilde \lambda_4$, the constraint on $\bar c_\gamma$ from the LHC restricts the values of $\tilde \lambda_3$. The most interesting case is when we compare constraints from LEP and LHC which set limits on the coupling $\lambda_4$ and the scale of new physics. Indeed, one can see this interplay between the two sources of data in Figure \ref{fig:combinedl4}, where we present results in the $\tilde \lambda_4$ and $\tilde \mu_2$ plane. The red-dotted lines enclose the region allowed by constraints on the $S$ parameter, and the black-dashed lines are the corresponding LHC limits from \cite{ESY2014b}. The purple-solid region corresponds to the combined limits and it is quite constraining as the two regions have little overlap. 
 Note that the preferred region at 95\% C.L. is consistent with the presence of light particles. This is due to the slightly non-zero positive value of $\bar c_{HW}$ in the global LHC fit and the tension with LEP limits, which favour negative values of  $\bar c_{HW}$. This {\it hint} for new physics is gone at the level of 3$\sigma$.

 Finally, we comment on the 2HDM and the impact on the EW phase transition. As recently shown in \cite{Dorsch:2014qja}, a first order EW phase transition strongly favours a large $m_{A^0} - m_{H^0}$ splitting, which via (\ref{modifiedcouplingsalignment}) implies $\tilde{\lambda}_5 < 0$ and sizable. Since this implies that it is not possible to simultaneously have vanishing
 $\Delta S$ and $\Delta T$ unless $\tilde \mu_2 \to \infty$, it would be possible to observe a deviation from the SM in $\Delta S$ or $\Delta T$ as the experimental precision increases (see Figure 
 \ref{fig:Singlet_EWPOEllipse}).

 %%%%%%%%%%%%%%%%%%%%%%%%%%%%%%
\subsection{A Dilaton/Radion Scenario}
%%%%%%%%%%%%%%%%%%%%%%%%%%%%%%
 \label{dilaton_section}
 So far we have discussed scenarios where the operators involving the Higgs and massive vector bosons, such as  $\bar c_{W}$, are generated at loop level. In this section we present a case where they appear at tree-level, through the exchange of a dilaton or a radion scalar particle. We consider two equivalent scenarios:
 
 \begin{itemize}
  \item {\bf The extra-dimensional radion: }
A radion $r$ is the excitation of the graviton perturbation in extra-dimensions along the direction of the extra-dimension. The interactions of the radion with SM particles are then obtained by expanding the metric at linear order in $r$. We will consider conformal metrics of the form
\bea
d s^2 = w(z)^2 (\eta_{\mu\nu} d x^\mu d x^\nu - d z^2)
\eea
where $w(z)$ is the warp factor, which depends on the extra-dimension $z$. Minkowski space corresponds to $w=1$ and in Anti-deSitter metrics $w=1/(k z)$, with $k$ the curvature in the extra-dimension.  The extra-dimension is compactified, with $z$ stretching between two points which we will call the UV and IR branes, i.e. $z\in [z_{UV},z_{IR}]$. The interaction then reads
 \bea
{\cal S} = \int d^d x \sqrt{-g} {\cal L} \supset \int d^d x \sqrt{-g} \, w^2(z) \,  2 r T_\mu^\mu \label{L5d}
\eea
where $T_\mu^\mu$ is the trace of the stress-energy tensor. The dimensional reduction of the interaction in (\ref{L5d}) to a four-dimensional (4D) interaction depends on the localization of the radion and the SM particles, and generally speaking it would look as
\bea
{\cal L}_{4D} = \frac{c_i}{\sqrt{6} \, \Lambda} \, r \, T^i \label{L4Dr}
\eea
 where now $r$ and $T^i=Tr(T_{\mu\nu}^i)$ are the dimensionally-reduced  radion and trace of the stress-energy tensor of species $i$.  $\Lambda$ corresponds to the compactification scale, and the coefficients $c_i$ account for the overlap of different bulk species $i$ with the radion. 
 
 In warped extra-dimensions, the radion is a field localized near the IR brane at $z_{IR}$. Therefore, for fields localized near the IR brane, $c_i  \simeq {\cal O}(1)$, for fields delocalized in the bulk (with a flat profile) one finds $c_i  \simeq \frac{1}{\int_{z_{IR}}^{z_{UV}} w(z) \, d z}  $ and for fields on the UV brane, $c_i  \left(\frac{z_{UV}}{z_{IR}}\right)^{a}$ with $a$ a positive number~\cite{gap-metrics}. As an example, in Anti-deSitter (AdS) models, the suppression for bulk fields is $\int w(z) d z = \log\frac{z_{UV}}{z_{IR}}$, which in Randall-Sundrum translates into $\log(\mathrm{M}_{\mathrm{P}}/\mathrm{TeV})\simeq {\cal O}$(30).

Although massless gauge fields do not contribute to the trace of the stress-energy tensor at tree-level, loop-level couplings are generated by the trace anomaly
\be
T^\mu_{\mu,anom}=-\sum_{i} \frac{b_i \alpha_i}{8 \pi}\,F^i_{\mu\nu} F^{i\mu\nu} , \label{anom}
\ee
where $i$ runs over the gauge groups of the SM, $SU(3)_c\times SU(2)_L\times U(1)_Y$, $\alpha_i=g_i^2/4\pi$ and $g_i$ are the respective coupling constants. The $b_i$ are the $\beta$ function coefficients leading to anomalous scale-invariance violations (dilaton) or, equivalently, contributions from fields localized near the IR brane or bulk running (radion)~\cite{Csaki-Lee, GMDM}. The values of $b_i$ depend on the degree of compositeness of fermions in the SM and possible new CFT contributions. For simplicity we neglect these and use the pure CFT value, $b^{CFT}_i=8 \pi^2/(g_i^2 \log(z_{UV}/z_{IR}))$. In this limit the coefficient of the anomaly is then independent of the value of the gauge coupling
\be
T^\mu_{\mu,anom}=-\sum_{i} \frac{\log(z_{IR}/z_{UV})}{4}\,F^i_{\mu\nu} F^{i\mu\nu} , \label{anom2}
\ee

  \item {\bf The dilaton of spontaneous breaking of scale invariance: } Assume the existence of a new sector whose couplings are scale invariant, but this scale invariance is spontaneously broken by some dynamics. For example, in QCD scale invariance is spontaneously broken by confinement dynamics, leading to quark condensates. In our example, we are considering a new strongly coupled sector, possibly with similar dynamics as QCD. The Goldstone boson of this spontaneous breaking is a dilaton. Let us denote the dilaton by $r$ as in the radion case. Non-derivative couplings of the dilaton to SM fields are proportional to the explicit breaking of conformal invariance, namely they are of the form
 \bea
 \frac{r}{f} \partial_\mu J^\mu
 \eea 
  with $f$ the scale of the spontaneous breaking of dilatation symmetry and $J^\mu$ the global conserved current for this symmetry,  $J^\mu=T^{\mu\nu} x_\nu$. Therefore the coupling of the dilaton to the SM fields take the same form as those of the radion (\ref{L4Dr})~\cite{dilaton}. We stress that the quadratic couplings to SM particles do not have the same form for the radion and dilaton~\cite{dilaton}, but this difference will not affect the matching of dimension-six coefficients.
\end{itemize} 
 
The radion/dilaton mass is a model-dependent parameter, related to the mechanism of stabilization of the extra-dimension or the explicit breaking of dilatation symmetry. In absence of stabilization the radion is massless but one could stabilize this dilaton/radion in several ways, and no definite prediction of the mass can be drawn unless we specify the mechanism. For example, in the Goldberger-Wise mechanism~\cite{GW} in warped extra-dimensions the radion mass is a function of the {\it vev} and mass of the stabilizing field~\cite{GW-pheno} and could be very light as well as around the scale of compactification (or spontaneous breaking) $f$. 

\vspace{2mm}

We proceed to integrate out the dilaton assuming its mass $m_r$  is larger than the scales we are probing with colliders. The effective Lagrangian has the form
 \bea
 {\cal L}_{eff} = - \frac{1}{f^2} \frac{1}{m_r^2} T^2
 \eea  
  where $T$ is the trace of the stress-energy tensor. For the Higgs and gauge bosons,
  \bea
  T \subset -2 \, |D_\mu \Phi|^2 + 4 \, V(\Phi^\dagger\Phi) - \frac{b_i \alpha_i}{8 \pi}\,F^i_{\mu\nu} F^{i\mu\nu}
  \eea
where $V=-m_h^2 |\Phi|^2+\lambda |\Phi|^4$. One can then extract the Wilson coefficients of the effective operators

\begin{empheq}[box=\fbox]{align}
&  \bar c_H = -\bar c_6 =  8 \, \frac{m_h^2 v^2}{f^2 m_r^2}  \nonumber \\
&  \bar c_{HW} = - \bar c_W   =  - \frac{b_2 \alpha_2}{4}  \, \frac{m_h^2 v^2}{f^2 m_r^2} \nonumber \\
&  \bar c_{HB} = - \bar c_B   =  - \frac{b_1 \alpha_1}{4}  \, \frac{m_h^2 v^2}{f^2 m_r^2} \nonumber \\
\label{cradion}
&  \bar c_{\gamma} =  - \frac{b_1 \alpha_1}{4\pi}  \, \frac{g^2\, m_h^2 v^2}{g'^2\,f^2 m_r^2} \quad \quad 
\bar c_{g} = - \frac{b_3 \alpha_3}{4\pi}  \, \, \frac{g^2\, m_h^2 v^2}{g_s^2\,f^2 m_r^2} 
\end{empheq}

We stress again that for a pure CFT value $b^{CFT}_i$, $b_1 \alpha_1 = b_2 \alpha_2 =  b_3 \alpha_3$.
We could also consider a more general situation, where the coupling of the dilaton/radion is not universal. This amounts to setting different values of the coefficients $c_i$ in (\ref{L4Dr}) and would lead to a prefactor in the coefficients of the effective operators (\ref{cradion}), dependent of the degree of overlap of the wavefunctions in the bulk (radion) or participation on the composite dynamics of the species (dilaton).

\begin{figure}[ht!]
\begin{center}
\includegraphics[width=0.77\textwidth]{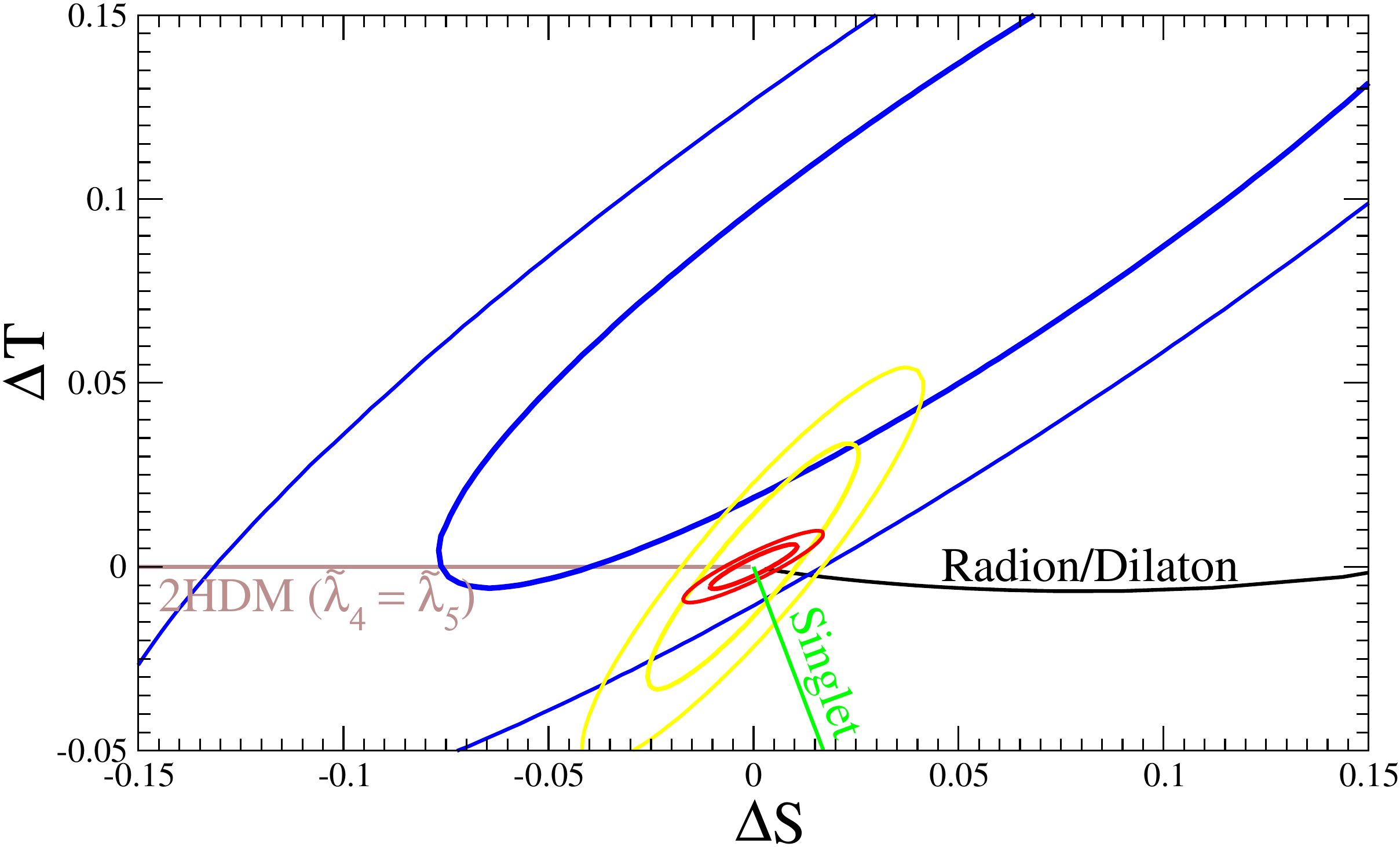} 
\includegraphics[width=0.77\textwidth]{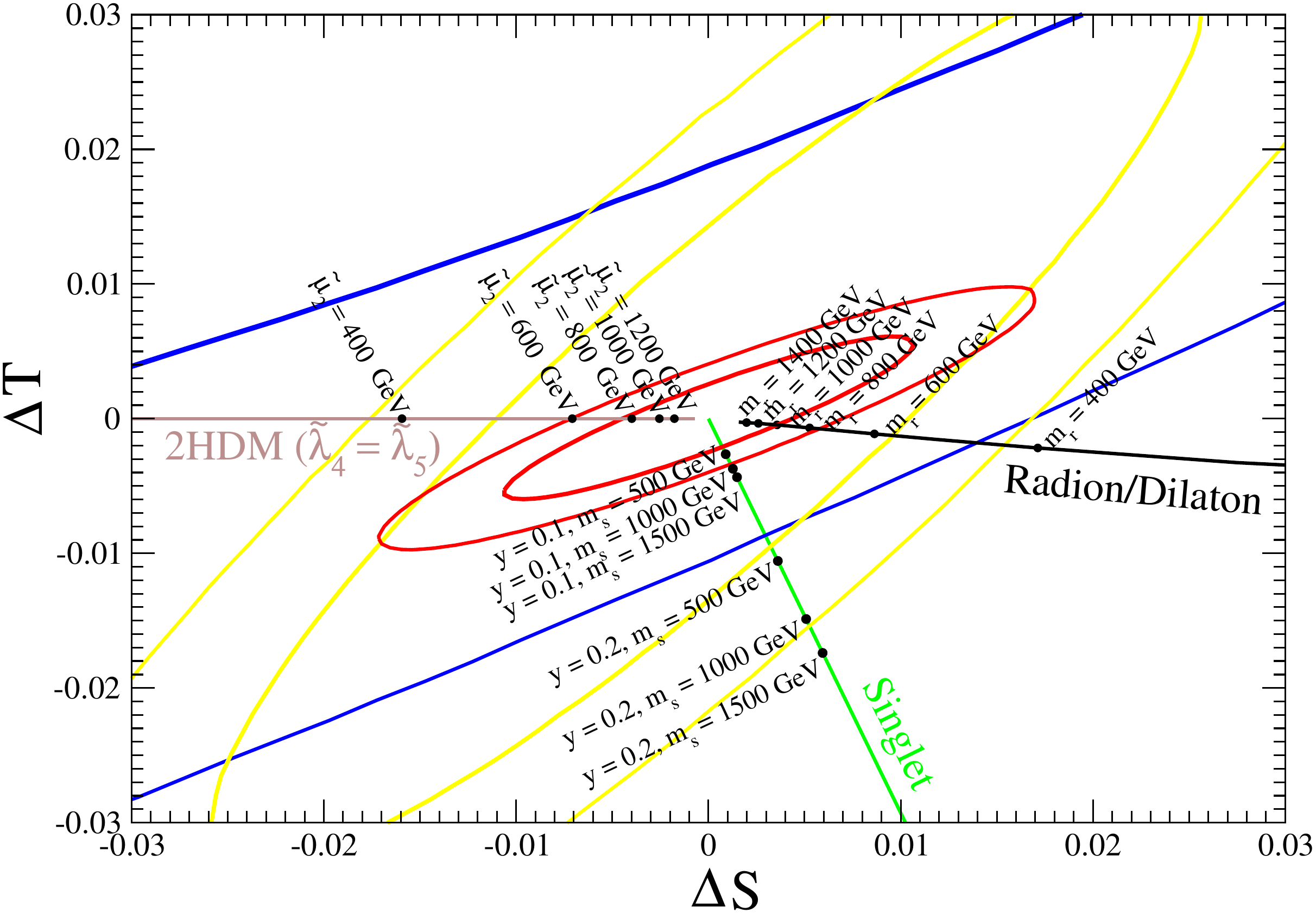} 
\caption{$\Delta S - \Delta T$ constraints on the radion/dilaton (black), singlet-Higgs portal (green) and 2HDM w. custodial symmetry (brown) scenarios. The thick (thin) ellipses correspond to the 68\% C.L. (95\% C.L.) allowed region for LEP (blue) and the projected ILC Giga-Z (yellow) and TLEP TeraZ (red), see Section~\ref{singlet_section} for more details.  The zoom on the percent region (Bottom) includes points for the Higgs portal with various values of mixing parameter $y$ and mass $m_s$, for the radion/dilaton with $f = 2$ TeV and various values of the radion mass $m_r$, and for the 2HDM with 
$\tilde \lambda_4 = \tilde \lambda_5 = -\pi$ and several values of $\tilde \mu_2$.}
\label{fig:Singlet_EWPOEllipse}
\end{center}
\end{figure}

\vspace{2mm}

While the contributions to $\Delta T$ vanish in this case at tree-level \cite{Csaki:2000zn,Gunion:2003px}, they will be generated both at 1-loop 
(this is also the case for $\Delta S$) and through operator mixing due to RG running. The 1-loop contribution to $\Delta T$ and $\Delta S$ 
is given by \cite{Csaki:2000zn}
\be \bar c_T   = \frac{g^2\,s_W^2}{64\, \pi^2\,c_W^2} \, \frac{v^2}{f^2} \, \mathrm{log}\left(\frac{f}{m_r}\right)  \quad, \quad  \quad
\bar c_W + \bar c_B  = - \frac{g^2\,v^2}{(24 \pi)^2\,f^2} \, \mathrm{log}\left(\frac{f}{m_r}\right)  \label{1loopEWPO_Dilaton}
\ee
The corresponding contributions to $\bar c_T(m_Z)$ and $\bar c_W(m_Z)+\bar c_B(m_Z)$ from the RG running of $\bar c_H$ are 
given by (\ref{1loopRG_2HDM1}-\ref{1loopRG_2HDM2}) with the substitution $\tilde{\mu}_2 \to m_r$ 
and using $\bar c_H (m_r)$ from (\ref{cradion}). We note that the RG running gives a contribution to $\Delta T$ and $\Delta S$ which
despite begin suppressed by $m_h^2/m_r^2$ \textit{w.r.t} the 1-loop contribution (\ref{1loopEWPO_Dilaton}), may become dominant  
due to a much smaller numerical suppression. The fact that the two contributions have opposite signs leads to a partial 
cancellation effect. In Figure \ref{fig:Singlet_EWPOEllipse} we show the results on the $\Delta S - \Delta T$ plane, and 
compare the correlation in the oblique parameters with that appearing in the ones encountered respectively in the 
Higgs portal and 2HDM scenarios from Section \ref{singlet_section} and \ref{2HDMsec}.  

 %%%%%%%%%%%%%%%%%%%%%%%%%%%%%%
\section{The validity of the EFT: from operators to total rates and distributions}
%%%%%%%%%%%%%%%%%%%%%%%%%%%%%%
 \label{validitysec}
 The EFT approach, where higher-dimensional operators are written as a way to encode effects of New Physics, is a good approximation at 
 low energies and should not be used in an arbitrary range of energies. Indeed, as these operators are suppressed by some New Physics scale, parton level cross-sections would diverge with the parton energy.

\begin{figure}[h!]
\begin{center}
\includegraphics[width=0.75\textwidth]{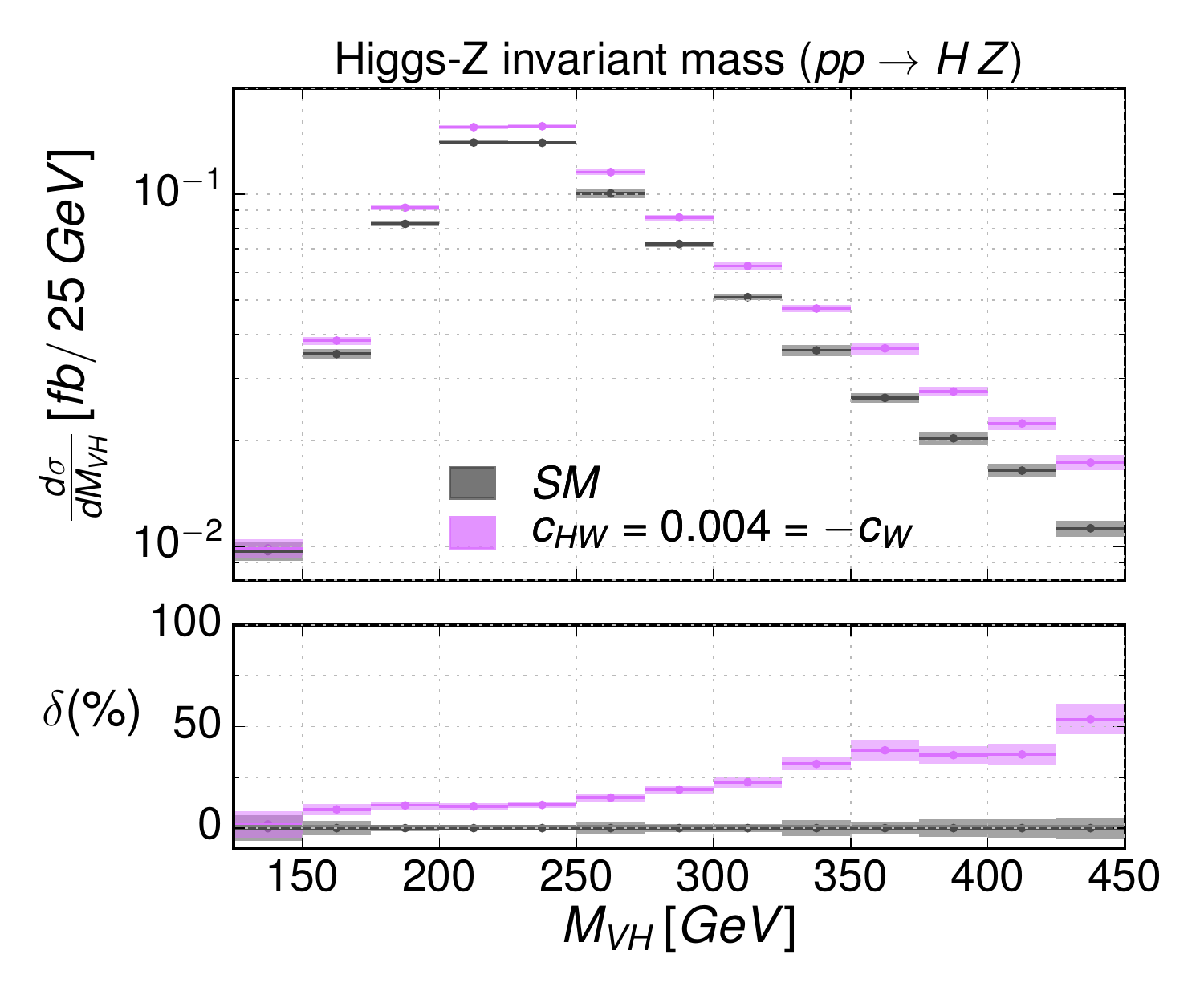} 
\caption{Higgs produced in association with a vector boson: Invariant mass distribution.  Grey and pink distributions correspond to SM and EFT with $\bar c_{HW} = -\bar c_{W} = 0.004$. The width correspond to varying the factorization scale  $\mu=m_h+m_Z$ in a range of $2 \mu$ and $\mu/2$. The value of $\bar c_W$ is consistent with Run1 data and will be explored in the Run2 LHC. The value $\delta$ is the deviation respect to the SM central value.}
\label{MVHdist}
\end{center}
\end{figure}

This growth with energy is a signature that the EFT approach breaks down when one is able to probe the dynamics of the heavy particles one integrates out. In hadron colliders, this corresponds to the moment when the partonic energy $\sqrt{\hat s}$ is comparable with the masses in the loop. Specifically, for a tree-level exchange as in Figure~\ref{fig:c6singlet},
\bea
\sqrt{\hat s} \simeq M 
\eea
The same argument applies to the dilaton exchange, and to fermion couplings in 2HDM as in Figure~\ref{Figure:Feynman_fermion}. 
On the other hand, in loop-induced processes as in Figure~\ref{fig:Feynman} the validity of the EFT extends to the threshold to produce a pair of new particles, namely
\bea
\sqrt{\hat s} \simeq 2 \, M \ .
\label{EFTbreak}
\eea

How this energy is related to the strength of the dimension-six deviation from the SM depends on the model. Let us focus on operators of the type $\bar c_{W}$ and  $\bar c_{HW}$ which contain derivatives on the Higgs field. We have found the matching with the 2HDM and radion-exchange is roughly speaking
\bea
\frac{\bar c}{m_W^2} \sim \frac{\lambda}{192 \pi^2} \frac{1}{M^2} \textrm{ {\it (2HDM)}, and } \frac{m_h^2}{\Lambda^2} \, \frac{1}{M^2} \textrm{ \it (radion/dilaton)} ,
\eea
where $M$ plays the role of the mass of the extra Higgses and dilaton, $\lambda$ denotes a combination of quartic couplings in the Higgs potential, and $\Lambda$ an effective scale suppression for the dimension-five interaction between the dilaton and the SM particles, $\Lambda > M$. 

At energies around the mass threshold , the EFT description should be substituted by the full UV theory including the resonance dynamics. The question we would like to ask is what is the deviation of the EFT predictions from the full UV model, which depends on the process and distribution one is looking at. A full comparison of the UV theory and the EFT results is beyond the scope of this paper, but in this section we would like to discuss the qualitative features of this deviation. 

To do so, we present an example of a specially useful distribution~\cite{ESYbefore} to probe New Physics, namely the invariant mass of the Higgs and vector boson system in Higgs associated production, $M_{VH}$, as it corresponds to the parton energy, $\hat s = M^2_{VH}$. 
In Figure~\ref{MVHdist} we show how the invariant mass distribution changes when dimension-six operators are included. Grey and pink distributions correspond to SM and EFT with $\bar c_{HW} = -\bar c_{W} = 0.004$. The width correspond to varying the factorization scale  $\mu=m_h+m_Z$ in a range of $2 \mu$ and $\mu/2$. The value of $\bar c_W$ is consistent with Run1 data and will be explored in the Run2 LHC. The value $\delta$ is the deviation respect to the SM central value. 

With the rough estimate given in Eq.~\ref{EFTbreak}, one can see that for a 2HDM, with $\lambda = 3 \times 4 \, \pi$ the breakdown occurs at around $M_{VH}\simeq 360$ GeV. For the dilaton, in the case of $\Lambda$ = 500 GeV, the breakdown occurs at about 320 GeV. These numbers highlight the importance to relate EFT distributions with specific UV models.  
  
The distributions are the result of a simulation including PDF effects ({\tt NN23LO1}) and parton-showering and hadronization performed using an implementation from {\tt MCFM}~\cite{mcfm}  into {\tt POWHEG}~\cite{powheg} at NLO in QCD. We have checked that these results are consistent with a parallel implementation in  {\tt MG5 MC@NLO}~\cite{MG5} using the model 
implementation in \cite{Benj} into NLO QCD, based on {\tt Feynrules} \cite{feynrules} and the {\tt UFO}~\cite{ufo} format. Note that the EFT distribution does not 
grow with $\sqrt{\hat s}$, but falls down due to the PDF effects.

Besides this distribution, one can find others sensitive to New Physics effects, e.g. $m_{jj}$ and $\Delta y_{jj}$ in Vector Boson Fusion~\cite{rohini-djouadi} 
(see also \cite{Edezhath:2015lga}), or $\Delta R_{hh}$ and $p_{T}^{hh}$ in di-Higgs production~\cite{LHproc}.

 %%%%%%%%%%%%%%%%%%%%%%%%%%%%%%
\section{Discussion and Summary}
%%%%%%%%%%%%%%%%%%%%%%%%%%%%%%
\label{conclusion}

During Run2 of the LHC, interpretation of data using the Effective Field Theory (EFT) approach will become a standard way to communicate results 
to theorists, as the translation to UV models is more direct than, for example, the use of form factors. 

The approach based on anomalous couplings and the one considered here are related. New Physics affects the behaviour of SM  particles, inducing anomalous 
couplings, for example involving  the Higgs and electroweak bosons. In the context of heavy New Physics, the anomalous couplings can be translated into combinations 
of higher-dimensional operators as shown in Tables~\ref{Table1} and \ref{Table2}.  

The use of the EFT approach, however, is limited to heavy new physics, whereas current LHC limits strive to reach the TeV region, leading to a delicate balance between 
using as much data as possible and the model-independent EFT approach. The question of what LHC data is suitable to constrain EFTs depends both on the UV completion and the 
type of signal one is looking at. Benchmark models allow to address these issues, and also to draw correlations with direct searches and other non-LHC sources of indirect 
constraints such as flavour physics and LEP. In this paper we have taken a first step towards this program by studying the matching between UV completions encompassing 
extensions of the Higgs sector and the EFT. 

This matching allowed us to study the suitability of LHC data to constrain EFTs. For example, one can consider UV completions with large couplings where the validity of the 
EFT is improved, such as the 2HDM in Section~\ref{2HDMsec} with large quartic couplings. But even for weakly-coupled UV completions, the use of the EFT may be justified as 
long as the LHC data one uses is restricted to small values of $\sqrt{\hat s}$, below the masses of new particles, as discussed in Section~\ref{validitysec}.

We have shown the matching of the UV theory with the low-energy coefficients in detail. This matching is straightforward 
in the case of the tree-level dilaton exchange (Section~\ref{dilaton_section}) or mixing with a singlet (Section~\ref{singlet_section}). For 
loop-induced dimension-six operators, however,  the interplay of  with higher order terms (dimension-eight operators) needs to be handled with care as shown in Section~\ref{Sec_MSSM}. 

Another advantage of the comparison with UV completions is to address correlations among the EFT coefficients which are present in models and 
reduce the number of free parameters in a global analysis. For example, in our benchmarks specific relations among operators can be traced back to the 
(limited) Lorentz structures one can build up from scalar fields, see discussion around (\ref{Cgamma_UB}).

Note that we have not discussed the very interesting possibility of CP-violating effects from, for example, complex parameters in the 2HDM. This deserves further study, 
as their kinematic distributions and dependence with energy have to be simulated as close as possible to the actual cuts applied by the LHC collaborations.
 
\vspace{3mm}

\noindent {\bf Note added:} As this work was being submitted for publication, we became aware of \cite{Henning:2014wua}, which also discusses UV completions of 
Higgs Effective Field Theory via extended Higgs sectors finding results similar to ours. We stress that both works are complementary: we focus on the connection 
between the EFT and UV models as a way to assess the range of validity of the EFT, whereas \cite{Henning:2014wua} discusses the connection between the EFT and 
UV models in the context of a systematic framework for the obtention of the EFT Wilson coefficients from an arbitrary UV theory.

\acknowledgments

We would like to thank Ken Mimasu for help on the QCD NLO simulation. V.S and J.M.N. acknowledge support from the Science and Technology Facilities Council (STFC) through the grants 
ST/J000477/1 and ST/L000504/1. The work of M.G. is partially supported by the STFC consolidated grant ST/L000431/1. J.M.N. is supported by the 
People Programme (Marie curie Actions) of the European Union Seventh Framework Programme (FP7/2007-2013) under REA grant agreement PIEF-GA-2013-625809.     
     
\appendix     
     
%%%%%%%%%%%%%%%%%%%%%%%%%%%%%%
\section{2HDM Results and Conventions}
%\addcontentsline{toc}{section}{A $\ $Aspects of 2HDM}
%\setcounter{equation}{0}
%\renewcommand{\theequation}{A.\arabic{equation}}
%%%%%%%%%%%%%%%%%%%%%%%%%%%%%%  

\subsection{Scalar Potential and Mass Spectrum}
\label{Appendix1}

Let us recall the scalar potential for a 2HDM with a softly-broken $\mathbb{Z}_2$-symmetry in the CP-conserving case, given by (\ref{2HDM_potential}). 
After EW symmetry breaking, the scalar mass eigenstates can be written in terms of the original fields in (\ref{2HDMminima}) as
\begin{eqnarray}
\label{rotation_states}
G^{\pm}=\cos\beta\ \varphi_1^{\pm} + \sin\beta\ \varphi_2^{\pm} & \hspace{1.2cm} & H^{\pm}=-\sin\beta\ \varphi_1^{\pm} + \cos\beta\ \varphi_2^{\pm} \nonumber \\
G^0=\cos\beta\ \eta_1 + \sin\beta\ \eta_2 \hspace{3.4mm} & \hspace{1.2cm} & A^0=-\sin\beta\ \eta_1 + \cos\beta\ \eta_2 \\
h=-\sin\alpha\ h_1 + \cos\alpha\ h_2  & \hspace{1.2cm} & H^0=-\cos\alpha\ h_1 - \sin\alpha\ h_2 \nonumber
\end{eqnarray}
with $H^{\pm}$, $A^0$, $H^0$, $h$ being the physical states of the theory and $G^{\pm}$, $G^0$ the Goldstone bosons from the breaking of EW symmetry. 
Requiring that the scalar potential be bounded from below yields 
\begin{equation}
\label{Stability}
\lambda_1 > 0\,, \quad \lambda_2 > 0\,, \quad \lambda_3 > -\sqrt{\lambda_1 \lambda_2}\,, \quad \lambda_3 + \lambda_4 - |\lambda_5|> -\sqrt{\lambda_1 \lambda_2}\,.
\end{equation}
Obtaining the correct EW vacuum as the minimum of the scalar potential (\ref{2HDM_potential}) imposes the relations
\begin{equation}\begin{split}
	&\mu_1^2=M^2 \,s^2_{\beta} -\frac{v^2}{2}\left(\lambda_1 \,c^2_{\beta}+\lambda_{345}\, s^2_{\beta}\right),\\
	&\mu_2^2=M^2 \,c^{2}_{\beta}- \frac{v^2}{2} \left(\lambda_2 \,s^2_{\beta}+\lambda_{345}\, c^2_{\beta}\right),
\end{split}\end{equation}
with $\lambda_{345}\equiv\lambda_3+\lambda_4+\lambda_5$, $s_{\beta} \equiv \sin\beta$, $c_{\beta} \equiv \cos\beta$ and $M^2 = \mu^2/s_{\beta}c_{\beta}$. 
Similarly, the quartic couplings $\lambda_{1-5}$ in (\ref{2HDM_potential}) may be re-expressed in terms of the masses of the physical states $m_{H^\pm}$, $m_{A^0}$, $m_{H^0}$, $m_{h}$, the mixing angles  $\alpha$, $\beta$ and $M^2$ as
\bea
\label{couplings1}
\lambda_1&=&\frac{1}{v^2\, c^2_{\beta}} \left(-M^2\, s^2_{\beta} + m_{h}^2 \,s^2_{\alpha}+m_{H^0}^2 \,c^2_{\alpha}\right),\\
\lambda_2&=&\frac{1}{v^2\, s^2_{\beta}} \left(-M^2\, c^2_{\beta}+m_{h}^2 \,c^2_{\alpha}+m_{H^0}^2 \,s^2_{\alpha}\right),\\
\lambda_3&=&\frac{1}{v^2} \Big[-M^2 +2 m_{H^\pm}^2 + \left(m_{H^0}^2-m_{h}^2\right)\frac{s_{2\alpha}}{s_{2\beta}}\Big],\\
\label{couplings4}
\lambda_4&=&\frac{1}{v^2} (M^2+m_{A^0}^2-2 m_{H^\pm}^2),\\
\label{couplings5}
\lambda_5&=&\frac{1}{v^2}\left(M^2 - m_{A^0}^2\right).
\eea
We stress that since the masses of the physical states $m_{H^\pm}$, $m_{A^0}$, $m_{H^0}$, $m_{h}$ and the mixing angles $\alpha$, $\beta$ are obtained upon EW symmetry breaking, the relations (\ref{couplings1}-\ref{couplings5}) only hold in the EW broken theory. 

\vspace{3mm}
  
We now perform an $SU(2)$ rotation from the field basis $\Phi_1$, $\Phi_2$ to a field basis $H_1$, $H_2$ in which only $H_1$ takes a {\it vev}: $\langle H_1\rangle = \frac{v}{\sqrt{2}}$, $\langle H_2\rangle = 0$. This rotation is given by 
\begin{equation}
\label{Higgsbasis}
H_1= c_{\beta}\,  \Phi_1 + s_{\beta} \, \Phi_2 \quad \quad \quad
H_2= - s_{\beta} \, \Phi_1 + c_{\beta} \, \Phi_2 
\end{equation}
As seen from (\ref{rotation_states}), for $\alpha = \beta - \pi/2$, we obtain after the field rotation
\begin{equation}
\label{2HDM_H12}
H_{1} = \Phi_{\mathrm{SM}} = \left( \begin{array}{c} G^+ \\ \frac{v + h + i\, G^0}{\sqrt{2}}  \end{array} \right),\hspace{0.5cm}
H_{2} = \left( \begin{array}{c} H^+ \\ \frac{H^0 + i\, A^0}{\sqrt{2}}  \end{array} \right),
\end{equation}
This is the {\it alignment limit} $c_{\beta-\alpha} = 0$, in which $H_{1}$ corresponds exactly to the SM Higgs doublet. Away from this limit there is mixing of the neutral CP-even physical states
$h,H^0$ between $H_{1,2}$.

\vspace{3mm}

\noindent The scalar potential for $H_{1,2}$ is given by
\begin{eqnarray}	
\label{2HDM_potential2}
V_{\rm tree}(H_1,H_2)= \tilde{\mu}^2_1 \left|H_1\right|^2 + \tilde{\mu}^2_2\left|H_2\right|^2 - \tilde{\mu}^2 \left[H_1^{\dagger}H_2+\mathrm{H.c.}\right] +\frac{\tilde{\lambda}_1}{2}\left|H_1\right|^4 \nonumber \\
 +\frac{\tilde{\lambda}_2}{2}\left|H_2\right|^4 + \tilde{\lambda}_3 \left|H_1\right|^2\left|H_2\right|^2
+\tilde{\lambda}_4 \left|H_1^{\dagger}H_2\right|^2+
 \frac{\tilde{\lambda}_5}{2}\left[\left(H_1^{\dagger}H_2\right)^2+\mathrm{H.c.}\right] \nonumber \\
 + \tilde{\lambda}_6 \left[  \left|H_1\right|^2 H_1^{\dagger}H_2 +\mathrm{H.c.}\right] + \tilde{\lambda}_7 \left[  \left|H_2\right|^2 H_1^{\dagger}H_2 +\mathrm{H.c.}\right]\,.
\end{eqnarray}
The modified mass parameters $\tilde{\mu}^2_1$, $\tilde{\mu}^2_2$, $\tilde{\mu}^2$ and quartic couplings $\tilde{\lambda}_{1-5}$ are expressed in terms of $m^2_{H^\pm}$, $m^2_{A^0}$, $m^2_{H^0}$, $m^2_{h}$, $M^2$, $\alpha$ and $\beta$ as
\begin{equation}\begin{split}
\label{modifiedmasses} 
	&\tilde{\mu}_1^2 = -\frac{1}{2}\left[ m_{H^0}^2 c^2_{\beta-\alpha}  + m_{h}^2 s^2_{\beta-\alpha} \right] < 0 \\
	&\tilde{\mu}_2^2 = M^2  - \frac{1}{2} \left[  \left(m_{H^0}^2-m_{h}^2\right) \frac{s_{2\alpha}}{s_{2\beta}} + m_{H^0}^2 s^2_{\beta-\alpha}  + m_{h}^2 c^2_{\beta-\alpha}  \right] \\
	&\tilde{\mu}^2 =  - \frac{1}{2} \left(m_{H^0}^2-m_{h}^2\right) s_{2(\beta-\alpha)} 
\end{split}\end{equation}
\begin{equation}\begin{split}
	\label{modifiedcouplings} 
	&\tilde{\lambda}_1=- \frac{\tilde{\mu}_1^2}{2 \, v^2} > 0	\\
	&\tilde{\lambda}_2=\frac{\left[-M^2 \left(t^2_{\beta}+ t^{-2}_{\beta} - 2 \right) + m_{H^0}^2 \left(- s^2_{\beta-\alpha} + c^2_{\alpha} t^2_{\beta} + s^2_{\alpha} t^{-2}_{\beta} \right) +m_{h}^2 \left(- c^2_{\beta-\alpha} + s^2_{\alpha} t^2_{\beta} + c^2_{\alpha} t^{-2}_{\beta} \right)\right]}{v^2}
		   \\
	&\tilde{\lambda}_3=\frac{\Big[-2 M^2 +2 m_{H^\pm}^2 + \left(m_{H^0}^2-m_{h}^2\right)\frac{s_{2\alpha}}{s_{2\beta}} + m_{H^0}^2 s^2_{\beta-\alpha}  + m_{h}^2 c^2_{\beta-\alpha}\Big]}{v^2}
		   \\
	&\tilde{\lambda}_4=\frac{\left[m_{A^0}^2-2 m_{H^\pm}^2+m_{H^0}^2 s^2_{\beta-\alpha}  + m_{h}^2 c^2_{\beta-\alpha}\right]}{v^2} \\
	&\tilde{\lambda}_5=\frac{\left[ - m_{A^0}^2 + m_{H^0}^2 s^2_{\beta-\alpha}  + m_{h}^2 c^2_{\beta-\alpha} \right]}{v^2} \\
	&\tilde{\lambda}_6= \frac{\tilde{\mu}^2}{v^2} 	\\
	&\tilde{\lambda}_7=\frac{s_{2\beta} \left[M^2 \left(t^2_{\beta}- t^{-2}_{\beta}\right) + m_{H^0}^2 \left(\frac{c_{\alpha} t_{\beta}}{c_{\beta}} - \frac{s_{\alpha} t^{-1}_{\beta}}{s_{\beta}}\right) s_{\beta-\alpha} + m_{h}^2 \left(\frac{c_{\alpha} t^{-1}_{\beta}}{s_{\beta}} - \frac{s_{\alpha} t_{\beta}}{c_{\beta}} \right) c_{\beta-\alpha} \right]}{2\,v^2}
\end{split}\end{equation}  
The relations $\tilde{\mu}_1^2 = -2\,\tilde{\lambda}_1\, v^2$ and $\tilde{\mu}^2 = \tilde{\lambda}_6\, v^2$ are necessary to obtain correct {\it vev} for $H_{1,2}$ upon minimization of 
(\ref{2HDM_potential2}). The relations among the masses of the new scalar states may be written in terms of $x_0 \equiv m^2_{H^0}/m^2_{H^{\pm}}$ and $x_A \equiv m^2_{A^0}/m^2_{H^{\pm}}$
as
\bea
\label{x0xA1}
1-\frac{m_{A^0}^2}{m_{H^\pm}^2} \equiv 1-x_{A} & =& \frac{v^2}{2\, m_{H^\pm}^2} \, (\tilde{\lambda}_5- \tilde{\lambda}_4) \\
\label{x0xA2}
1-\frac{m_{H^0}^2}{m_{H^\pm}^2} \equiv 1-x_{0} & =& - \frac{v^2}{2\,m_{H^\pm}^2} \, (\tilde{\lambda}_5 + \tilde{\lambda}_4) + c^2_{\beta-\alpha} \left[ \frac{m_h^2}{m_{H^\pm}^2} - x_0 \right]
\eea
In the alignment limit, the above expressions (\ref{modifiedmasses}-\ref{modifiedcouplings}) simplify considerably, yielding
\be
\label{modifiedmassesalignment} 
- \tilde{\mu}_1^2 = \frac{m_{h}^2}{2} \quad , \quad \tilde{\mu}_2^2 = M^2  - \frac{m_{h}^2}{2} \quad , \quad \tilde{\mu}^2 = 0
\ee
\begin{equation}\begin{split}
	\label{modifiedcouplingsalignment} 
	&\tilde{\lambda}_1= \frac{m_{h}^2}{4 \, v^2} \quad , \quad 	\tilde{\lambda}_2=\frac{-M^2 \left(t^2_{\beta}+ t^{-2}_{\beta} - 2 \right) + m_{H^0}^2 \left(- 1 + s^2_{\beta} t^2_{\beta} + c^2_{\beta} t^{-2}_{\beta} \right) +m_{h}^2}{v^2} \, ,\\
	&\tilde{\lambda}_3=\frac{-2\, M^2 +2\, m_{H^\pm}^2 + m_{h}^2}{v^2} \quad , \quad \tilde{\lambda}_4=\frac{m_{H^0}^2+m_{A^0}^2-2 m_{H^\pm}^2}{v^2} \quad,  \quad \tilde{\lambda}_5=\frac{m_{H^0}^2 - m_{A^0}^2}{v^2}, \\
	&\tilde{\lambda}_6= 0 \quad , \quad \tilde{\lambda}_7= s_{2\beta}\,\left(t^2_{\beta}- t^{-2}_{\beta}\right) \frac{M^2  - m_{H^0}^2}{2\,v^2}\,.
\end{split}\end{equation}  
The interaction vertices $g_{hH^{+}H^{-}}$,  $g_{hA^{0}A^{0}}$ and $g_{hH^{0}H^{0}}$ are in this limit proportional respectively to $\tilde{\lambda}_3$, $\tilde{\lambda}_3 + \tilde{\lambda}_4 -\tilde{\lambda}_5$ and $\tilde{\lambda}_3 +\tilde{\lambda}_4+\tilde{\lambda}_5$.

\subsection{Electroweak Precision Observables}
\label{Appendix1bis}

We turn now to the discussion of EW precision constraints in the 2HDM. The non-SM contributions to the oblique parameters $S$ and $T$ are given by (\ref{EWPO_2HDM_S}-\ref{EWPO_2HDM_T}) (the general expressions for the various oblique parameters in models with an arbitrary number of scalar doublets may be found in \cite{Grimus:2008nb}), with the functions $F_{A,B}$, $G_{A,B,C}$, $\hat G_{A,B}$ being 
\bea
F_{A,B} &=& \frac{m^2_A + m^2_B}{2} - \frac{m^2_A \, m^2_B}{m^2_A - m^2_B} \mathrm{log} \left(\frac{m^2_A}{m^2_B}\right) \nonumber \\
G_{A,B,C} &=& -\frac{16}{3} + 5 \frac{m^2_A + m^2_B}{m^2_C}-2\frac{(m^2_A - m^2_B)^2}{m^4_C} \nonumber \\
& & +\frac{3}{m^2_C}\left[\frac{m^4_A + m^4_B}{m^2_A - m^2_B}+\frac{m^4_A - m^4_B}{m^2_C}+\frac{(m^2_A - m^2_B)^3}{3\,m^4_C} \right]\mathrm{log}\left(\frac{m^2_A}{m^2_B}\right) \\
& & +\frac{m^4_C + m^2_C \,(m^2_A + m^2_B) + (m^2_A - m^2_B)^2}{m^6_C} \, f(r_{A,B,C},t_{A,B,C})\nonumber \\
\hat G_{A,B} &=& G_{A,B,B} \, - 24 + 12\left( \frac{m^2_A - m^2_B}{m^2_B} - \frac{m^2_A + m^2_B}{m^2_A - m^2_B}\right) \mathrm{log}\left(\frac{m^2_A}{m^2_B}\right) + 12 \,\frac{f(r_{A,B,B},t_{A,B,B})}{m^2_B}  \nonumber
\eea
with
\bea
r_{A,B,C} &=&  m^4_C + m^2_C \,(m^2_A + m^2_B) + (m^2_A - m^2_B)^2\nonumber \\
t_{A,B,C} &=& m^2_A + m^2_B - m^2_C \nonumber \\
f(r,t) &=& \left\lbrace \begin{array}{cc}
                         \sqrt{r} \,\mathrm{log}\left(\left|\frac{t-\sqrt{r}}{t+\sqrt{r}} \right| \right)& r>0\\
                         0 & r=0 \\
                         2\, \sqrt{-r}\,\mathrm{arctan}\left(\frac{\sqrt{-r}}{t} \right) & r<0
                        \end{array}\right.
\eea
Finally, we stress that the potential (\ref{2HDM_potential2}) preserves custodial symmetry in the limit $\tilde{\lambda}_4 = \tilde{\lambda}_5$ (see {\it e.g.} \cite{Kanemura:1996eq}), which from (\ref{x0xA1}) corresponds to $m_{A^0} = m_{H^\pm}$. To see this, instead of expressing the 2HDM scalar potential in terms of $H_{1,2}$, we can introduce the $2\times2$ matrices ${\bf\Phi}_{1}=(i\sigma _{2}H_{1}^{*}, H_{1}), \, {\bf\Phi}_{2}=(i\sigma _{2}H_{2}^{*}, H_{2})$. The scalar potential for the 2HDM then reads
\begin{equation}
 \label{Potential_2times2}
\begin{split}
V = & -\frac{\tilde{\mu}_1^2}{2}\,\mathrm{Tr}\left[{\bf\Phi}^{\dagger}_{1} {\bf\Phi}_{1} \right] + \frac{\tilde{\mu}_2^2}{2}\,\mathrm{Tr}\left[{\bf\Phi}^{\dagger}_{2} {\bf\Phi}_{2} \right] + 
\frac{\tilde{\mu}^2}{2}\,\left(\mathrm{Tr}\left[{\bf\Phi}^{\dagger}_{1} {\bf\Phi}_{2} \right] + \mathrm{Tr}\left[{\bf\Phi}^{\dagger}_{2} {\bf\Phi}_{1} \right]\right)
\\ &
+ \frac{\tilde{\lambda}_1}{4}\,\left(\mathrm{Tr}\left[{\bf\Phi}^{\dagger}_{1} {\bf\Phi}_{1} \right] \right)^2 
+ \frac{\tilde{\lambda}_2}{4}\,\left(\mathrm{Tr}\left[{\bf\Phi}^{\dagger}_{2} {\bf\Phi}_{2} \right] \right)^2  
+ \frac{\tilde{\lambda}_3}{4}\, \mathrm{Tr}\left[{\bf\Phi}^{\dagger}_{1} {\bf\Phi}_{1} \right] \, \mathrm{Tr}\left[{\bf\Phi}^{\dagger}_{2} {\bf\Phi}_{2} \right] \\ &
+ \frac{\tilde{\lambda}_4 + \tilde{\lambda}_5}{16}\, \left( \mathrm{Tr}\left[{\bf\Phi}^{\dagger}_{1} {\bf\Phi}_{2} \right] + \mathrm{Tr}\left[{\bf\Phi}^{\dagger}_{2} {\bf\Phi}_{1} \right] \right)^2 
- \frac{\tilde{\lambda}_4 - \tilde{\lambda}_5}{16}\, \left( \mathrm{Tr}\left[{\bf\Phi}^{\dagger}_{1} {\bf\Phi}_{2} \sigma_3\right] - \mathrm{Tr}\left[\sigma_3{\bf\Phi}^{\dagger}_{2} {\bf\Phi}_{1} \right] \right)^2 \\ &
+ \frac{1}{16}\, \left(\tilde{\lambda}_6\, \mathrm{Tr}\left[{\bf\Phi}^{\dagger}_{1} {\bf\Phi}_{1} \right] + \tilde{\lambda}_7\,\mathrm{Tr}\left[{\bf\Phi}^{\dagger}_{2} {\bf\Phi}_{2} \right] \right) 
\left( \mathrm{Tr}\left[{\bf\Phi}^{\dagger}_{1} {\bf\Phi}_{2} \right] + \mathrm{Tr}\left[{\bf\Phi}^{\dagger}_{2} {\bf\Phi}_{1} \right] \right) 
\end{split}
\end{equation}
Both ${\bf\Phi}_{1,2}$ transform as bi-doublets of a global symmetry $SU(2)_L \times SU(2)_R$: ${\bf\Phi}_{i} \to L{\bf\Phi}_{i}R$ (with $L \in SU(2)_L$ and $R \in SU(2)_R$). The potential 
(\ref{Potential_2times2}) is then invariant under a {\it custodial} $SU(2)_L \times SU(2)_R$ in the absence of the term proportional to  $\tilde{\lambda}_4 - \tilde{\lambda}_5$.

%, yielding a vanishing $T$ parameter: $\Delta T = 0$.

%Moreover, we can define ${\tilde{\bf\Phi}}_{2} = {\bf\Phi}_{2} \sigma_3$, which also transforms as a bi-doublet under $SU(2)_L \times SU(2)_R$. By recasting the potential (\ref{Potential_2times2})
%in terms of ${\tilde{\bf\Phi}}_{2}$ and ${\bf\Phi}_{1}$, we see that the $SU(2)_L \times SU(2)_R$ global symmetry is also preserved for $\lambda_4 = -\lambda_5$, which would yield 
%$\Delta m^2_{+} = 0$. Then both cases $\lambda_4 = \pm\lambda_5$ preserve a {\it custodial} $SU(2)_L \times SU(2)_R$ symmetry and yield a vanishing $T$ parameter: $\Delta T = 0$.

%%%%%%%%%%%%%%%%%%%%%%%%%%%%%%
%\section*{\label{Appendixc}Appendix C: Effective Theory matching in the broken phase}
%\addcontentsline{toc}{section}{C $\ $Effective Theory matching in the broken phase}
%\setcounter{equation}{0}
%\renewcommand{\theequation}{B.\arabic{equation}}
%%%%%%%%%%%%%%%%%%%%%%%%%%%%%% 

\section{Effective Theory Matching in the EW Broken Phase}
\label{Appendixc}

We now relate the Wilson coefficients from the $D = 6$ operators for the SM $SU(2)_L \times U(1)_Y$ invariant effective field theory given in (\ref{eq:silh}), (\ref{eq:lagG}) and (\ref{eq:silhCPodd}) to the SM effective Lagrangian after EW symmetry breaking. The complete set of relations from the full $D \leq 6$ effective Lagrangian (\ref{eq:effL}) may be found in \cite{benj}. In the following we 
present the generic relations for the 3- and 4-point interactions involving Higgs/gauge bosons. 
%Higgs effective lagrangian in the $ $ unbroken theory to (\ref{eq:silh}), (\ref{eq:lagG}) and (\ref{eq:silhCPodd}) to the lagrangian after EW symmetry breaking, we follow the discussion in \cite{benj}. Several operators in (\ref{eq:silh}) give contributions to the kinetic terms of SM fields ({\it e.g.} $\bar c_{\sss H}$ modifies the kinetic term for the Higgs field), and we will discuss their impact in the next section, in the context of concrete BSM scenarios. In the following, we consider the $n$-point interactions with $n = 3,4$. 
The CP-conserving 3-point interactions involving at least one light Higgs scalar $h$ are obtained from ${\cal L}_{\rm SILH}$ and read 
\be\bsp
  {\cal L}_{3h} = &\ 
    - \frac{m_{\sss H}^2\, g^{(1)}_{\sss hhh}}{2 v} \,h^3 + \frac{g^{(2)}_{\sss hhh}}{2} \, h\partial_\mu h \partial^\mu h - \frac{g_{\sss hgg}}{4} \, G^a_{\mu\nu} G_a^{\mu\nu} h
     - \frac{g_{\sss h\gamma\gamma}}{4} \, F_{\mu\nu} F^{\mu\nu} h
\\ &\
    - \frac{g_{\sss hww}^{(1)}}{2} \, W^{\mu\nu} W^\dag_{\mu\nu} h - \Big[g_{\sss hww}^{(2)} W^\nu \partial^\mu W^\dag_{\mu\nu} h + {\rm h.c.} \Big]
    +  g_{\sss hww}^{(3)} W_\mu^\dag  W^\mu h
\\ &\ 
    - \frac{g_{\sss hzz}^{(1)}}{4} \, Z_{\mu\nu} Z^{\mu\nu} h
    - g_{\sss hzz}^{(2)} \, Z_\nu \partial_\mu Z^{\mu\nu} h
    + \frac{g_{\sss hzz}^{(3)}}{2} \, Z_\mu Z^\mu h
\\ &\
    - \frac{g_{\sss haz}^{(1)}}{2}\, Z_{\mu\nu} F^{\mu\nu} h
    - g_{\sss haz}^{(2)} \, Z_\nu \partial_\mu F^{\mu\nu} h \ .
\esp\label{eq:massbasis}
\ee
where we have introduced abelian field-strength tensors $W_{\mu\nu}$, $Z_{\mu\nu}$ and $F_{\mu\nu}$ for the $W$-boson, $Z$-boson and photon respectively, and $G^a_{\mu\nu}$ is still
the non-abelian gluon field-strength tensor. The CP-conserving 4-point interactions involving at least one light Higgs scalar $h$ similarly read 
\be\bsp
  {\cal L}_{4h} = &\
    - \frac{m_{\sss H}^2\, g_{\sss hhhh}^{(1)}}{8 v^2}\, h^4 + \frac{g_{\sss hhhh}^{(2)}}{2} \, h^2 \partial_\mu h\partial^\mu h - \frac{g_{\sss hhgg}}{8} \, G^a_{\mu\nu} G_a^{\mu\nu} h^2
    - \frac{g_{\sss hh\gamma\gamma}}{8} \, F_{\mu\nu} F^{\mu\nu} h^2    
\\ &\ 
    - \frac{g_{\sss hhww}^{(1)}}{4} \,  W^{\mu\nu} W^\dag_{\mu\nu} h^2 - \frac12 \Big[g_{\sss hhww}^{(2)} W^\nu \partial^\mu W^\dag_{\mu\nu} h^2 + {\rm h.c.} \Big]
    + \frac{g_{\sss hhww}^{(3)}}{2}  \, W_\mu^\dag  W^\mu h^2
\\ &\
    - \frac{g_{\sss hhzz}^{(1)}}{8}\,  Z_{\mu\nu} Z^{\mu\nu} h^2 - \frac{g_{\sss hhzz}^{(2)}}{2} \, Z_\nu \partial_\mu Z^{\mu\nu} h^2
    + \frac{g_{\sss hhzz}^{(3)}}{4} \, Z_\mu Z^\mu h^2 - \frac{g_{\sss hhaz}^{(1)}}{4} \, Z_{\mu\nu} F^{\mu\nu} h^2
\\ &\
     - \frac{g_{\sss hhaz}^{(2)}}{2} \, Z_\nu \partial_\mu F^{\mu\nu} h^2 - i g_{\sss haww}^{(1)} F^{\mu\nu} W_\mu W^\dag_\nu h + \Big[i g_{\sss haww}^{(2)} W^{\mu\nu} A_\mu W^\dag_\nu h + {\rm h.c.} \Big]
\\ &\
    + i g_{\sss haww}^{(3)} A_\mu W_\nu W^\dag_\rho \big[ \eta^{\mu\rho}\partial^\nu h - \eta^{\mu\nu}\partial^\rho h\big] 
    - i g_{\sss hzww}^{(1)} Z^{\mu\nu} W_\mu W^\dag_\nu  h
\\ &\
        + \Big[i g_{\sss hzww}^{(2)} W^{\mu\nu} Z_\mu W^\dag_\nu h + {\rm h.c.} \Big]
        - i g_{\sss hzww}^{(3)} Z_\mu W_\nu W^\dag_\rho \big[ \eta^{\mu\rho}\partial^\nu h - \eta^{\mu\nu}\partial^\rho h\big] \ .
\esp\label{eq:massbasis4}\ee
We note that the 4-point interactions involving one Higgs $h$ and three gluons (relevant, {\it e.g.} for Higgs production in association with a jet via gluon fusion) 
are already taken into account by the $g_{\sss hgg}$ term in (\ref{eq:massbasis}). In a similar fashion to (\ref{eq:massbasis}) and (\ref{eq:massbasis4}), the CP-odd 3- and 4-point interactions involving at least one light Higgs scalar $h$ are obtained from ${\cal L}_{\rm CP}$, yielding 
\be\bsp
  \tilde{{\cal L}}_{h} = &\ 
    - \frac{\tilde g_{\sss hgg}}{4}  G^a_{\mu\nu} \tilde G^{\mu\nu} h 
    - \frac{\tilde g_{\sss h\gamma\gamma}}{4} F_{\mu\nu} \tilde F^{\mu\nu} h - \frac{\tilde g_{\sss hww}}{2}  W^{\mu\nu} \tilde W^\dag_{\mu\nu} h - \frac{\tilde g_{\sss hzz}}{4} Z_{\mu\nu} \tilde Z^{\mu\nu} h  - \frac{\tilde g_{\sss haz}}{2}  Z_{\mu\nu} \tilde F^{\mu\nu} h
\\ &\ 
       - \frac{\tilde g_{\sss hhgg}}{8}  G^a_{\mu\nu} \tilde G_a^{\mu\nu} h^2 
     - \frac{\tilde g_{\sss hh\gamma\gamma}}{8}  F_{\mu\nu} \tilde F^{\mu\nu} h^2\ - \frac{\tilde g_{\sss hhww}}{4} W^{\mu\nu} \tilde W^\dag_{\mu\nu} h^2 - \frac{\tilde g_{\sss hhzz}}{8}  Z_{\mu\nu} \tilde Z^{\mu\nu} h^2  
\\ &\
     - \frac{\tilde g_{\sss hhaz}}{4}  Z_{\mu\nu} \tilde F^{\mu\nu} h^2 + i \tilde g_{\sss haww}^{(1)} \tilde F^{\mu\nu} W_\mu W^\dag_\nu h + \Big[i \tilde g_{\sss haww}^{(2)} \tilde W^{\mu\nu} A_\mu W^\dag_\nu h + {\rm h.c.} \Big]
\\ &\
    + i \tilde g_{\sss hzww}^{(1)} \tilde Z^{\mu\nu} W_\mu W^\dag_\nu  h - \Big[i \tilde g_{\sss hzww}^{(2)} \tilde W^{\mu\nu} Z_\mu W^\dag_\nu h + {\rm h.c.} \Big]\,.
\esp\label{eq:massbasisCPodd}
\ee
Finally, the 3- and 4-point gauge boson self-interactions receive CP-even contributions from both ${\cal L}_{\rm SILH}$ and ${\cal L}_{G}$ (they also receive CP-odd contributions from ${\cal L}_{\rm CP}$, which we do not include here). We adopt here the parametrization of triple gauge couplings (TGCs) from \cite{Hagiwara:1986vm,Hagiwara:1993ck}, which yields
\be\bsp
  {\cal L}_{3V} &= e \,g_{1}^{\gamma}\Big[ i \,W^\dag_{\mu\nu} A^\mu W^\nu \hc \Big] + g\,\cW\, g_{1}^{\sss Z} \Big[ i \, W^\dag_{\mu\nu} Z^\mu W^\nu \hc \Big]
\\&\
     + e\,\kappa_{\gamma} \, \left( i\, F_{\mu\nu} W^\mu W^{\nu\dag}\right) + g\,\cW\, \kappa_{\sss Z} \left( i\, Z_{\mu\nu} W^\mu W^{\nu\dag}\right) 
\\&\ 
     + \frac{e\,\lambda_{\gamma}}{\mW^2}\, i \,W_{\mu\nu}W^\dag_{\nu\rho}F_{\rho\mu} + \frac{g\,\cW\,\lambda_{\sss Z}}{\mW^2}\, i \,W_{\mu\nu}W^\dag_{\nu\rho}Z_{\rho\mu} \ ,
\esp\label{eq:l3v}\ee
with $g_{1}^{\gamma} = g_{1}^{\sss Z} = \kappa_{\gamma} = \kappa_{\sss Z} = 1$ and $\lambda_{\gamma} = \lambda_{\sss Z} = 0$ for the SM. 
$U(1)_{\mathrm{EM}}$ gauge invariance imposes $g_{1}^{\gamma} = 1$, while the (spontaneously broken) $SU(2)_L \times U(1)_{Y}$ gauge invariance combined with the restriction to 
$D \leq 6$ effective operators in ${\cal L}_{\rm Eff}$ lead to the relations \cite{Hagiwara:1993ck}
\be
g_{1}^{\sss Z} = \kappa_{\sss Z} + \frac{s^2_{\sss W}}{c^2_{\sss W}}\left( \kappa_{\gamma} -1 \right) \quad , \quad \quad \lambda_{\gamma} = \lambda_{\sss Z} \,.
\ee

\begin{table}[h] 
\begin{center}
\renewcommand{\arraystretch}{2} 
\begin{tabular}{c} 
{\bf{\large${\cal L}_{3h}$} Couplings} {\it vs}  {\bf$SU(2)_L \times U(1)_Y$  ($D \leq 6$) Wilson Coefficients} \\
\hline
$g^{(1)}_{\sss hhh} = 1 + \frac{5}{2} \, \bar c_{\sss 6}$\, , \quad $g^{(2)}_{\sss hhh} = \frac{g}{m_W} \, \bar c_{\sss H}$\,, \quad 
$g_{\sss hgg} = g^{\mathrm{SM}}_{\sss hgg} -  \frac{4\,g^2_s\,v\,\bar c_g}{m^2_W}$\, , \quad 
$g_{\sss h\gamma\gamma} = g^{\mathrm{SM}}_{\sss h\gamma\gamma} -  \frac{8\,g\,s^2_W\,\bar c_\gamma}{m_W}$ \\

$g_{\sss hww}^{(1)} = \frac{2 g}{\mW} \bar c_{\sss HW}$\, , \quad $g_{\sss hzz}^{(1)} = g_{\sss hww}^{(1)} + \frac{2 g}{\cW^2 \mW} \Big[ \bar c_{\sss HB} \sW^2 - 4 \bar c_{\sss \gamma} \sW^4\Big]$\, , \quad $g_{\sss hww}^{(2)} = \frac{g}{2\,\mW} \Big[ \bar c_{\sss W} + \bar c_{\sss HW} \Big]$ \\
 
$g_{\sss hzz}^{(2)} = 2\,g_{\sss hww}^{(2)} + \frac{g\,\sW^2}{\cW^2 \mW} \Big[(\bar c_{\sss B} + \bar c_{\sss HB})\Big]$\, ,  \quad $g_{\sss hww}^{(3)} = g\,\mW$\, , \quad 
$g_{\sss hzz}^{(3)}=\frac{g_{\sss hww}^{(3)}}{\cW^2}  (1- 2\, \bar c_{T})$ \\ 
$g_{\sss haz}^{(1)} = \frac{g \,\sW}{\cW\,\mW} \Big[ \bar c_{\sss HW} - \bar c_{\sss HB} + 8\, \bar c_{\gamma}\, \sW^2 \Big]$\, , \quad  
$g_{\sss haz}^{(2)}=\frac{g \,\sW}{\cW\,\mW} \Big[ \bar c_{\sss HW} - \bar c_{\sss HB} - \bar c_{\sss B}  + \bar c_{\sss W} \Big]$\\

\hline 
{\bf{\large${\cal L}_{4h}$} Couplings} {\it vs}  {\bf$SU(2)_L \times U(1)_Y$  ($D \leq 6$) Wilson Coefficients} \\
\hline

$g^{(1)}_{\sss hhhh} = 1 + \frac{15}{2} \, \bar c_{\sss 6}$\,  , \quad $g^{(2)}_{\sss hhhh} = \frac{g^2}{4\,m^2_W} \, \bar c_{\sss H}$\, , \quad  
$g_{\sss hhgg} = - \frac{4\,g^2_s\,\bar c_g}{m^2_W}$\, , \quad  
$g_{\sss hh\gamma\gamma} = -  \frac{4\,g^2\,s^2_W\,\bar c_\gamma}{m^2_W}$ \\

$g^{(1,2)}_{\sss hhxy} = \frac{g}{2\,\mW}\, g^{(1,2)}_{\sss hxy}$ \quad $\left(\,x,y = W,Z,\gamma\right)$\,  , \quad $g_{\sss hhww}^{(3)} = \frac{g^2}{2}$\, , \quad $g_{\sss hhzz}^{(3)} = \frac{g_{\sss hhww}^{(3)}}{\cW^2}  (1- 6\, \bar c_{T})$  \\ 

$g_{\sss haww}^{(1)} = \frac{g^2 \,\sW}{\mW} \Big[2\,\bar c_{\sss W} + \bar c_{\sss HW} + \bar c_{\sss HB}\Big]$\, , \quad $g_{\sss hzww}^{(1)} = \frac{g^2 }{\cW\,\mW} \Big[\cW^2 \,\bar c_{\sss HW} - \sW^2 \bar c_{\sss HB} + (3- 2\sW^2)\,\bar c_{\sss W} \Big]$ \\ 
$g_{\sss haww}^{(2)} = \frac{2\,g^2 \,\sW}{\mW} \,\bar c_{\sss W}$\, , \quad $g_{\sss hzww}^{(2)} = \frac{g^2 }{\cW\,\mW} \Big[\bar c_{\sss HW} + (3- 2\sW^2)\,\bar c_{\sss W} \Big]$ 
\\

$g_{\sss haww}^{(3)} = \frac{g^2 \,\sW}{\mW} \Big[\bar c_{\sss W} + \bar c_{\sss HW}\Big]$\, , \quad
$g_{\sss hzww}^{(3)} = \frac{\sW}{\cW}\,g_{\sss haww}^{(3)} $\\

\hline 
{\bf{\large$\tilde {\cal L}_{h}$} Couplings} {\it vs}  {\bf$SU(2)_L \times U(1)_Y$  ($D \leq 6$) Wilson Coefficients} \\
\hline

$\tilde g_{\sss hgg} =-  \frac{4\,g^2_s\,v\,\tilde c_g}{m^2_W}$\, , \quad $\tilde g_{\sss h\gamma\gamma} = -  \frac{8\,g\,s^2_W\,\tilde c_\gamma}{m_W}$\,  ,  \quad 
$\tilde g_{\sss hhgg} = -  \frac{4\,g^2_s\,\tilde c_g}{m^2_W}$\, , \quad 
$\tilde g_{\sss hh\gamma\gamma} = -  \frac{4\,g^2\,s^2_W\,\tilde c_\gamma}{m^2_W}$ \\

$\tilde g_{\sss hww} = \frac{2 g}{\mW} \tilde c_{\sss HW}$\, , \quad $\tilde g_{\sss hzz} = \tilde g_{\sss hww} + \frac{2 g}{\cW^2 \mW} \Big[ \tilde c_{\sss HB} \sW^2 - 4 \tilde c_{\sss \gamma} \sW^4 \Big]$  \\ 

$\tilde g_{\sss hhxy} = \frac{g}{2\,\mW}\, g_{\sss hxy}$ \quad  $\left(\,x,y = W,Z,\gamma\right)$\, ,  \quad $\tilde g_{\sss haww}^{(1)} = \frac{g^2 \,\sW}{\mW} \Big[\tilde c_{\sss HW} - \tilde c_{\sss HB} \Big]$ \\ 

$\tilde g_{\sss hzww}^{(1)} = \frac{g^2 }{\cW\,\mW} \Big[(2 - \sW^2) \,\tilde c_{\sss HW} + \sW^2 \tilde c_{\sss HB}\Big]$\, , \quad $g_{\sss hzww}^{(2)} = \frac{g^2 }{\cW\,\mW} \Big[\bar c_{\sss HW} + (3- 2\sW^2)\,\bar c_{\sss W} \Big]$ 
\\

$\tilde g_{\sss haww}^{(2)} = \frac{g^2 \,\sW}{\mW} \,\tilde c_{\sss HW}$\, , \quad
$\tilde g_{\sss hzww}^{(2)} = \frac{2 \cW }{\sW} \tilde g_{\sss haww}^{(2)}$\, ,  \quad $\tilde g_{\sss haz} = \frac{g \,\sW}{\cW\,\mW} \Big[ \tilde c_{\sss HW} - \tilde c_{\sss HB} + 8\, \tilde c_{\gamma}\, \sW^2 \Big]$\\
\hline
\end{tabular}
\end{center}
%\etb
%\captionsetup{singlelinecheck=off}
\caption[ . ]{Relations between the different couplings appearing in ${\cal L}_{3h}$, ${\cal L}_{4h}$, $\tilde {\cal L}_{h}$ and the Wilson coefficients for the $D \leq 6$ effective operators in $\mathcal{L}_{\mathrm{Eff}}$.}
\label{Table1}
\end{table}

\begin{table}[h] 
\begin{center}
\renewcommand{\arraystretch}{2} 
\begin{tabular}{c} 
{\bf{\large${\cal L}_{3V}$} Couplings} {\it vs}  {\bf$SU(2)_L \times U(1)_Y$  ($D \leq 6$) Wilson Coefficients} \\
\hline
$g_{1}^{\sss Z} = 1 - \frac{1}{\cW^2} \Big[ \bar c_{\sss HW} - (2\sW^2 -3) \bar c_{\sss W}\Big]$\, ,  \quad 
$\kappa_{\sss Z} = 1 - \frac{1}{\cW^2} \Big[ \cW^2 \bar c_{\sss HW} - \sW^2 \bar c_{\sss HB} - (2\sW^2 -3) \bar c_{\sss W} \Big]$ \\

$g_{1}^{\gamma} = 1 $\, ,  \quad $\kappa_{\gamma} = 1 -2 \,\bar c_{\sss W}  - \bar c_{\sss HW} - \bar c_{\sss HB}$\, ,  \quad $\lambda_{\gamma} = \lambda_{\sss Z} = 3 \,g^2 \,\bar c_{\sss 3W} $\\

\hline 
{\bf{\large${\cal L}_{4V}$} Couplings} {\it vs}  {\bf$SU(2)_L \times U(1)_Y$  ($D \leq 6$) Wilson Coefficients} \\
\hline
$g_{2}^{\sss W} = 1 - 2 \,\bar c_{\sss HW} - 4 \,\bar c_{\sss W}$\, , \quad   
$g_{2}^{\sss Z} = 1 - \frac{1}{\cW^2} \Big[ 2\, \bar c_{\sss HW} + 2 \,(2 -\sW^2) \,\bar c_{\sss W} \Big]$ \\

$g_{2}^{\gamma} = 1 $\, , \quad   
$g_{2}^{\gamma\sss Z} = 1 - \frac{1}{\cW^2} \Big[ \bar c_{\sss HW} + (3 -2\sW^2) \,\bar c_{\sss W} \Big]$ \\

$\lambda_{\sss W} = \lambda_{\gamma \sss W}  = \lambda_{\gamma \sss Z} = \lambda_{\sss WZ}  = 6 \,g^2 \,\bar c_{\sss 3W}$ \\
\hline
\end{tabular}
\end{center}
%\etb
%\captionsetup{singlelinecheck=off}
\caption[ . ]{The Trilinear and Quartic Gauge Couplings appearing in ${\cal L}_{3V}$ and ${\cal L}_{4V}$ as a function of the Wilson coefficients for the $D \leq 6$ effective operators in $\mathcal{L}_{\mathrm{Eff}}$.}
\label{Table2}
\end{table}  

These degeneracies
%\footnote[3]{$D = 8$ effective operators such as $(D_\mu \Phi)^\dag (D_\nu \Phi)\,  \Phi^\dag \big[D^\mu, D^\nu \big] \Phi$ break these degeneracies \cite{Hagiwara:1993ck}. However, since these operators will be suppressed by extra powers of the cut-off scale  of the effective theory $\Lambda$, the degeneracies are expected to hold to a very good approximation.} 
imply that there can be at most three independent contributions to anomalous TCGs from ${\cal L}_{\rm SILH}$ and ${\cal L}_{G}$.
The operators proportional to $g_{1}^{\sss Z}$, $\kappa_{\sss Z}$ and $\kappa_{\gamma}$ in (\ref{eq:l3v}) contain one derivative and are obtained from ${\cal L}_{\rm SILH}$, 
which then gives rise at most to two independent contributions to anomalous TGCs. The last two operators in (\ref{eq:l3v}) contain three derivatives and are obtained from ${\cal L}_{G}$, which then gives rise to a sole linearly independent contribution to anomalous TGCs. This is clear for the operator $\bar c_{\sss 3W}\mathcal{O}_{3W}$, as it is constructed from three gauge field-strengths and only their abelian parts can contribute to TGCs (after EW symmetry breaking it directly leads to the last two operators in (\ref{eq:l3v}) with $\lambda_{\gamma} = \lambda_{\sss Z}$). The 
operator $\bar c_{\sss 2W}\mathcal{O}_{DW}$ also leads after EW symmetry breaking to those operators with $\lambda_{\gamma} = \lambda_{\sss Z}$, but also produces operators of the form  
\be
g \, \epsilon_{ijk} \left(\square W^i_{\mu\nu}\right) W^j_{\mu}  W^k_{\nu}
\ee
These operators however either vanish or reduce to operators from (\ref{eq:l3v}) for on-shell gauge bosons (as well as for the virtual photon). 
%Also, while $\mathcal{O}_{3W}$ only contributes to $n$-point gauge interactions with $n \geq 3$, the operator $\mathcal{O}_{DW}$ also gives a tree-level contribution to the gauge boson propagators via a term $W_{\mu\nu}^a \square W_a^{\mu\nu}$, and the size of its Wilson coefficient $\bar c_{\sss 2W}$ is severely constrained from LEP1 measurements \cite{De Rujula:1991se}. 
Therefore, the parametrization of TGCs (\ref{eq:l3v}) holds. Moreover, we can safely disregard the contribution of 
$\bar c_{\sss 2W}$ to $\lambda_{\gamma} = \lambda_{\sss Z}$ in the following discussion, since this operator can be re-casted as (very constrained) fermionic currents and the operators involving the Higgs we have already considered~\cite{De Rujula:1991se}. 
For the case of quartic gauge couplings (QGCs) we have
\begin{small}
\bea
\label{eq:l4v}
  {\cal L}_{4V}& = & \frac{g^2 g_2^{\sss W}}{2} \Big[ W_\mu W^\mu W^\dag_\nu W^{\nu\dag} - W^\mu W^{\dag}_{\mu} W^\nu W^\dag_\nu  \Big]
    + e^2 g_{2}^{\gamma} \Big[A_\mu A^\nu W^\dag_\nu W^\mu - A^\mu A_\mu W^\nu W^\dag_\nu  \Big] \nonumber \\
  & + & 2\, e\, g\, \cW g_{2}^{\gamma \sss Z} \Big[A_\mu Z^\nu W^\dag_\nu W^\mu - A^\mu Z_\mu W^\nu W^\dag_\nu  \Big]
  + g^2 \cW^2  g_{2}^{\sss Z} \Big[Z_\mu Z^\nu  W^\dag_\nu W^\mu - Z^\mu Z_\mu W^\nu W^\dag_\nu   \Big] \nonumber \\
  & + &\frac{e\, g\, \cW \lambda_{\gamma\sss Z}}{\mW^2} \Big[ W_\mu W^{\mu\rho\dag} \left(A_\nu Z^{\nu\rho} + Z_\nu F^{\nu\rho} \right) - W_\mu W^{\nu\rho\dag} \left(A_\nu Z^{\mu\rho} + Z_\nu F^{\mu\rho} \right)  \Big] \nonumber \\
  & + &\frac{g^2 \lambda_{\sss W}}{2\,\mW^2} \Big[ W_\mu W^{\mu\rho\dag} W^\dag_\nu W^{\nu\rho} - W_\mu W^{\nu\rho\dag} W^\dag_\nu W^{\mu\rho}  \Big]\\
  & + &\frac{e^2 \lambda_{\gamma\sss W}}{\mW^2}  \Big[ W_\mu W^{\mu\rho\dag} A_\nu F^{\nu\rho} - W_\mu W^{\nu\rho\dag} A_\nu F^{\mu\rho}  \Big] \nonumber \\ 
  & + &\frac{g^2 \cW^2 \lambda_{\sss WZ}}{\mW^2} \Big[ W_\mu W^{\mu\rho\dag} Z_\nu Z^{\nu\rho} - W_\mu W^{\nu\rho\dag} Z_\nu Z^{\mu\rho}  \Big] \nonumber
\eea
\end{small}

\noindent where again $g_2^{\sss W} = g_2^{\gamma} = g_2^{\sss Z} = g_2^{\gamma \sss Z} = 1$ and $\lambda_{\sss W} = \lambda_{\gamma\sss W} = \lambda_{\gamma\sss Z} = \lambda_{\sss WZ} = 0$ corresponds to the SM case.

\vspace{3mm}

The correspondence between the Wilson coefficients $\bar c_i$ for the effective operators from (\ref{eq:silh}), (\ref{eq:lagG}), (\ref{eq:silhCPodd}) and the various couplings arising after EW symmetry breaking in (\ref{eq:massbasis}-\ref{eq:l3v}), (\ref{eq:l4v}) is shown in Tables \ref{Table1} and \ref{Table2}. In the next section, we compute the values of these couplings by direct matching in the 
broken EW theory, for the specific setup of Benchmark A in \ref{BenchmarkA}: the 2HDM in the alignment limit $c_{\beta -\alpha} = 0$. 

\subsection{A Specific Example: 2HDM in the Alignment Limit $c_{\beta -\alpha} = 0$}

\begin{figure}[ht!]
\begin{center}
\includegraphics[width=0.23\textwidth]{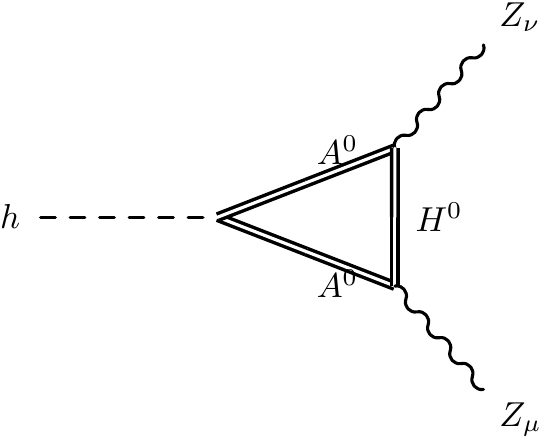} \hspace{4mm}
\includegraphics[width=0.23\textwidth]{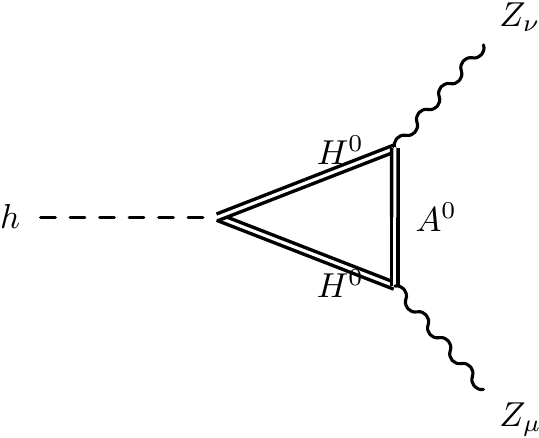} \hspace{4mm}
\includegraphics[width=0.23\textwidth]{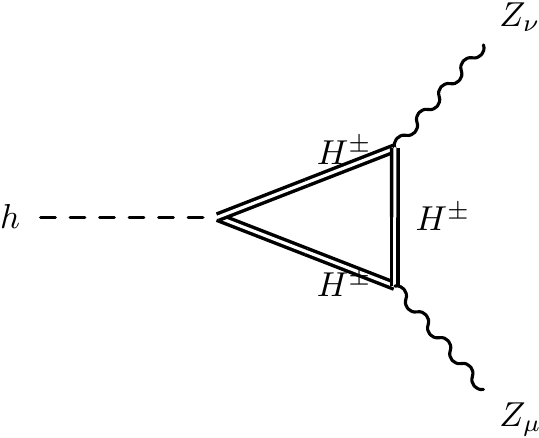}  
\includegraphics[width=0.23\textwidth]{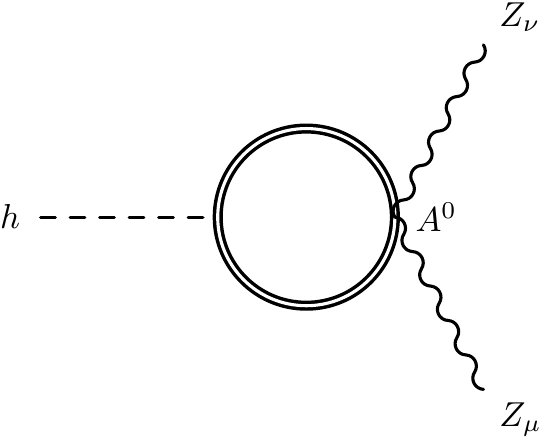} \hspace{4mm}
\includegraphics[width=0.23\textwidth]{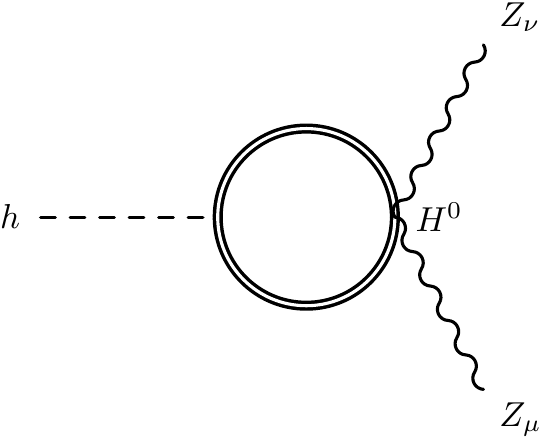} \hspace{4mm}
\includegraphics[width=0.23\textwidth]{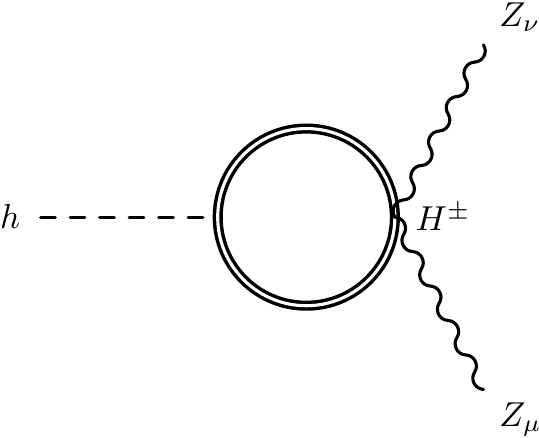}
\caption{1-loop Feynman diagrams involving the $H^{\pm},A^0,H^0$ scalars and contributing to the $Z_{\mu}Z_{\nu}\,h$ vertex in the alignment limit $c_{\beta-\alpha} = 0$}
\label{fig:Feynman}
\end{center}
\end{figure}

Here we will obtain the values of the various couplings from (\ref{eq:massbasis}) for the 2HDM in the alignment limit $c_{\beta -\alpha} = 0$, by direct matching in the 
broken EW theory. We focus on the $V_{\mu}V_{\nu}\,h$ interaction vertices, which receive contributions from loops of scalars $H^{\pm},A^0,H^0$ (these are shown in Figure \ref{fig:Feynman} 
for the case $V_{\mu} = Z_{\mu}$). We note that loops involving only SM particles cancel in the matching between full and effective theories and may then be disregarded, while loops involving both SM particles and $H^{\pm},A^0,H^0$ scalars vanish in the alignment limit. 

We compute the 1-loop contributions to $V_{\mu}V_{\nu}\,h$ and expand the result in powers of the 4-momenta of the $Z$-bosons $p_1,p_2$.  
By truncating the expansion at order $\mathcal{O}(p^2)$, we obtain
\bea
\Delta \mathcal{L} = \Delta \mathcal{L}_{W} + \Delta \mathcal{L}_{Z} + \Delta \mathcal{L}_{\gamma}   
\eea
with 
\bea
\label{LWWh}
\Delta \mathcal{L}_{W} &=& -\frac{g^{(1)}_{hww}}{2}\, W^{\mu\nu}W^{\dagger}_{\mu\nu} h - \left[ g^{(2)}_{hww}\,W^{\nu}\partial^{\mu}W^{\dagger}_{\mu\nu}h + \mathrm{h.c.}\right] + g^{(3)}_{hww} \,W^{\mu}W^{\dagger}_{\mu}h \nonumber \\
&+& g^{(4)}_{hww} \, (\partial_{\mu}W^{\dagger}_{\nu})(\partial^{\mu}W^{\nu}) h + \left[ g^{(5)}_{hww}\, W^{\nu}\partial^{\mu}\partial_{\mu}W^{\dagger}_{\nu}h + \mathrm{h.c.}\right] \\
&+& g^{(6)}_{hww}\, \partial^{\mu}W_{\mu}\partial^{\nu}W^{\dagger}_{\nu} h \nonumber
\eea
\bea
\label{LZZh}
\Delta \mathcal{L}_{Z} & = &-\frac{g^{(1)}_{hzz}}{4}\, Z_{\mu\nu}Z^{\mu\nu} h - g^{(2)}_{hzz}\,Z_{\nu}\partial_{\mu}Z^{\mu\nu}h + \frac{g^{(3)}_{hzz}}{2} \,Z_{\mu}Z^{\mu}h \nonumber \\
& + &g^{(4)}_{hzz} \, (\partial_{\mu}Z_{\nu})(\partial^{\mu}Z^{\nu}) h + g^{(5)}_{hzz}\, Z_{\nu}\partial_{\mu}\partial^{\mu}Z^{\nu}h + g^{(6)}_{hzz}\, \partial_{\mu}Z^{\mu}\partial_{\nu}Z^{\nu} h
\eea
\bea
\label{Lgammagammah}
\Delta \mathcal{L}_{\gamma} = -\frac{g^{(1)}_{h\gamma\gamma}}{4}\, F_{\mu\nu}F^{\mu\nu} h -\frac{g^{(1)}_{h\gamma z}}{2}\, Z_{\mu\nu}F^{\mu\nu} h - g^{(2)}_{h\gamma z}\,Z_{\nu}\partial_{\mu}F^{\mu\nu}h
\eea
with the last three terms in both (\ref{LWWh}) and (\ref{LZZh}) corresponding to {\it e.o.m.}-vanishing terms, which are nevertheless generated by the off-shell 1-loop corrections.
Performing an expansion to linear order in $1- m^2_{H^0}/m^2_{H^{\pm}} = 1-x_0$ and $1- m^2_{A^0}/m^2_{H^{\pm}} =1-x_A$ yields for $\Delta \mathcal{L}_{W}$ 
\bea
\label{gWWh1}
g^{(1)}_{hww} & = &\frac{-g^2\,v}{192 \, \pi^2\,m^2_{H^{\pm}}} \left[\frac{g_0 + g_A + 2 g_+}{2}+ (1-x_0)\frac{4g_0 + g_+}{10} + (1-x_A)\frac{4g_A + g_+}{10} 
\right] \\
g^{(2,6)}_{hww} & = & \frac{g^{(4)}_{hww}}{2}  = \frac{2\,g^{(5)}_{hww}}{5}  = \frac{g^2\,v}{192 \, \pi^2\,m^2_{H^{\pm}}} \left[(1-x_0)\frac{g_0 - g_+}{20} + (1-x_A)\frac{g_A - g_+}{20}
\right] \\
\label{gWWh3}
g^{(3)}_{hww} & = & \frac{g^2\,v}{192 \, \pi^2} \left[(1-x_0) (g_+- g_0) + (1-x_A) (g_+- g_A)
\right] 
\eea
with $g_0 \equiv g_{hH^0H^0}/v$, $g_A \equiv g_{hA^0A^0}/v$ and $g_+ \equiv g_{hH^+H^-}/v$, and the trilinear scalar couplings $g_{hH^+H^-}$, $g_{hH^0H^0}$ and $g_{hA^0A^0}$ given in the EW broken 
theory with $c_{\beta -\alpha} = 0$ by
\bea 
\label{trilinears}
g_{hH^+H^-} &= &\frac{\left(-2\,M^2 + 2\, m_{H^{\pm}}^2  + m^2_{h} \right)}{v} = \tilde{\lambda}_3\, v \nonumber \\
g_{hH^0H^0} &= &\frac{\left(-2\,M^2 + 2\, m_{H^{0}}^2  + m^2_{h} \right)}{v} = (\tilde{\lambda}_3 + \tilde{\lambda}_4 +\tilde{\lambda}_5)\, v \\
g_{hA^0A^0} &= &\frac{\left(-2\,M^2 + 2\, m_{A^{0}}^2  + m^2_{h} \right)}{v} = (\tilde{\lambda}_3 + \tilde{\lambda}_4 -\tilde{\lambda}_5)\, v \nonumber
\eea
By means of (\ref{trilinears}), the relations (\ref{gWWh1} - \ref{gWWh3}) may be written as 
\bea
\label{gWWh1_lambda}
g^{(1)}_{hww} & = &\frac{-g^2\,v}{192 \, \pi^2\,m^2_{H^{\pm}}} \left[(2 \tilde{\lambda}_3 + \tilde{\lambda}_4) + \mathcal{O}\left[(1-x_0),(1-x_A)\right]  \right] \\
g^{(2,4-6)}_{hww} & \sim & \mathcal{O}\left[(1-x_0)^2,(1-x_A)^2\right] \\
\label{gWWh3_lambda}
g^{(3)}_{hww} & = & \frac{g^2\,v}{192 \, \pi^2} \left[(1-x_0)(\tilde{\lambda}_4 + \tilde{\lambda}_5) + (1-x_A)(\tilde{\lambda}_4 - \tilde{\lambda}_5)\right]
\eea
Performing a similar expansion in $\Delta \mathcal{L}_{Z}$ then yields 
\bea
\label{gZZh1}
g^{(1)}_{hzz} & = &\frac{-g^2\,v}{192 \, \pi^2\,m^2_{H^{\pm}}} \left[\frac{g_0 + g_A}{2\,c^2_W} + g_+ (c^2_W - 2\,s^2_W) + (1-x_0)\frac{4g_0 + g_+}{10} \right. \nonumber \\
& & + \left. (1-x_A)\frac{4g_A + g_+}{10} 
\right] \\
g^{(2,6)}_{hzz} & = &  \frac{g^{(4)}_{hzz}}{2}  = \frac{2\,g^{(5)}_{hzz}}{5}  = \frac{g^2\,v}{192 \, \pi^2\,m^2_{H^{\pm}}} \left[\frac{(x_A-x_0)(g_0 - g_A)}{40\,c^2_W} \right] \\
\label{gZZh3}
g^{(3)}_{hzz} & = & \frac{g^2\,v}{192 \, \pi^2\, c^2_W} \left[(x_0-x_A)(g_0 - g_A)
\right]
\eea
while for $\Delta \mathcal{L}_{\gamma}$ we obtain 
\be
\label{gaah1}
g^{(1)}_{h\gamma\gamma} = \frac{g^2\, v \,s^2_W\,g_{+}}{64\, \pi^2\, m^2_{H^{\pm}}}\,,\quad \quad g^{(1)}_{h\gamma z} = 
\frac{g^2\, v \,s_W \,(c^2_W - s^2_W)\, g_{+}}{128\, \pi^2\, m^2_{H^{\pm}}\, c_W} 
\, , \quad \quad g^{(2)}_{h\gamma z} = 0
\ee
Use of (\ref{trilinears}) in (\ref{gZZh1} - \ref{gaah1}) yields
\bea
\label{gZZh1_lambda}
g^{(1)}_{hzz} & = & g^{(1)}_{hww} - \frac{g^2\,v}{192 \, \pi^2\,m^2_{H^{\pm}}} \left[ \frac{(-2 \tilde{\lambda}_3 + \tilde{\lambda}_4)\, s^2_W}{c^2_W} + 
\frac{3\, \tilde{\lambda}_3\, s^4_W}{c^2_W} \right] \\
g^{(2,4-6)}_{hzz} & \sim & \mathcal{O}\left[(1-x_0)^2,(1-x_A)^2\right] \\
\label{gZZh3_lambda}
g^{(3)}_{hzz} & = & \frac{g^2\,v}{192 \, \pi^2 \, c^2_W} \left[2\, (x_0-x_A)\, \tilde{\lambda}_5 
\right] \\
\label{gaah1_lambda}
g^{(1)}_{h\gamma\gamma} & = & g^{(1)}_{h\gamma z} \, t_{2W}= \frac{g^2\, v \,s^2_W\,\tilde{\lambda}_3 }{64\, \pi^2\, m^2_{H^{\pm}}} \\
\label{gaah2}
g^{(2)}_{h\gamma z} & = & 0
\eea
Altogether, these results show that $g^{(2,4-6)}_{hVV} \sim v^4/\tilde{\mu}_2^4$ at least, and only receive contributions from $D = 8$ effective operators. 
For the $D = 6$ effective operators, we use the leading order in (\ref{gWWh1_lambda} - \ref{gWWh3_lambda}) and (\ref{gZZh1_lambda} - \ref{gaah2}), and through the relations from 
Table \ref{Table1} we get
\be
\label{CgammaCWCHW}
\bar{c}_{\gamma} = \frac{m^2_W\, \tilde{\lambda}_3}{256 \, \pi^2\,m^2_{H^{\pm}}} \,,\quad \quad \bar{c}_{W} = - \bar{c}_{HW} = \frac{m^2_W\,(2\,\tilde{\lambda}_{3}+ \tilde{\lambda}_{4})}{192\, \pi^2\, m^2_{H^{\pm}}}
\ee
\be
\label{CHBCBCT}
\bar{c}_{B} = - \bar{c}_{HB} = \frac{m^2_W\,(-2\,\tilde{\lambda}_{3}+ \tilde{\lambda}_{4})}{192\, \pi^2\, m^2_{H^{\pm}}}
\,,\quad \quad
\bar{c}_W - \bar{c}_B = - (\bar{c}_{HW}-\bar{c}_{HB}) 
= \frac{8}{3} \,\bar{c}_{\gamma} 
\ee
\be
\label{CTbroken}
\bar{c}_{T} \equiv \frac{1}{2\,g\,m_W} \left(g^{(3)}_{hww} - c^2_W \,g^{(3)}_{hzz} \right) = (\tilde{\lambda}_4^2  - \tilde{\lambda}_5^2) \, \frac{v^2}{192 \, \pi^2 \,m^2_{H^{\pm}} }
\ee
which reproduce the results from (\ref{Cgamma_UB}) upon the substitution $m^2_{H^{\pm}} \rightarrow  \tilde{\mu}_2^2$.

%We now explore the different limits for the 2HDM: in the degenerate limit $m_{A^0} = m_{H^0} = m_{H^\pm} = M^2 \gg v^2$ (corresponding to $\lambda_3 - m^2_h/v^2 = \lambda_4 = \lambda_5 = 0$), the contribution in $\bar{C}_{HW}$ and 
%$\bar{C}_{HB}$ proportional to $(1-x_0)$ vanishes, while 
%
%\be
%\frac{g_+\,v}{m^2_{H^{\pm}}} = \frac{m^2_h}{m^2_{H^{\pm}}} \ll 1
%\ee
%
%In the $\mathbb{Z}_2$-symmetric case $M^2 = 0$, no decoupling limit exists \cite{Gunion:2002zf}, and $m^2_{\phi} \gg v^2$ implies  $\left|\lambda_{i}\right| \gg 1$. It is however possible to be in the non-decoupling regime and satisfy $1-x_0\ll1$ and $1-x_H\ll1$, such that the effective field theory holds. In this limit, we have 
%
%\be
%\frac{g_+\,v}{m^2_{H^{\pm}}} = \frac{(2\,m^2_{H^{\pm}}+ m^2_h)}{m^2_{H^{\pm}}} \sim 2
%\ee
%
%with the contribution proportional to $(1-x_0)$ being subdominant. In the intermediate regime, with $|m^2_{\phi} - M^2 | \lesssim v^2$ ($\left|\lambda_{i}\right| \lesssim 1$),
%
%
%\be
%\frac{g_+\,v}{m^2_{H^{\pm}}} \sim (1-x_0)
%\ee
%
%and both contributions to $\bar{C}_{HW}$ and $\bar{C}_{HB}$ are of similar size.

%\vspace{3mm}

%Considering now the $H^0,A^0,H^{\pm}$ contributions to trilinear and quartic gauge couplings at loop levelStill, since the interactions $V_{\mu}V_{\nu}\,\phi$ (with $\phi = H^0,A^0,H^{\pm}$)also vanish in the alignment limit, probing the $\Phi$ scalar states at LHC directly seems challenging, and the loop-induced effects in Higgs and gauge boson couplings could be a promising avenue for discovering these new states.

% \bibliography{mono}{}
 \providecommand{\href}[2]{#2}\begingroup\raggedright\endgroup

 \end{document}